\documentclass[aip,graphicx]{revtex4-1}

\usepackage{graphicx}
\usepackage{mathptmx}
\usepackage{amsmath}
\usepackage{physics}
\usepackage{dcolumn}
\usepackage{subcaption}
\usepackage{lineno}

\draft 

\begin{document}


\title{Reacting and Non-reacting, Three-dimensional Shear Layers with Spanwise Stretching} 



\author{Jonathan L. Palafoutas}
\email[]{jpalafou@princeton.edu}
\noaffiliation

\author{William A. Sirignano}
\noaffiliation

\affiliation{Department of Mechanical and Aerospace Engineering, University of California, Irvine}


\date{\today}

\begin{abstract}
A three-dimensional, steady, laminar shear-layer flow spatially developing under a boundary-layer approximation with mixing, chemical reaction, and imposed normal strain is analyzed. The purpose of the study is to determine conditions by which certain stretched vortex layers appearing in turbulent combustion  are the asymptotic result of a spatially developing shear flow with imposed compressive strain.    The imposed strain creates a counterflow that stretches the vorticity in the spanwise direction. The equations are reduced to a two-dimensional form for
three velocity components. The non-reactive and reactive cases of the two-dimensional form of the governing equations are solved numerically, with consideration of the several of parameter inputs such as Damk\"{o}hler number, Prandtl number, chemical composition, and free-stream velocity ratios. The analysis of the non-reactive case focuses on the mixing between hotter gaseous oxygen and cooler gaseous propane. The free-stream strain rate $\kappa^*$ is predicted by ordinary differential equations based upon the imposed spanwise pressure variation. One-step chemical kinetics are used to describe diffusion flames and multi-flame structures. The imposed normal strain rate has a significant effect on the width of downstream mixing layers as well as the burning rate. Asymptotically in the downstream direction, a constant width of the shear layer is obtained if imposed normal strain rate is constant. A similar solution with layer width growing with the square root of downstream distance is found when imposed strain rate decreases as the reciprocal of downstream distance. The reduced-order asymptotic solutions can provide useful guidance in developing flamelet models for simulations of turbulent combustion.
\end{abstract}


\maketitle 




%
%

%


\section{Introduction}

\subsection{The Role of Vortex Stretching in Turbulent Mixing and Combustion}

In order to  study turbulent combustion in practical engines via computational analysis, it is necessary to establish sub-grid flamelet models which can be coupled with large-eddy simulations.  The models must provide the burning rate based on the magnitudes of the strain rates and vorticity imposed by the larger  turbulent flow. Since the flames are known to occur at smaller scales than the largest eddies in the flow, they often cannot be resolved by direct numerical simulation, thereby requiring a separate analysis. Fortunately, the flames typically occur on small scales where laminar behavior may be assumed. The purpose of this study is to give some foundation to recent three-dimensional flamelet models \cite{Sirignano2022a, Sirignano2022b, Sirignano2022c} based on vortex stretching. Specifically, those models assume an asymptotic  form whereby the scalar variables and the three components of velocity vary with only one spatial variable. Our aim here is to show whether and how  a three-dimensional structure consisting of a shear layer with vortex stretching will asymptotically, with growing downstream distance,  yield a dependence on only the transverse position across the shear layer. Both reacting and non-reacting cases will be examined.

A useful flamelet model must have a statistically accurate representation of the relative orientations on this smallest scale of the vorticity vector, scalar gradients, and the directions of the three principal axes for strain rate.  Several studies with direct numerical simulations (DNS) exist that are helpful in understanding this important alignment issue for application to turbulent flows.  Certainly for incompressible flow and generally for variable-density flow, one principal strain rate $\gamma$ locally will be compressive, another principal strain rate $\alpha$ will be tensile, and the third can be either extensional or compressive and will have an intermediate  strain rate $\beta$ of lower magnitude than the other like strain rate. Specifically, $\alpha > \beta > \gamma , \;\alpha > 0 , \; \gamma <0,$ and, for incompressible flow, $\alpha + \beta + \gamma = 0$.  Betchov \cite{Betchov}  indicates that, for incompressible, homogeneous, isotropic turbulence, the case with $\beta > 0$ is most important for vorticity production and the turbulence energy cascade to smaller scales. Ashurst et al. \cite{Ashurst} and Nomura and Elghobashi \cite{Nomura1992} show that the vorticity alignment with the intermediate strain direction is most probable in both cases of homogeneous sheared turbulence and  isotropic turbulence but especially in the case with shear.  They find that the  intermediate strain rate is most likely to be extensive (positive). Dresselhaus \cite{Dresselhaus} predicts the tendency for alignment of the intermediate strain direction with the vorticity. Kerr \cite{Kerr} reports that large values of helicity are not found in the turbulence cascade process, thereby indicating that vorticity does not have strong alignment with the major compressive or major tensile strain direction.

Nomura and Elghobashi \cite{Nomura1993} find, for reacting turbulent flow,  that in regions of heat release and variable density, alignment of the vorticity with the most tensile strain direction can occur. As the strain rates increase, the intermediate direction becomes more favored for vorticity alignment.  The scalar gradient and the direction of compressive strain are commonly aligned \cite{Ashurst, Nomura1992, Nomura1993, Boratav1996, Boratav1998} .  There is wide agreement that the most common intermittent vortex structures in regions of high strain rate are sheets or ribbons rather than tubes.

Based on those understandings concerning vector orientations, one may construct a flamelet model based upon superposition of a strained counterflow and flow structure with vorticity. Such  models have recently been created  \cite{Sirignano2022a, Sirignano2022b, Sirignano2022c}. They are three-dimensional in the sense that three velocity components are determined, generally with some dependence on three spatial coordinates. However, the scalar variables in the counterflow configuration are dependent on one variable. Here, we attempt to provide support for the findings by showing it can easily occur as the downstream asymptote for a shear layer subject to a compressive strain in the transverse direction.  Following the guidance from the cited DNS studies, we choose a shear-layer configuration with the scalar gradient and direction of compressive normal strain aligned with each other and orthogonal to the vorticity vector. Any shear layer is also a vortex layer because vorticity is present in  the sheet-like configuration. The imposed counterflow in our situation results in a stretched vortex layer.

As noted above, there is evidence from the DNS literature that these vortex layers or sheets can exist in both non-reacting and reacting turbulent flows. Further evidence is provided by stability analyses for incompressible flows.  For non-reacting flows, Neu \cite{neu1984dynamics} works with Burgers stretched vortex sheet showing that it can be stable; however,  for sufficient vortex strength (or insufficient imposed strain rate), it becomes unstable causing formation of a periodic array of rolled-up concentrated vortices with stretched braids. Corcos and Sherman \cite{Corcos1} address stability of a purely two-dimensional viscous vortex sheet without describing it as Burgers stretched vortex sheet. The roll-up of the two-dimensional flow is examined. Corcos and Lin \cite{Corcos2} study the stability under three-dimensional perturbations. Lin and Corcos \cite{Corcos3} study the phenomenon further exploring streamwise vorticity. The general implication for turbulent combustion and flamelet theory is that a vortical layer or sheet can exist. However, it can also become unstable through the roll-up mechanism leading to larger vortex-tube structures. This is consistent with the observation from DNS \cite{Ashurst, Nomura1992, Nomura1993, Boratav1996, Boratav1998} and experiment \cite{Buch} that both vortex layers and vortex tubes can exist in the same turbulent flow field.

This study relates to some interesting classical work on temporal, viscous vortex layers and vortex tubes subject to normal strain. A finding in those studies was that a balance between the diffusion and advection of vorticity could be achieved, resulting in a steady-state solution. Burgers\cite{burgers1948mathematical}, followed by Rott\cite{rott1958viscous}, examined the axisymmetric behavior of a stretched vortex tube for incompressible flow. The stretching (extensive or tensile strain) in the direction aligned with the vorticity vector resulted in an inward swirling motion. The steady-state solution of the axisymmetric Navier-Stokes equation (known as Burgers stretched vortex tube) requires a matching of vorticity strength and viscosity such that radially outward diffusion of vorticity and radially inward advection of vorticity are in balance. The two-dimensional analog of the stretched vortex tube involves a viscous shear layer, which is simultaneously a vortex layer, subject to normal compressive strain in a direction orthogonal to the shear-layer stream direction and with the associated tensile strain aligned with the vortex vector. The solution of the steady-state configuration for this vortex layer has been attributed to unpublished work presented in lectures by Burgers\cite{burgers1951unpublished}. The two-dimensional analog is also mentioned without attribution by Batchelor\cite{batchelor1967introduction} where this layer is described as a vortex sheet. Neu\cite{neu1984dynamics} refers to this two-dimensional layer as the ``stretched Burgers vortex sheet''. Note that, in contrast, the description ``vortex sheet'' for an inviscid flow implies a mathematical discontinuity in velocity and a vortex sheet of zero thickness; see Saffman\cite{saffman1992vortex}.

The incompressible velocity field defined by Burgers stretched vortex sheet has $u_{x}$, $u_{y}$, and $u_{z}$ as the velocity components. The imposed normal strain rate $S$,
kinematic viscosity $\nu$, and free-stream velocity magnitude $U$ are taken as positive
constants. As $y \rightarrow \infty$, $u_{x}(x, y, z, t) \rightarrow U$, $\partial u_{y}/ \partial y
\rightarrow – S$, and $\partial u_{z}/ \partial z \rightarrow S$. And, as $y \rightarrow -\infty$,
$u_{x}(x, y, z, t) \rightarrow - U$, $\partial u_{y}/ \partial y \rightarrow – S$, and $\partial
u_{z}/ \partial z \rightarrow S$. Then, the exact steady-state solution to the Navier-Stokes equations is found for the velocity
components and the vorticity $\omega_{z}$ whereby

\begin{eqnarray}
u_{x} = U \erf \big(\sqrt{\frac{S}{2\nu}}y \big) \;\;\; ; \;\;\; u_{y} = - Sy \;\;\; ; \;\;\; u_{z} =
Sz \nonumber
\end{eqnarray} and
\begin{equation}
\omega_{z} = -\frac{\partial u_{x}}{\partial y}
= - U \sqrt{\frac{S}{2\pi \nu}}\exp\big(-\frac{Sy^2}{2 \nu}\big) \text{.} \nonumber
\end{equation}
Although the sheet is being stretched in the $z$-direction, diffusion of momentum and
vorticity in the $y$-direction allows a balance with advection in the $y$-direction that results in a steady solution.

Our analysis considers a steady, spatially developing shear layer in the $x$-direction with a two-dimensional imposed strain in the $yz$-plane. The imposed strain can affect the growth of shear-layer width with downstream distance. In principle, if the imposed strain rate is constant with $x$, an asymptote should be reached downstream where layer thickness becomes constant, which resembles Burgers stretched vortex sheet. Variable density with low Mach number will be considered. Both reacting and non-reacting flows
will be studied.

\subsection{Flamelet Modeling}

Laminar, two-dimensional shear layers with mixing of fuel and oxidizer and resulting diffusion
flames is a classical subject of study. See Williams\cite{williamscombustion}. Two-dimensional planar and axisymmetric counterflow configurations of laminar diffusion flames have been studied extensively: Linán\cite{linan1974asymptotic}, Peters\cite{peterscombustion}, Williams\cite{williams2000progress}, Pierce \& Moin\cite{pierce2004progress}. More recently, Rajamanickam et al.\cite{rajamanickam2019influences} considered a three-dimensional flame configuration with both counterflow and a mixing layer which resulted in a spanwise stretching; however, the use of a uniform Oseen velocity avoided the effects of shear and vorticity. In a series of papers, Sirignano\cite{sirignano2019counterflow, sirignano2021combustion, sirignano2021diffusion, sirignano2021mixing} has considered three-dimensional flame configuration with shear, counterflow (and associated stretching), mixing, and combustion. Here, we extend the work of Sirignano\cite{sirignano2021mixing} who considered similar and approximately similar solutions that reduced the order of the problem to one-dimensional equations. We will reduce the three-dimensional problem to a two-dimensional system of equations to consider spatial development of the layer.

Our formulation builds on some classical treatments of shear layers and boundary layers with variable density. Crocco\cite{crocco1932transmission} studied effects due to viscous heat generation for compressible flow over a two-dimensional flat plate. Howarth\cite{howarth1948concerning}, Illingworth\cite{illingworth1949steady}, and Stewartson\cite{stewartson1949correlated} found compressible solutions by using a modification factor for incompressible solutions for a suite of related boundary layer problems.

Some time later, combustion scientists addressed counterflows where fuel and oxidizer streams oppose each other. Linán\cite{linan1974asymptotic} found analytical solutions to such a counterflow problem using one-step Arrhenius chemistry. Bilger\cite{bilger1976structure} used a more robust expression for chemical rate. Linán \& Williams\cite{linan1993ignition} extended the problem to account for temporal variation. There is uniform agreement that the presence of strain due to counterflow inhibits flame growth and temperature. Essentially, residence time can be considered as the reciprocal of strain rate.

The recent study\cite{sirignano2021mixing} addresses three-dimensional, steady laminar flow structures with mixing, chemical reaction, normal strain and shear strain. The problem is reduced to a two-dimensional form. A one-dimensional similar solution is developed. This present study reconsiders the non-reactive and reactive case of the two-dimensional form of the governing equations. In particular, these equations are expressed as a stepwise algorithm to solve the two-dimensional flow numerically, which allows consideration of many different parameter cases related to Prandtl number, chemical composition, and free-stream velocity ratios. The analysis here of the non-reactive case focuses on the mixing between gaseous oxygen at a temperature of \(800\text{ K}\) and gaseous propane at \(\frac{2}{3}\times800 = 533\text{ K}\). The analysis of the reactive case considers both ambient streams to have a temperature of \(300\text{ K}\) with an initial peak ignitor temperature of \(2000\text{ K}\). Both cases include imposed normal strain and shear strain. Various Prandtl numbers, free-stream horizontal velocity ratios, and free-stream enthalpy ratios are considered.

In Section II, the analysis is presented. Results for three different configurations are presented in Section III with Conclusions following in Section IV.

\section{Analysis}
The selection and development of steady-state equations that govern the flow, including the continuity, momentum, energy, and species continuity equations, are discussed in the next subsection. The second subsection offers a description of the finite-difference approximations and marching scheme utilized in this study. The final subsection outlines this study's comparison with Sirignano's one-dimensional similar solution\cite{sirignano2021mixing}.

\subsection{Governing Equations}
We consider the governing equations with the boundary-layer approximation described by Sirignano\cite{sirignano2021mixing} where velocity component $u$, enthalpy $h$, and mass fraction $Y_m$ are constant with the spanwise coordinate $z$. The free-stream flows are primarily in the $x$-direction with compressive normal strain in the $y$-direction and extensional normal strain in the $z$-direction, as depicted in Fig. 1. There is no imposed pressure gradient in the $x$-direction. Any pressure gradient in the $y$-direction caused by the imposed strain will not be consequential for the $x$-momentum equation. It follows that

\begin{figure*}
    \centering
    \includegraphics[width=\textwidth]{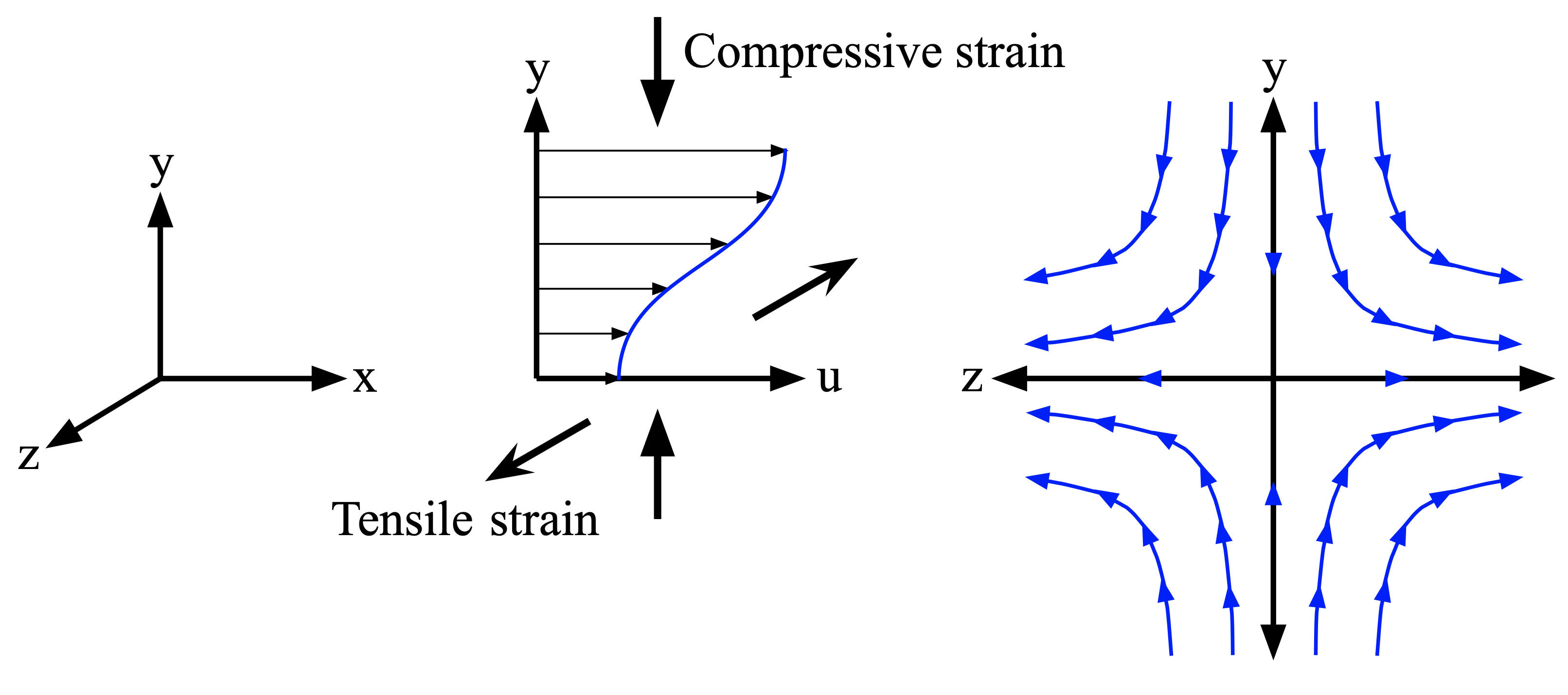}
    \caption{Schematic of shear mixing with imposed counterflow.}
    \label{fig:my_label}
\end{figure*}

\begin{equation}
\pdv{(\rho u)}{x} + \pdv{(\rho v)}{y} + \pdv{(\rho w)}{z} = 0
\end{equation}

\begin{equation}
\rho u \pdv{u}{x} + \rho v \pdv{u}{y} + \rho w \pdv{u}{z} = \frac{\partial}{\partial y}(\mu\pdv{u}{y})
\label{u}
\end{equation}

\begin{equation}
\rho u \pdv{w}{x} + \rho v \pdv{w}{y} + \rho w \pdv{w}{z} + \pdv{p}{z} =
\frac{\partial}{\partial y}(\mu\pdv{w}{y})
\end{equation}

\begin{equation}
\rho u \pdv{h}{x} + \rho v \pdv{h}{y} + \rho w \pdv{h}{z} = \frac{1}{Pr}\frac{\partial}{\partial y}(\mu\pdv{h}{y}) - \rho Q \dot{\omega}_F
\label{h}
\end{equation}

\begin{equation}
\rho u \pdv{Y_m}{x} + \rho v \pdv{Y_m}{y} + \rho w \pdv{Y_m}{z} = \frac{1}{Pr}\frac{\partial}{\partial y}(\mu\pdv{Y_m}{y}) + \rho \dot{\omega}_m;
\label{Y}
\end{equation}

\begin{equation*}
    m = \text{O}, \text{F}, \text{C}\text{O}_2, \text{H}_2\text{O} \nonumber
\end{equation*} where O and F correspond to oxidizer and fuel. $\rho\dot{\omega}_F$ gives the mass consumption rate per unit volume.

The Prandtl number ($Pr$) is assumed to be constant and equal to the Schmidt number ($Sc$), which results in a unity Lewis number ($Le = Sc/Pr$). Radiation and gravity are not considered. At the low speeds considered, heat generation via viscous dissipation is neglected.

The compressive counterflow imposed in the $y$-direction causes a symmetry for pressure and an anti-symmetry for the $w$-component of velocity in the $z$-direction. $w$ behaves according to the stagnation flow profile $w=\kappa z$ neglecting terms of $O(z^2)$ in the $yz$-plane\cite{sirignano2021mixing}. Positive values of $\kappa$ indicate that the flow field is stretched in the $z$-direction. To the same order, all other quantities are considered constant with $z$. We create a function

\begin{equation}
f(x) = \sqrt{-\pdv[2]{p}{z} \bigg|_{z=0}} > 0
\end{equation} to describe the variation of pressure in $z$ as a function of $x$. Under the boundary-layer approximation, pressure variation with $y$ is neglected in the $x$-momentum equation. So, $f$ is a function only of $x$. Therefore, the continuity equation (1) and $z$-momentum equation (3) become

\begin{equation}
\pdv{(\rho u)}{x} + \pdv{(\rho v)}{y} + \rho \kappa = 0
\label{cont}
\end{equation} and

\begin{equation}
\rho u \pdv{\kappa}{x} + \rho v \pdv{\kappa}{y} + \rho \kappa^2 - f(x)^2 =
\frac{\partial}{\partial y}(\mu\pdv{\kappa}{y})
\label{kappa}
\end{equation} respectively. $u$, $v$, $\rho$, and $\kappa$ vary with $x$ and $y$, but not with $z$.  Consistent with boundary-layer theory, Eq. (\ref{cont}) will be considered to govern velocity component $v$ while Eqs. (\ref{u}, \ref{h}, \ref{Y}, \ref{kappa}) govern $u, h, Y_m,$ and $\kappa$, respectively.

For the non-reactive case, $\dot{\omega}_F = \dot{\omega}_O = 0$. For the reactive case, a one-step chemical kinetics term appears in the energy and mass-fraction equations. One stream is composed of oxygen or a fuel-lean mixture while the other stream is composed of propane or a fuel-rich mixture. $m=\text{O}$ corresponds to $\text{O}_2$ and $m=\text{F}$ corresponds to $\text{C}_3 \text{H}_8$. The reaction rates of each gas are related by $\dot{\omega}_m = \dot{\omega}_F / \nu_m$, where $\nu_m$ and is the stoichiometric ratio between propane and gas $m$ by mass. $\nu$ is provided in Table I for each species. $Q$ is the heating value of propane per unit mass.

\begin{table*}
\caption{Chemical properties of all four species.}
\begin{ruledtabular}
\begin{tabular}{cccddd}
$m$&Molecular Formula&$M \text{ } [\frac{\text{g}}{\text{mol}}]$&\mbox{$T_c \text{ } [\text{K}]$}&\mbox{$V_c \text{ } [\frac{\text{cm}^3}{\text{g}}]$}&\mbox{$\nu$}\\
\hline
O & $\text{O}_2$ & 32 & 154.55 & 2.5 & 0.275\\

F & $\text{C}_3\text{H}_8$ & 44 & 369.15 & 4.5 & 1.\\

 & $\text{CO}_2$ & 44 & 304.15 & 2.1 & 0.334\\

 & $\text{H}_2\text{O}$ & 18 & 647.14 & 5.2 & 0.612\\

\end{tabular}
\end{ruledtabular}
\end{table*}

We follow the Westbrook \& Dryer\cite{westbrook1984chemical} one-step reaction rate for propane,

\begin{equation}
\dot{\omega}_F = - A \rho^{0.75} Y_O^{1.65} Y_F^{0.10} e^{-E_a/R_1 T_\infty h^*} \text{,}
\end{equation} where $A = 4.788 \times 10^8 \text{ }(\text{kg}/\text{m}^3)^{-0.75} / \text{s}$ is a reaction rate constant. $Y_O$ refers to the mass fraction of oxygen and $Y_F$ refers to propane. The activation energy of propane in oxygen is $E_a = 30.0 \text{ }\frac{\text{kcal}}{\text{mol}} $.

In order to make the results more general, the following normalized non-dimensional variables are created:

\begin{large}
\begin{center}
\begin{tabular}{ l l l l l l l }

$x^*=\frac{x}{x_0}$ &
; &
$u^*=\frac{u}{u_\infty}$ &
; &
$h^*=\frac{h}{h_\infty}$ &
; &
$\rho^*=\frac{\rho}{\rho_\infty}$ \\

$y^*=\frac{y}{\delta_0}$ &
; &
$v^*=\frac{\sqrt{Re_0}}{u_\infty}v$ &
; &
$\kappa^*=\frac{x^2_0}{\delta_0 u_\infty} \kappa$ &
; &
$\mu^*=\frac{\mu}{\mu_\infty}$ \\

$z^*=\frac{z}{\delta_0}$ &
; &
$w^*=\frac{\sqrt{Re_0}}{u_\infty}w$ &
; &
$f^*(x) = \frac{x_0}{u_\infty \sqrt{\rho_\infty}} f(x^*) $ &
; &
$Q^* = \frac{Q}{h_{\infty}}$\\

$Da = \frac{A x_0 \rho_\infty^{0.75}}{u_\infty}$ &
; &
$Re_0=\frac{\rho_\infty u_\infty x_0}{\mu_\infty}$ &
; &
$\delta_0 = \frac{x_0}{\sqrt{Re_0}} \text{.}$\\
\end{tabular}
\end{center}
\end{large} Upstream conditions are applied at $x=0$. $x_0$ is the positive value used to normalize the $x$ dimension. Based on the value of $x_0$, the Reynolds number is given as $Re_0$. $\delta_0$ is an estimate of the boundary-layer thickness at $x=x_0$ and is used to scale the results for $y$ and $z$. The subscripts \(\infty\) and \(-\infty\) are used to denote the free-stream behavior of a particular variable as \(y \rightarrow \infty\) and \(y \rightarrow -\infty\), respectively. A non-dimensional Damk\"{o}hler number $Da$ is created to define a normalized reaction rate of fuel. This number offers a ratio between the chemical rate to the transport rate within the flow. This number will be large (of the order \(10^6\)) because, in this study, chemical reactions occur much more rapidly than thermal or mass transport.

The governing non-dimensional equations become

\begin{equation}
\pdv{(\rho^* u^*)}{x^*} + \pdv{(\rho^* v^*)}{y^*} + \rho^* \kappa^* = 0
\end{equation}

\begin{equation}
\rho^* u^* \pdv{u^*}{x^*} + \rho^* v^* \pdv{u^*}{y^*} = \frac{\partial}{\partial y^*}(\mu^*\pdv{u^*}{y^*})
\end{equation}

\begin{equation}
\rho^* u^* \pdv{\kappa^*}{x^*} + \rho^* v^* \pdv{\kappa^*}{y^*} + \rho^* \kappa^{*2} - f^*(x^*)^2 =
\frac{\partial}{\partial y^*}(\mu^*\pdv{\kappa^*}{y^*})
\end{equation}

\begin{equation}
\rho^* u^* \pdv{h^*}{x^*} + \rho^* v^* \pdv{h^*}{y^*} = \frac{1}{Pr}\frac{\partial}{\partial y^*}(\mu^*\pdv{h^*}{y^*}) - \rho^* Q^* \dot{\omega}_F^*
\end{equation}

\begin{equation}
\rho^* u^* \pdv{Y_m}{x^*} + \rho^* v^* \pdv{Y_m}{y^*} = \frac{1}{Pr}\frac{\partial}{\partial y^*}(\mu^*\pdv{Y_m}{y^*}) + \rho^* \dot{\omega}_m^*;
\end{equation}

\begin{equation*}
    m = \text{O}, \text{F}, \text{C}\text{O}_2, \text{H}_2\text{O} \nonumber
\end{equation*} where

\begin{equation}
\dot{\omega}_F^* = - Da \rho^{*0.75} Y_O^{1.65} Y_F^{0.10} e^{-E_a/R_1 T_\infty h^*}\text{.}
\end{equation}

In addition to the equations that govern all quantities of interest through a numerical march in \(x\), upstream and boundary conditions must be established for all quantities of interest to allow the algorithms to propagate through the domain.

For \(u^*\), \(\kappa^*\), \(h^*\), and \(Y_m\) in the non-reactive case, the initial upstream condition is a monotonic variation with \(y^*\) at \(x^*=0\). This study uses a hyperbolic tangent curve in these cases. For example, $u^*(0,y^*) = \frac{1}{2}(1 + \frac{u_{-\infty}}{u_\infty}) + \frac{1}{2}(1 - \frac{u_{-\infty}}{u_\infty}) \tanh{(y^*)}$. For \(h^*\) in the reactive case, an inflow with a temperature peak is given to allow ignition. In particular, a Gaussian curve is taken. We take \(v^*=0\) at \(x^*=0\) and \(v^*=0\) at \(y^*=0\).

The free-stream boundary conditions for \(u^*\), \(\kappa^*\), \(h^*\), and \(Y_m\) are constant values. Later, \(\kappa_\infty\) and \(\kappa_{-\infty}\) will be considered as functions of \(x\).

Taking gaseous oxygen at \(300\) K and \(10\) bar where \(\rho_\infty = 4.798 \text{ } \frac{\text{kg}}{\text{m}^3}\), \(\mu_\infty = 42.43 \text{ } \frac{\mu\text{Ns}}{\text{m}^2}\), and \(u_\infty = 1 \text{ } \frac{\text{m}}{\text{s}}\), we find \(x_0 = 0.88 \text{ } \text{mm}\) for \(Re_0 = 100\). The estimate of the thickness of the shear layer $\delta_0$ under these conditions becomes $0.88/\sqrt{100} = 0.088 \text{ mm}$. In the two-dimensional case with no imposed counterflow, the actual shear layer thickness at $x=x_0$ is found to be $0.39 \text{ mm}$, which is of the same order as our estimate; $x_0$ is approximately $2.3$ times the shear layer width that occurs at that $x$-position.

The specific heat at constant pressure is taken to be constant throughout the domain and equal to \(1309 \frac{\text{J}}{\text{kg K}}\). This value is the average between the expected specific heat at constant pressure of gaseous oxygen (\(988 \frac{\text{J}}{\text{kg K}}\)) and gaseous propane (\(1630 \frac{\text{J}}{\text{kg K}}\)). Following the calorically-perfect-gas assumption (\(h=c_P T\)),
\begin{equation}
    T=T_{\infty} h^*
\end{equation} Following the ideal-gas assumption and using a uniform-pressure assumption,
\begin{equation}
    \rho^* = \frac{T_\infty}{T} \frac{W}{W_O}; \quad W = \frac{1}{\sum_{m=1}^N \frac{Y_m}{W_m}} \text{.}
\end{equation} \(\mu\) is determined by taking the weighted average of the fluid viscosities:

\begin{equation}
    \mu=\sum_{m=1}^N Y_m \mu_m \text{.}
\end{equation} The viscosity of a particular fluid component at a given temperature \(\mu_m(T)\) is found using the Chung, Ajlan, Lee and Starling\cite{chung1988generalized} relation for the viscosity of a dilute simple molecular gas, where

\begin{equation}
    \mu_m(T) = \frac{(4.0785 \times 10^{-5}) T^{1/2}}{M_m^{1/6} V_{c,m}^{2/3} \left( \frac{A}{T^{*B}} + \frac{C}{\exp(DT^*)} +\frac{E}{\exp(FT^*)} + GT^{*B}\sin(ST^{*W} - H) \right)} \text{.}
\end{equation} \(M\) is the species molecular weight in g/mol, \(V_c\) is the critical volume in cm\textsuperscript{3}/g, and the dimensionless temperature \(T^*\) can be found as

\begin{equation}
    T^* = 1.2593 \frac{T}{T_{c,n}}
\end{equation} where \(T_c\) is the critical temperature in K. The empirical constants are given as \(A = 1.16145\), \(B = 0.14874\), \(C = 0.52487\), \(D = 0.77320\), \(E = 2.16178\), \(F = 2.43787\), \(G = -6.435 \times 10^{-4}\), \(H = 7.27371\), \(S = 18.0323\), and \(W = -0.76830\). This relation along with Eq. (19) determines \(\mu\) at all points; \(\mu^*\) is determined simply with its definition.

The molecular weight \(M\), critical temperature \(T_c\), and critical volume \(V_c\) of each fluid is summarized in Table I.

\subsection{Computational Scheme}
The computational domain lies on the normalized \(x\) and \(y\) plane, where

\begin{equation}
   \label{eq}
    x^*\in[0,L] \quad \text{and} \quad y^*\in[-H,H] \text{.}
\end{equation} \(L\) and \(2H\) are the length of the computational domain in \(x^*\) and \(y^*\), respectively. We choose \(H > 1\) to ensure that \(y\) is everywhere wider than the mixing-layer thickness at \(x=x_0\).

The domain is discretized into a rectilinear set containing \(N_x\) samples in \(x^*\) and \(N_y\) samples in \(y^*\) with the intention of developing stepwise algorithms to solve for quantities of interest downstream (\(x^*>0\)). Finite-difference approximations of the differential terms are used, following Ferziger and Perić\cite{ferziger_peric_1996}.

For \(x\)-derivative terms, first-order forward-difference approximations are used, while for \(y\)-derivative terms, second-order central-difference approximations are taken. For the \(y\)-derivative of a product of \(\mu^*\) and another \(y\)-derivative, the following second-order approximation is employed:
\begin{equation}
    \frac{\partial}{\partial y^*}(\mu^* \pdv{a}{y^*}) \approx \frac{1}{\Delta y^{*2}}(\mu^{*k}_{n+\frac{1}{2}} a^{k}_{n+1} - (\mu^{*k}_{n+\frac{1}{2}} + \mu^{*k}_{n-\frac{1}{2}}) a^k_n + \mu^{*k}_{n-\frac{1}{2}} a^{k}_{n-1})
\end{equation} where \(k\) and \(n\) are indices for \(x\) and \(y\) positions, respectively. $n \pm \frac{1}{2}$ denotes a half step between two grid-points. For $\mu^*$,

\begin{equation}
    \mu^{*k}_{n-\frac{1}{2}} = \frac{\mu^{*k}_{n-1} + \mu^{*k}_n}{2}, \quad \mu^{*k}_{n+\frac{1}{2}} = \frac{\mu^{*k}_{n+1} + \mu^{*k}_n}{2} \text{.}
\end{equation}

Fig. 2 illustrates mesh-size independence for the reactive case. $\dot{\omega}^*_F$ is used to test numerical error because it is the most sensitive to changes in other parameters, thus ensuring mesh-size independence for all other parameters.

\begin{figure*}
    \centering
    \includegraphics[width=\textwidth]{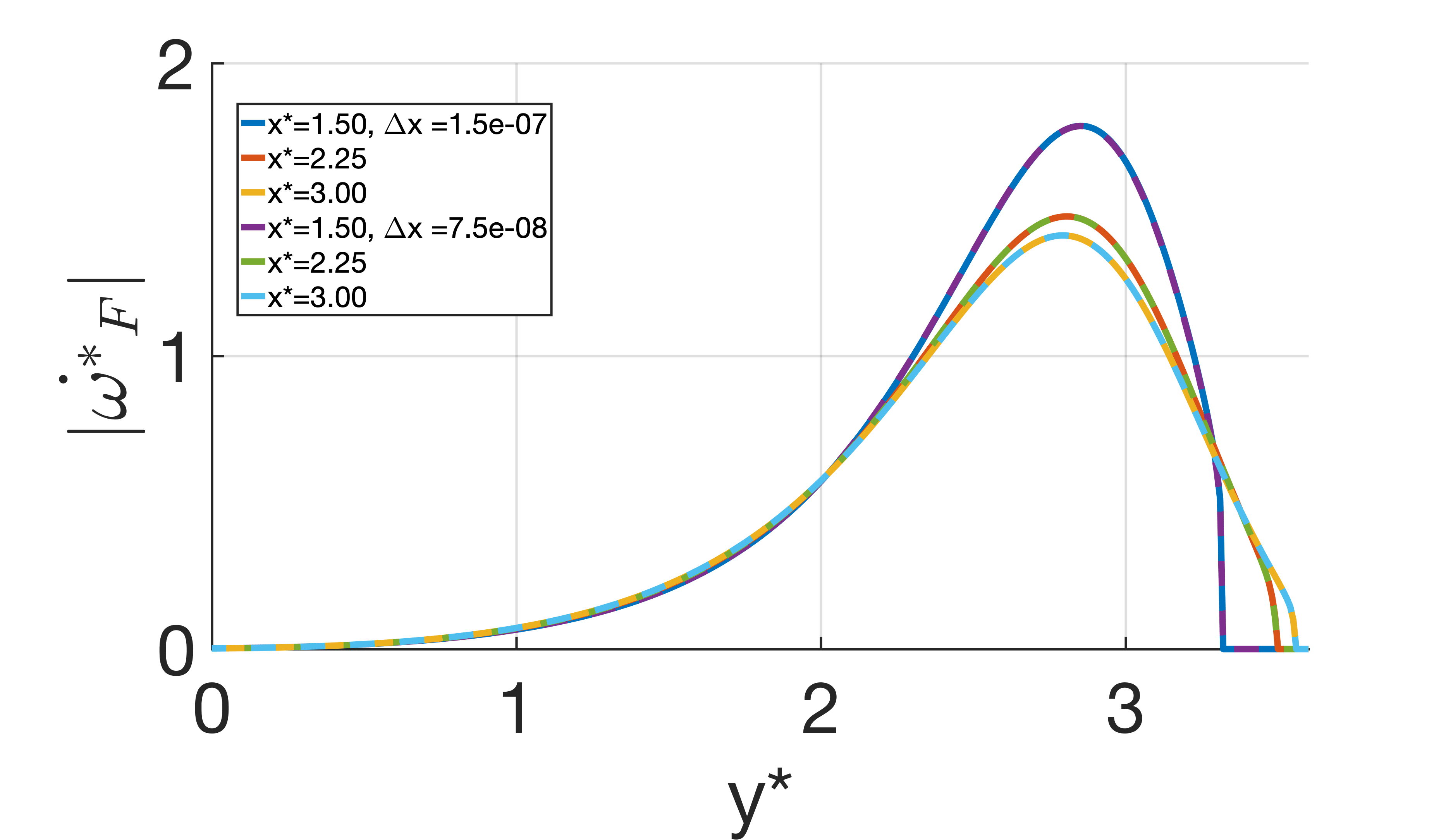}
    \caption{\(\abs{\dot{\omega}^*_F(y^*)}\) at multiple positions in \(x^*\) for different mesh sizes in the case where reactions occur in the mixing layer.}
    \label{fig:my_label}
\end{figure*}

\(N_x\) and \(N_y\) cannot be chosen arbitrarily. The stability of our finite-difference scheme depends on the relationship between \(\Delta x\) and \(\Delta y\) which can be recovered from the two-dimensional Navier-Stokes equations with negligible convection\cite{ferziger_peric_1996}. The criterion for stable computational results is given as

\begin{equation}
\frac{\Delta x^*}{u^*} < \frac{1}{2}\frac{\rho^*}{\mu^*} \Delta y^{*2} \text{.}
\end{equation}

In order to ensure that changes due to the reaction rate do not happen more rapidly than changes due to flow advection, the reaction rate of fuel $\dot{\omega}^*_F$ must satisfy

\begin{equation}
    \abs{\dot{\omega}_F^*} < \frac{u^*}{\Delta x^*} \text{.}
\end{equation} Combining Eqs. (24) and (25) results in the following stability criterion:

\begin{equation}
    \abs{\dot{\omega}^*_F} < \frac{2 \mu^*}{\rho^*} \frac{1}{\Delta y^{*2}} < \frac{u^*}{\Delta x^*} \text{.}
\end{equation}

\subsection{Comparing to the Similar Solution}
Sirignano\cite{sirignano2021mixing} devised a similar solution for the non-reactive case where $\kappa$ is allowed to vary as $1/x$ in the free-streams. Consequently, a function \(G(x)\) will remain constant in each of the free-streams where

\begin{equation}
    G=\frac{2 \kappa x}{u_\infty} =2 \kappa^* x^* \text{.}
\end{equation}

Let us consider the ODEs that describe \(\kappa\) in the free-streams. They result from setting the \(y\)-derivatives to zero in Eq. (8).

For \(y^* \rightarrow +\infty\),
\begin{equation}
\pdv{\kappa_\infty^*}{x^*} + \kappa_\infty^{*2} - f^*(x^*)^2 = 0 \text{.}
\end{equation}

For \(y^* \rightarrow -\infty\),
\begin{equation}
\rho_{-\infty}^* u_{-\infty}^* \pdv{\kappa_{-\infty}^*}{x^*} + \rho_{-\infty}^* \kappa_{-\infty}^{*2} - f^*(x^*)^2 = 0 \text{.}
\end{equation} With \(\rho^*_{-\infty} = \frac{1}{\rho_\infty / \rho_{-\infty}}\), the last ODE becomes

\begin{equation}
\pdv{\kappa_{-\infty}^*}{x^*} + \frac{u_\infty}{u_{-\infty}} \kappa_{-\infty}^{*2} - \frac{u_\infty}{u_{-\infty}} \frac{\rho_\infty}{\rho_{-\infty}} f^*(x^*)^2 = 0 \text{.}
\end{equation}

Suppose \(f^*\) varies as
\begin{equation}
    f^*(x^*)=\frac{F}{x^*}
\end{equation} where \(F\) is some given positive constant. Thus, \(\kappa^*\) varies in the free-streams as

\begin{equation}
    \kappa^*_\infty = \frac{A_\infty}{x^*}
\end{equation} and

\begin{equation}
    \kappa^*_{-\infty} = \frac{A_{-\infty}}{x^*}
\end{equation}  which is the desired similar form of \(\kappa\). This free-stream behavior in \(\kappa\) can be created after substitution of relations (31), (32), and (33) into (28) and (29). The \(A\) coefficients are determined by solution of quadratic equations:

\begin{equation}
    A_\infty = \frac{1 \pm \sqrt{1+4F^2}}{2}
\end{equation}

and

\begin{equation}
    A_{-\infty} = \frac{\frac{1}{u_\infty / u_{-\infty}} \pm \sqrt{(\frac{1}{u_\infty / u_{-\infty}})^2+4(\frac{\rho_\infty}{\rho_{-\infty}})F^2}}{2} \text{.}
\end{equation} Four combinations of solutions for \(A_\infty\) and \(A_{-\infty}\) become possible. Here, only the cases where both coefficients have the same sign will be considered, leaving two solutions. The positive (negative) values give a compressive (extensional) strain in the \(y\)-direction and an extensional (compressive) strain in the \(z\)-direction.

Note that here we differ from Sirignano\cite{sirignano2021mixing}, who set

\begin{equation}
    \kappa^*_\infty = \sqrt{\rho^*_{-\infty}}\kappa^*_{-\infty} = f^*(x^*)
\end{equation} which becomes approximately correct only if $F >> 1$.

In a different case where \(\kappa^*\) does not change along \(x\) in the free-streams (\(f^* = C\)),

\begin{equation}
    f^* = \kappa^*_\infty = C
\end{equation} and

\begin{equation}
    \kappa^*_{-\infty} = \sqrt{\frac{\rho_\infty}{\rho_{-\infty}}}C \text{.}
\end{equation}

Three cases are investigated: \(\kappa_\infty\) (\(\kappa_{-\infty}\)) that is constant, \(\kappa_\infty\) (\(\kappa_{-\infty}\)) that varies as a positive constant divided by \(x^*\), and \(\kappa_\infty\) (\(\kappa_{-\infty}\)) that varies as a negative constant divided by \(x^*\).

We define a position variable $\eta$ to demonstrate far downstream similarity. Following Sirignano\cite{sirignano2021mixing},

\begin{equation}
    \bar{y} = \int_0^y \rho(y) dy \text{.}
\end{equation} and

\begin{equation}
    \eta = \frac{\bar{y}}{\sqrt{\frac{2 \rho_\infty \mu_\infty x}{u_\infty}}} \text{.}
\end{equation}

Substituting our definitions for normalized variables gives

\begin{equation}
    \eta = \frac{\int_0^{y^*} \rho^*(y^*) dy^*}{\sqrt{2 x^*}} \text{.}
\end{equation}

\section{Results}
The visualizations of the non-reactive flow when \(\kappa^*\) is held constant in the free-streams are discussed in the next subsection. The following subsection discusses the non-reactive case where \(\kappa^*\) varies with \(x\) in the free-streams. The final subsections discuss the results of the reactive case, including an investigation of multiflame structures.

Table II summarizes the non-reactive flow cases, in which \(h_\infty / h_{-\infty} = 1.5\), \(Da=0\), and \(T_\infty = 800 \text{ K}\). \(\kappa^*_\infty\) in Cases 5a and 5b is found according to Eqs. (32) and (34). We consider only same-sign solutions for \(\kappa^*\) in the free-streams.

\begin{table}
\caption{Parameter definitions for the non-reactive cases.}
\begin{ruledtabular}
\begin{tabular}{cccdc}

 \textbf{Case} & \(f^*\)& \(u_\infty / u_{-\infty}\) & \mbox{\(Pr\)} & \(\kappa^*_\infty\) \\

 \hline

 \textbf{1} & \(1\) & \(4\) & 1. & \(1\) \\

 \textbf{2a} & \(0\) & \(4\) & 1. & \(0\) \\

 \textbf{2b} & \(2\) & \(4\) & 1. & \(2\) \\

 \textbf{3a} & \(1\) & \(2\) & 1. & \(1\) \\

 \textbf{3b} & \(1\) & \(8\) & 1. & \(1\) \\

 \textbf{4a} & \(1\) & \(4\) & 0.7 & \(1\) \\

 \textbf{4b} & \(1\) & \(4\) & 1.3 & \(1\) \\

 \textbf{5a} & \(1/x^*\) & \(4\) & 1. & \(1.6180/x^*\) \\

 \textbf{5b} & \(-1/x^*\) & \(4\) & 1. & \(-0.6180/x^*\)

\end{tabular}
\end{ruledtabular}
\end{table}

Table III summarizes the reactive flow cases, in which \(\kappa^*_\infty = f^*\), \(h_{\text{peak}}/h_\infty =6.67\), and \(T_\infty = 300 \text{ K}\). \(h_{\text{peak}}\) denotes the peak enthalpy of the ignitor at \(x^*=0\) and \(y^*=0\).

\begin{table}
\caption{Parameter definitions for the reactive cases.}
\begin{ruledtabular}

\begin{tabular}{cccdd}

\textbf{Case} & \(f^*\) & \(u_\infty / u_{-\infty}\) & \mbox{\(Pr\)} &  \mbox{\(Da\) } \\

\hline

\textbf{6} & \(1\) & \(4\) & 1. & 1,500,000 \\

\textbf{7a} & \(0\) & \(4\) & 1. & 1,500,000 \\

\textbf{7b} & \(2\) & \(4\) & 1. & 1,500,000 \\

\textbf{8a} & \(1\) & \(2\) & 1. & 1,500,000 \\

\textbf{8b} & \(1\) & \(8\) & 1. & 1,500,000 \\

\textbf{9a} & \(1\) & \(4\) & 0.7 & 1,500,000 \\

\textbf{9b} & \(1\) & \(4\) & 1.3 & 1,500,000 \\

\textbf{10a} & \(1\) & \(4\) & 1. & 0 \\

\textbf{10b} & \(1\) & \(4\) & 1. & 750,000 \\

\textbf{11a} & \(0\) & \(0.25\) & 1. & 1,500,000 \\

\textbf{11b} & \(1\) & \(0.25\) & 1. & 1,500,000 \\

\textbf{11c} & \(2\) & \(0.25\) & 1. & 1,500,000 \\

\textbf{12} & \(1\) & \(4\) & 1. & 3,000,000

\end{tabular}
\end{ruledtabular}
\end{table}

\clearpage
\subsection{Non-reactive Layer with Constant $\kappa_\infty$ and $\kappa_{-\infty}$}

Fig. 3 compares the effects of two-dimensional flow (\(f^*=0\) with only two velocity components) to three-dimensional flow (\(f^* \neq 0\) with three velocity components). Figs. 3(a), 3(d), and 3(e) show that increasing the flux and strain rate of the counterflow, through an increasing \(f^*\), decreases the width of the mixing layer for the scalar quantities. Thereby, the gradients and the vorticity are increased. Fig. 3(b) shows that this effect also increases the inward velocity of the flow in the \(y\)-direction. In Fig. 3(c), we see the increasing value of strain rate \(\kappa^*\) as a result of increasing \(f^*\). As $f^*$ increases, the flow field is stretched  more strongly in the $z$-direction. Accordingly, the vortex stretching is greater since the same integral of vorticity occurs in a thinner layer.

\begin{figure*}
     \centering
     \begin{subfigure}[b]{0.49\textwidth}
         \centering
         \includegraphics[width=\textwidth]{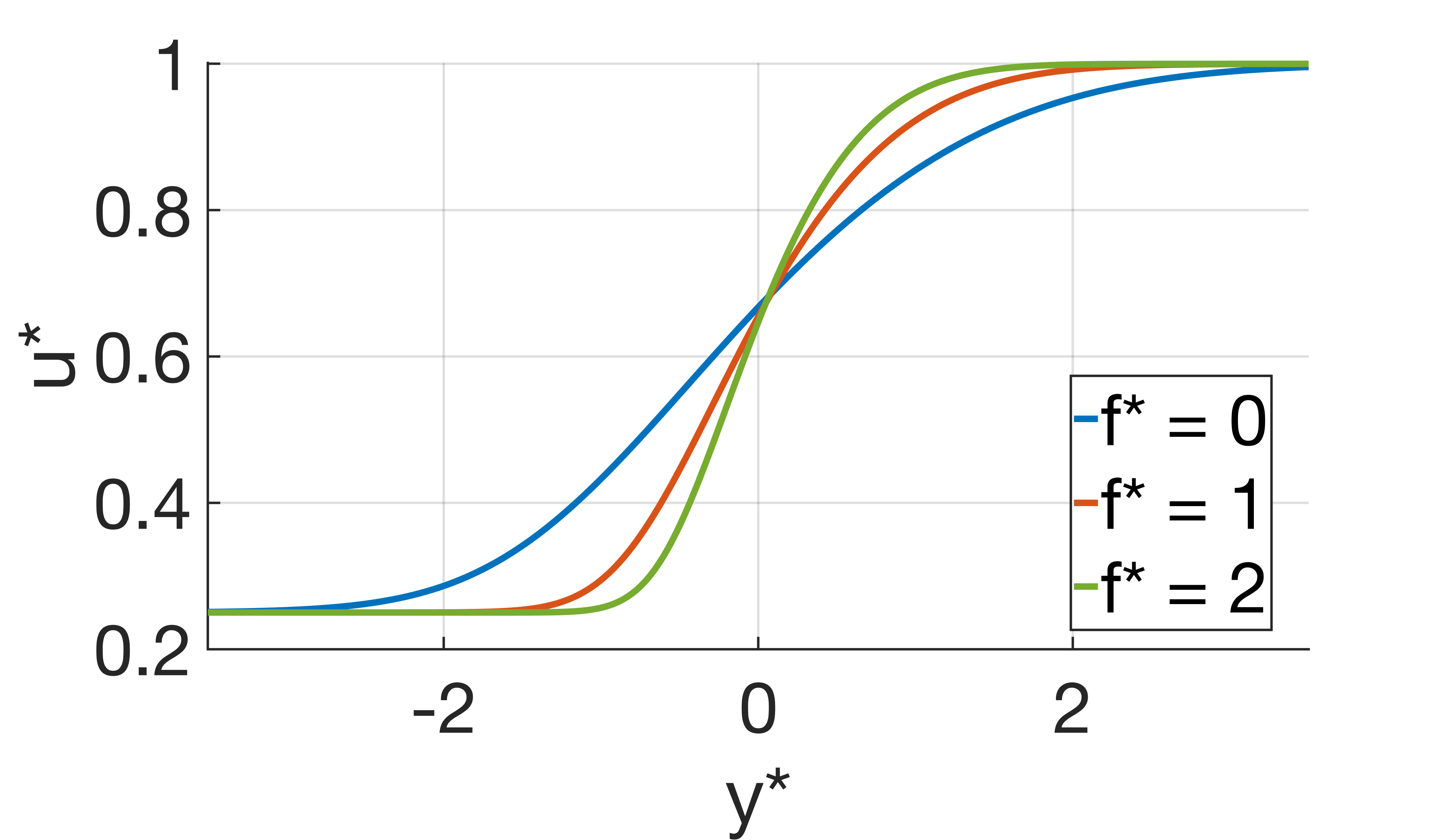}
         \caption{\(u^*(y^*)\)}
         \label{fig:y equals x}
     \end{subfigure}
     \hfill
     \begin{subfigure}[b]{0.49\textwidth}
         \centering
         \includegraphics[width=\textwidth]{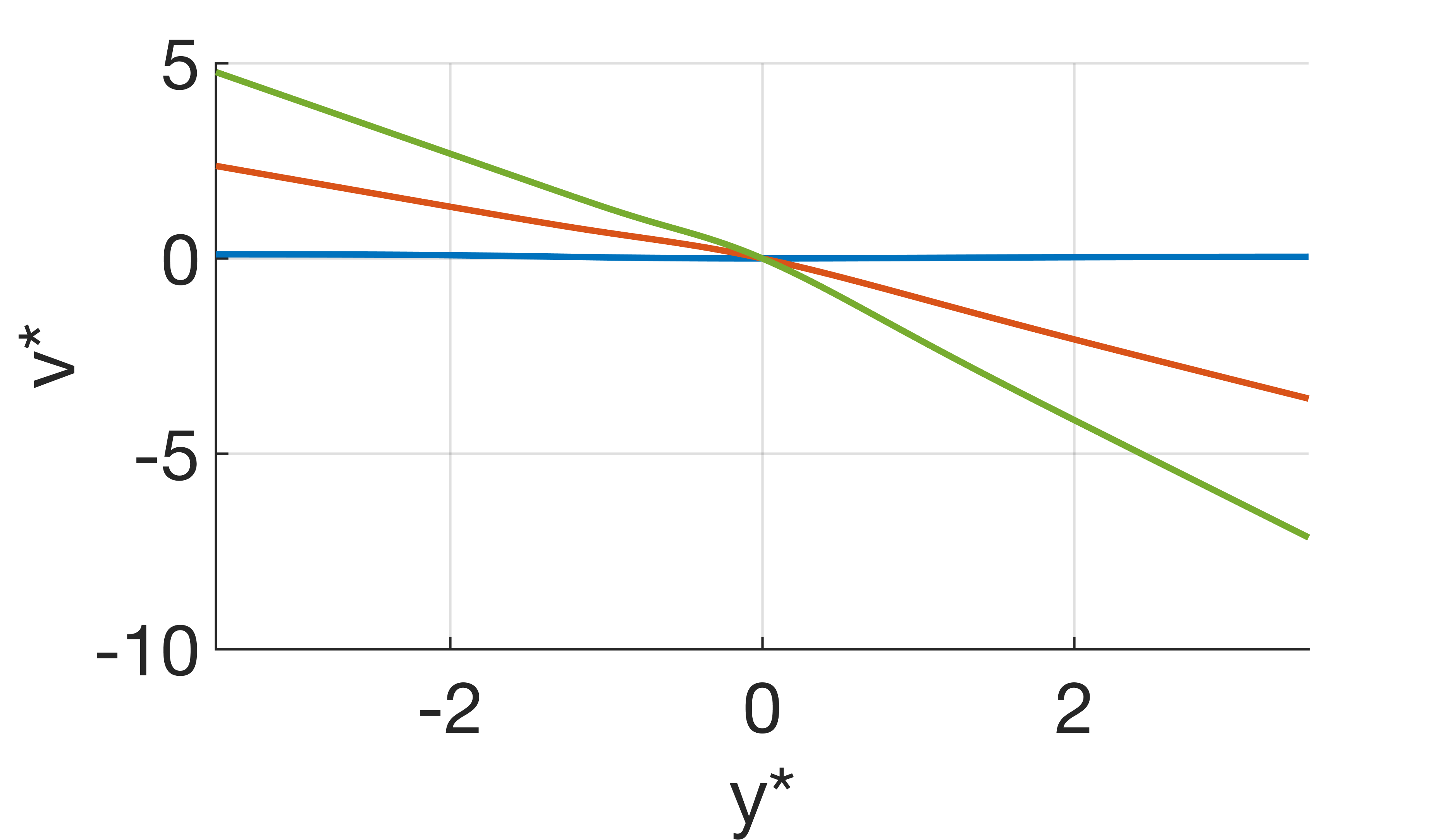}
         \caption{\(v^*(y^*)\)}
         \label{fig:three sin x}
     \end{subfigure}
     \hfill
     \begin{subfigure}[b]{0.49\textwidth}
         \centering
         \includegraphics[width=\textwidth]{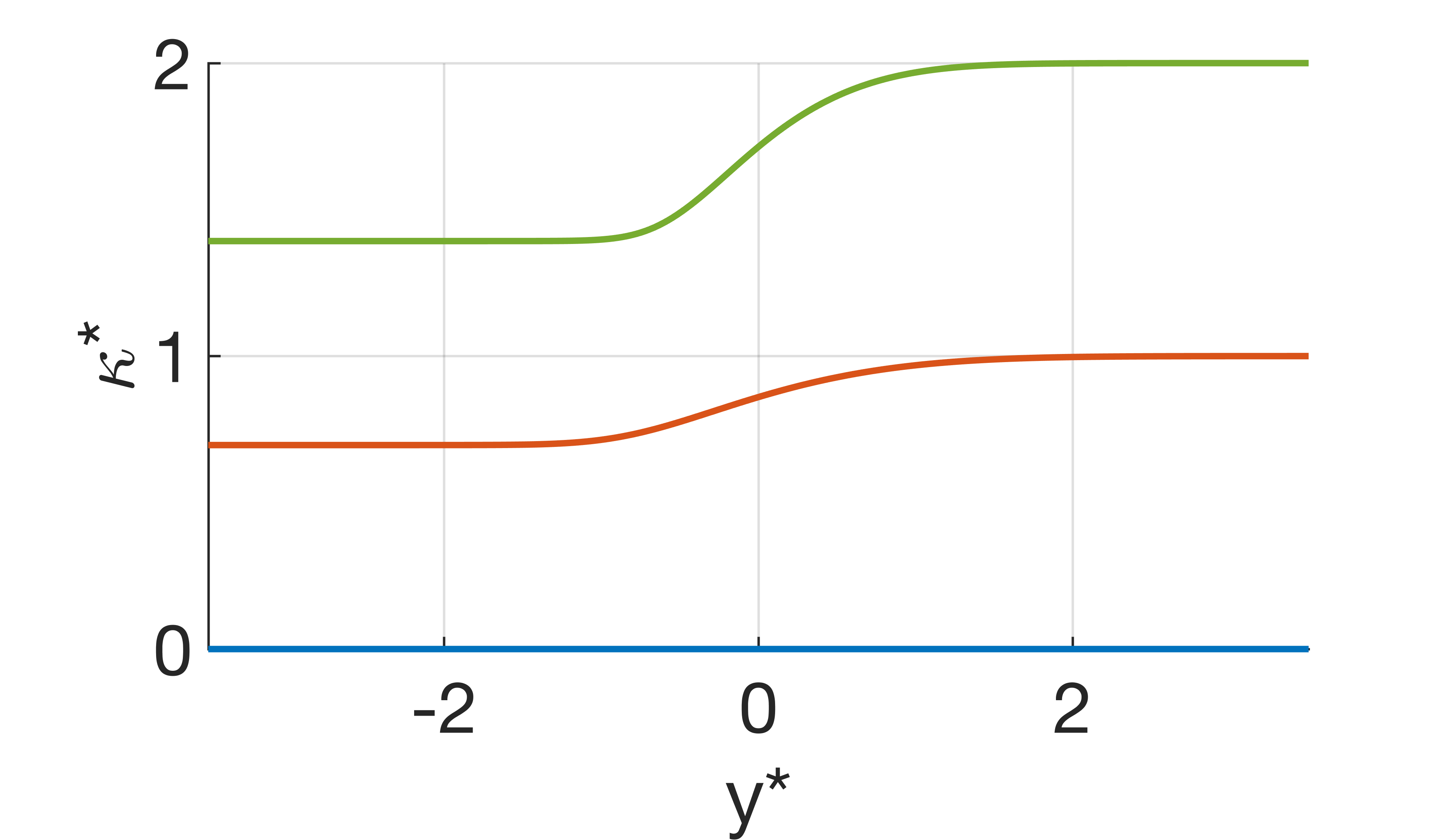}
         \caption{\(\kappa^*(y^*)\)}
         \label{fig:five over x}
     \end{subfigure}
     \begin{subfigure}[b]{0.49\textwidth}
         \centering
         \includegraphics[width=\textwidth]{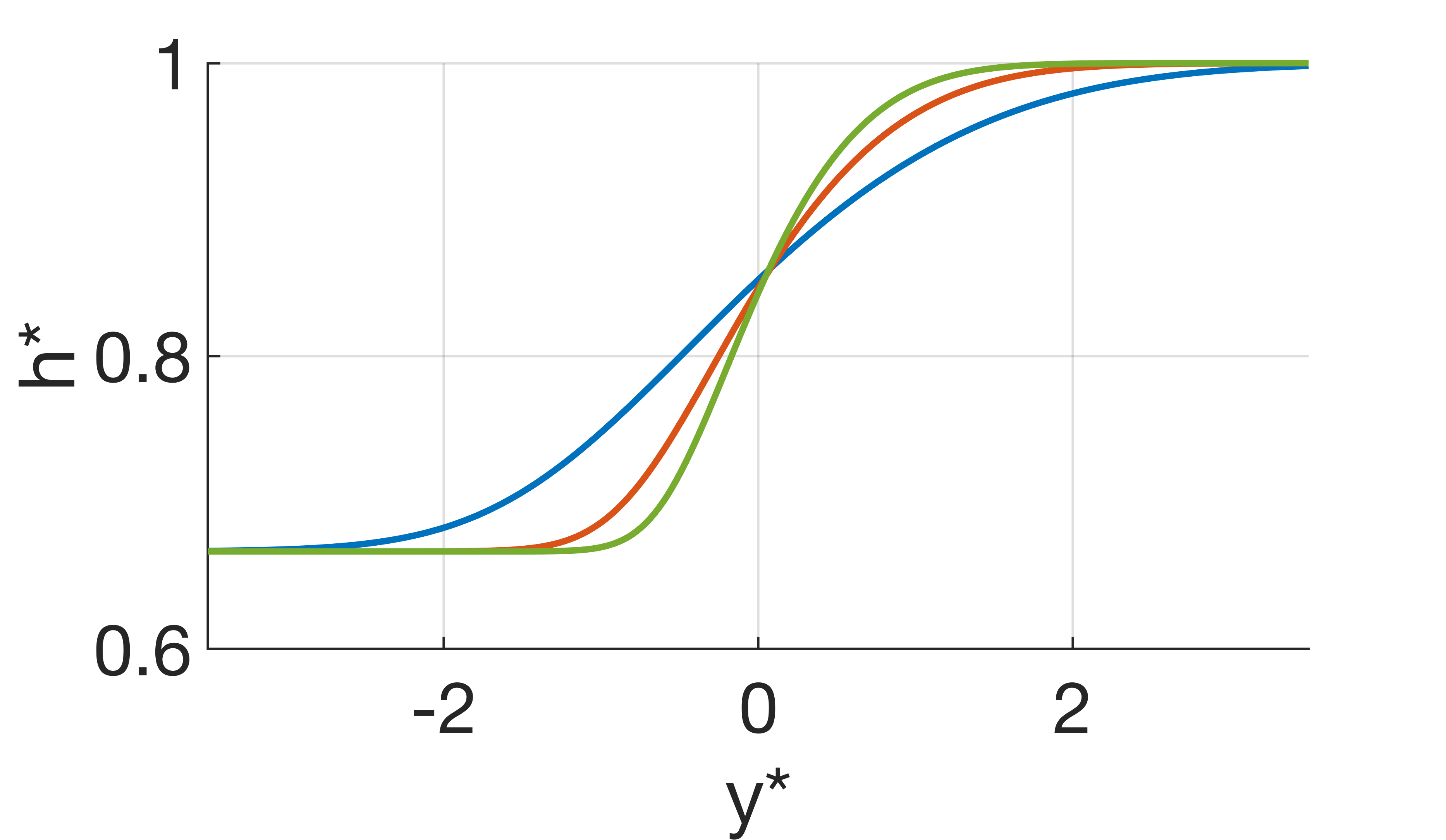}
         \caption{\(h^*(y^*)\)}
         \label{fig:five over x}
     \end{subfigure}
     \begin{subfigure}[b]{0.49\textwidth}
         \centering
         \includegraphics[width=\textwidth]{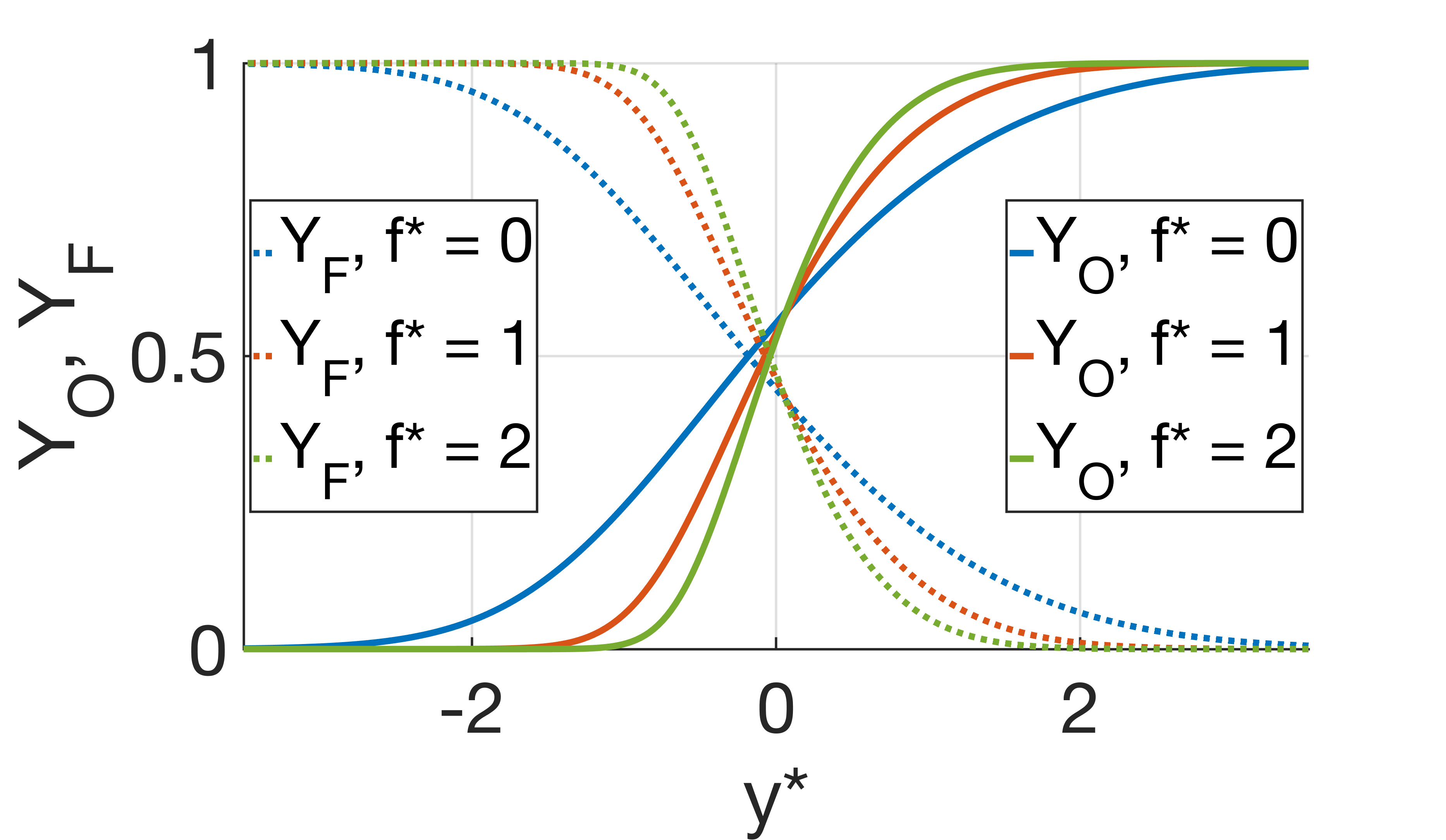}
         \caption{\(Y_O(y^*), Y_F(y^*)\)}
         \label{fig:five over x}
     \end{subfigure}
        \caption{\(u^*\), \(v^*\), \(h^*\), \(\kappa^*\), \(Y_O\), and \(Y_F\) at \(x^*=1\) for non-reactive Cases 2a, 1, and 2b. Ambient counterflow strain rate is constant with $x^*$ and varies between $0$ and $2$.}
        \label{fig:three graphs}
\end{figure*}

Only $u^*$ is shown in Fig. 4 because the ratio of the free-stream \(x\)-component velocities has a very minor effect on the downstream behavior of the other scalar quantities ($v^*$, $\kappa^*$, $h^*$, $Y_O$, $Y_F$) when all else is held constant. The use of \(u_\infty\) in the normalization results in a similarity of the behavior for many variables. The layer width is not affected by the velocity ratio; thus, the $u$-velocity gradient and the vorticity increase as the velocity ratio increases. However, the scalar gradients and the normal strain rates in the $y$- and $z$-directions are not affected significantly.

\begin{figure*}
    \centering
    \includegraphics[width=\textwidth]{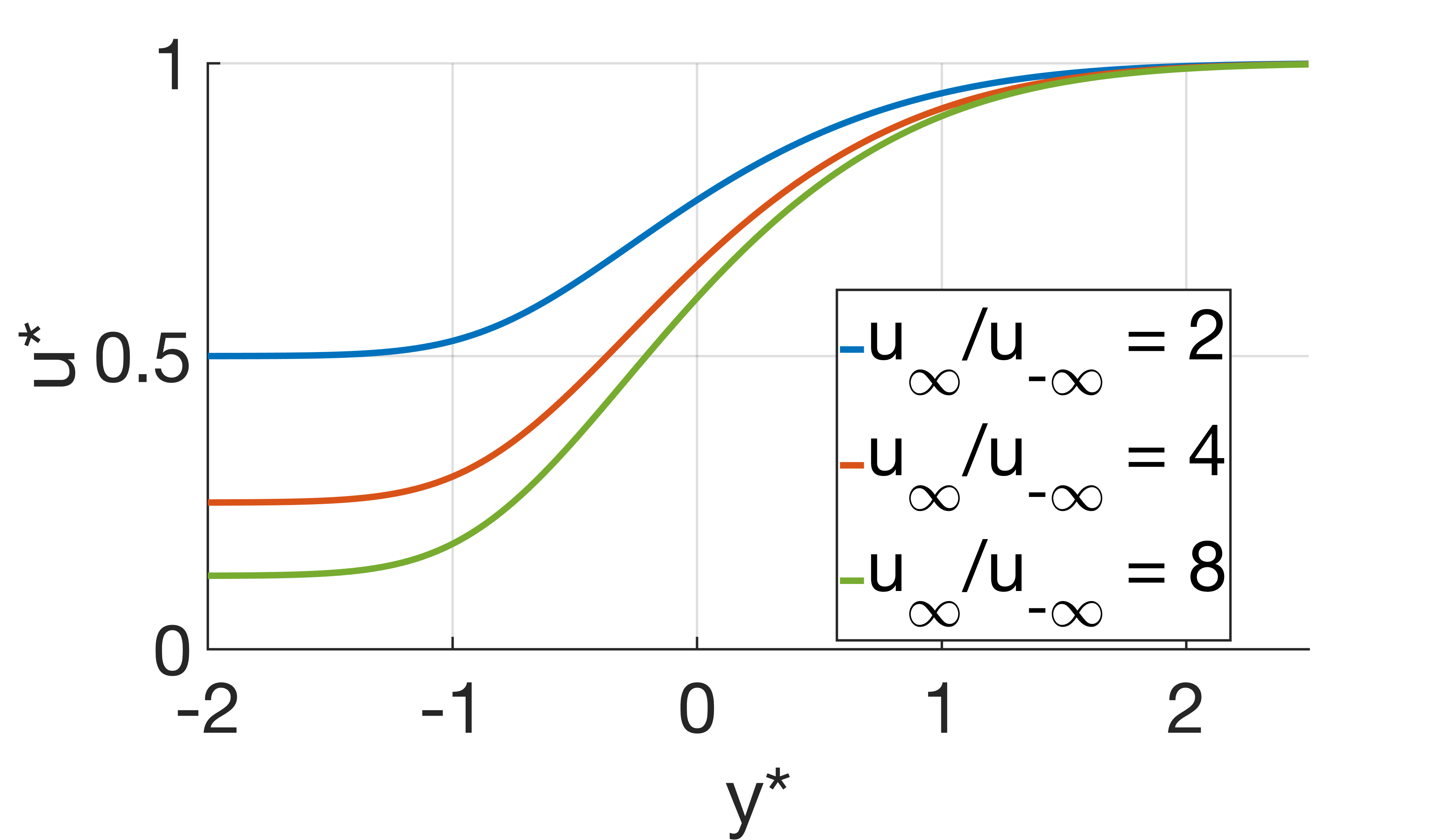}
    \caption{\(u^*\) at \(x^*=1\) for non-reactive Cases 3a, 1, and 3b. Free-stream velocity ratio varies from $2$ to $8$. Ambient counterflow strain rate is constant with $x^*$.}
    \label{fig:my_label}
\end{figure*}

Figs. 5(c) through 5(e) reveal that as \(Pr\) increases, the mixing layers for the scalar quantities become thinner since thermal conductivity and mass diffusivity decrease and large scalar gradients are regained. Fig. 5(f) shows that a unitary \(Pr\) results in a linear relationship between the normalized enthalpy and the normalized horizontal free-stream velocity; i.e., a Crocco integral forms.

\begin{figure*}
     \centering
     \begin{subfigure}[b]{0.49\textwidth}
         \centering
    \includegraphics[width=\textwidth]{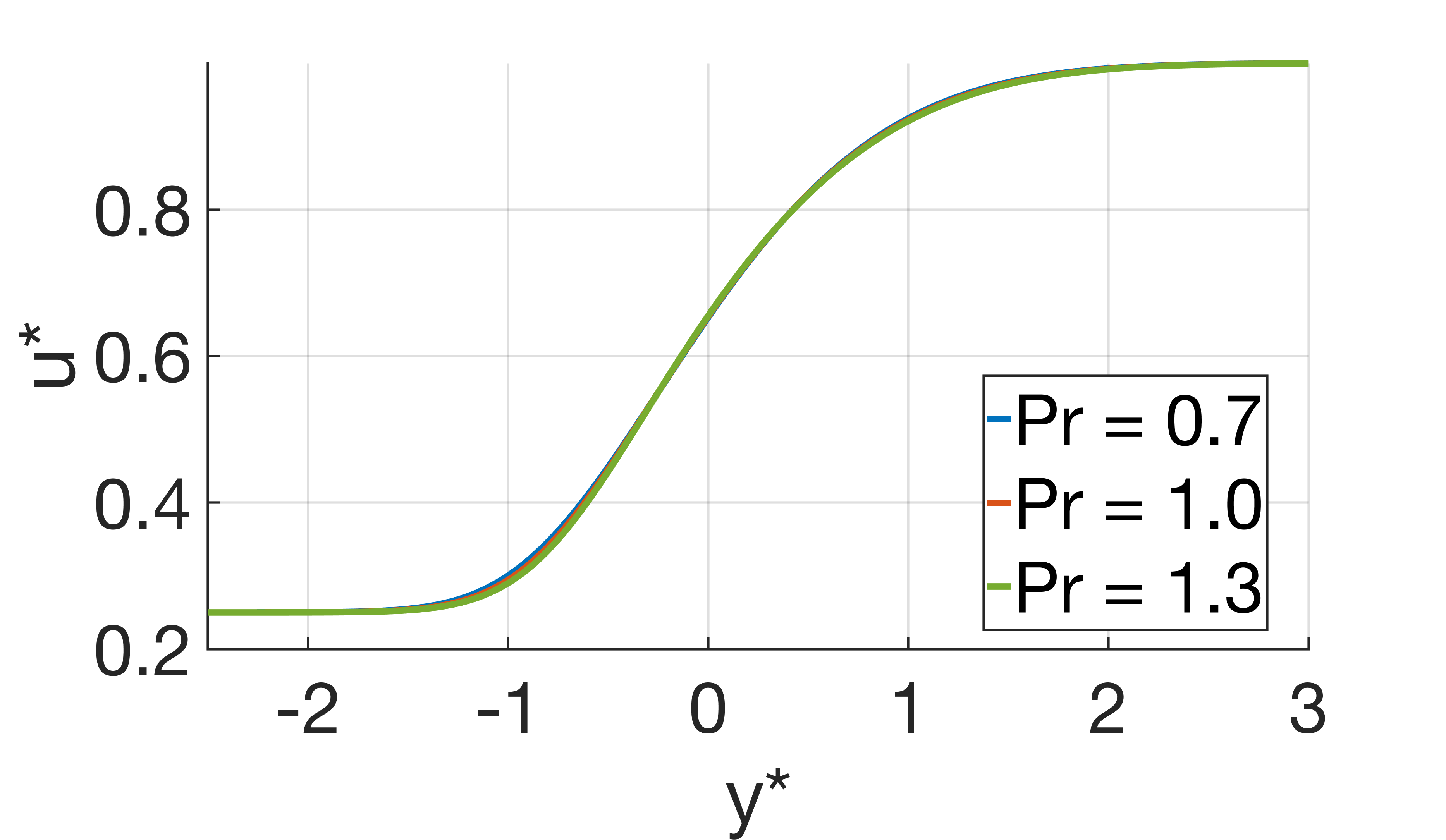}
         \caption{\(u^*(y^*)\)}
         \label{fig:y equals x}
     \end{subfigure}
     \hfill
     \begin{subfigure}[b]{0.49\textwidth}
         \centering
         \includegraphics[width=\textwidth]{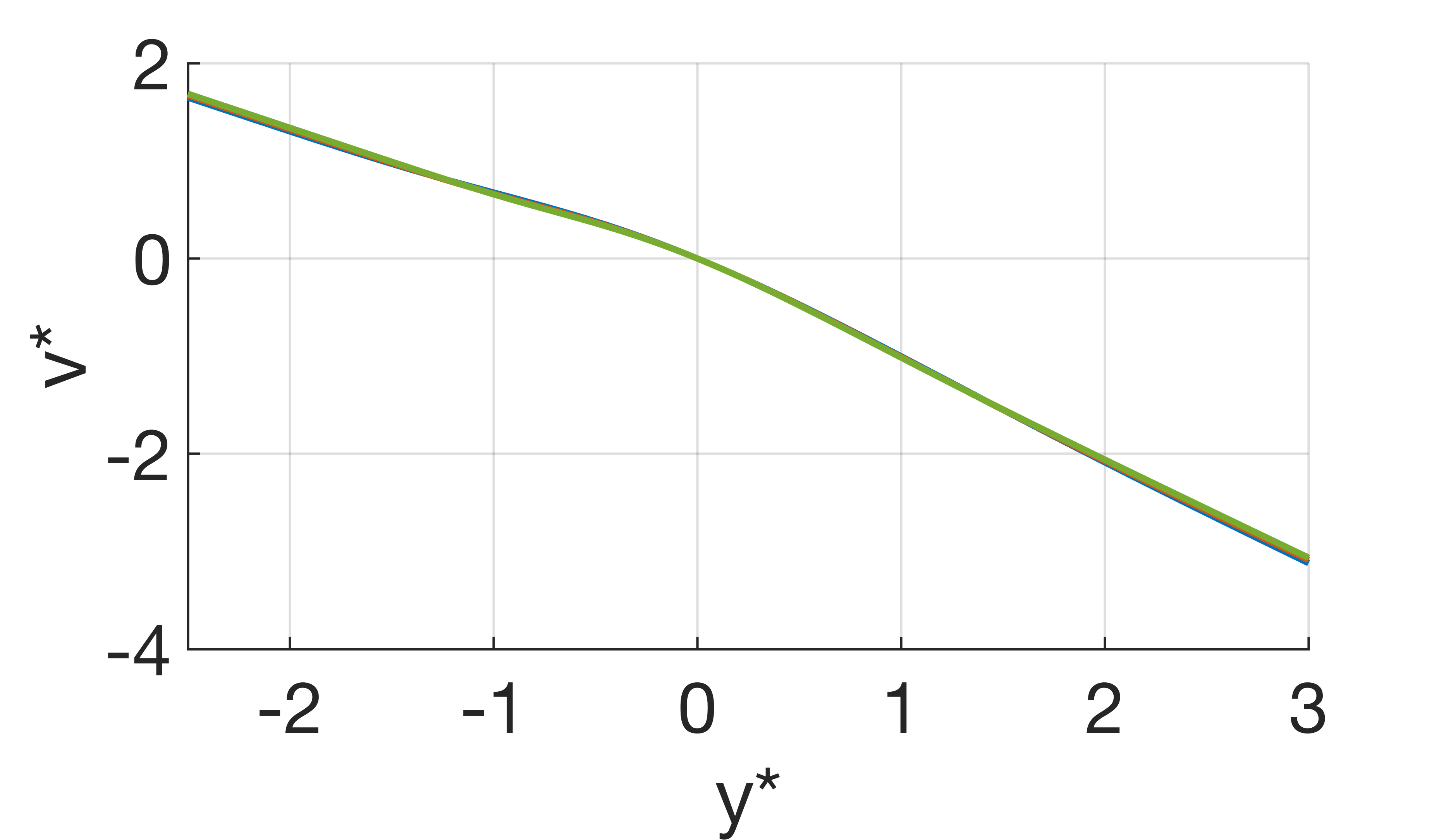}
         \caption{\(v^*(y^*)\)}
         \label{fig:three sin x}
     \end{subfigure}
     \hfill
     \begin{subfigure}[b]{0.49\textwidth}
         \centering
         \includegraphics[width=\textwidth]{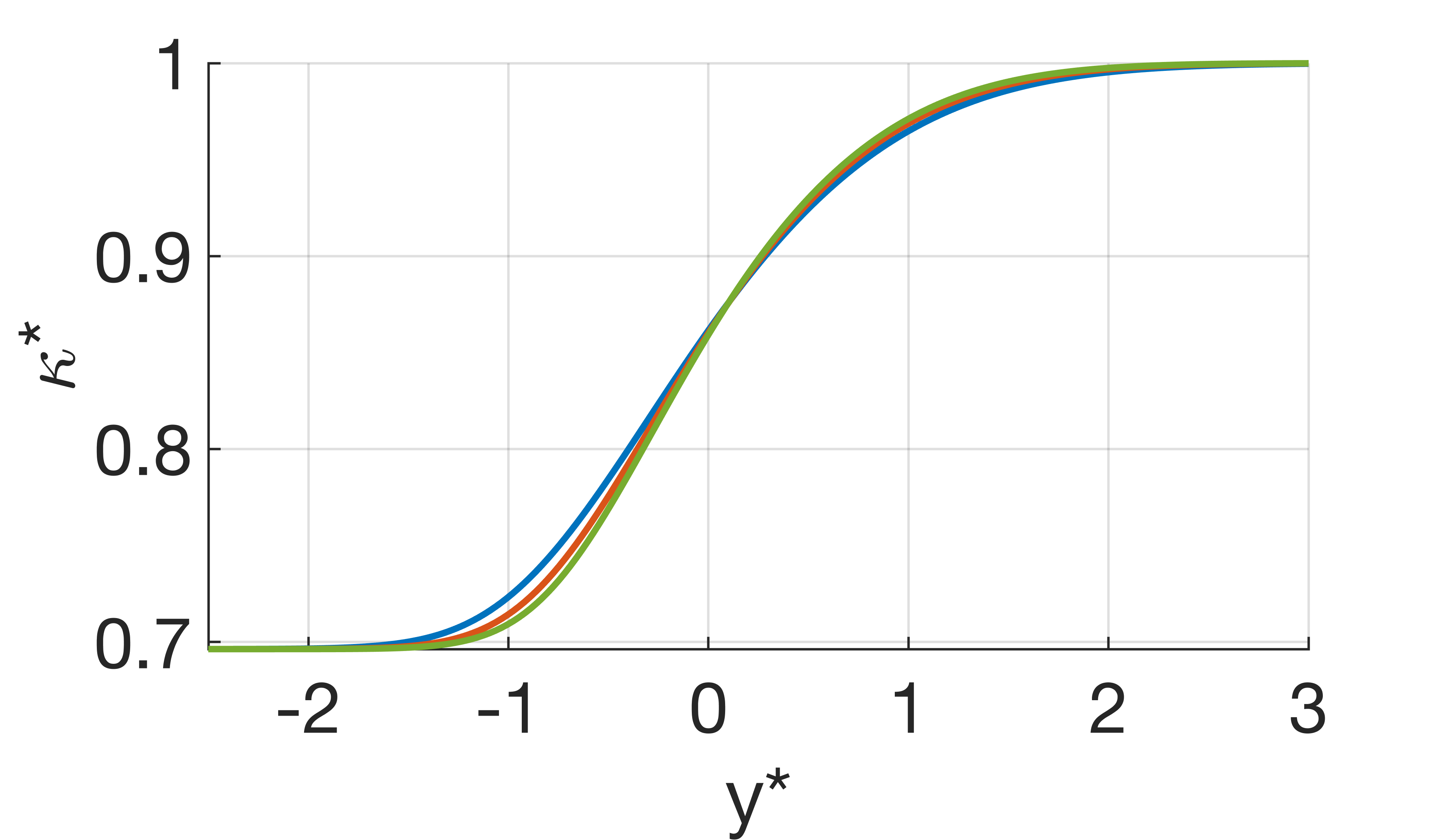}
         \caption{\(\kappa^*(y^*)\)}
         \label{fig:five over x}
     \end{subfigure}
     \begin{subfigure}[b]{0.49\textwidth}
         \centering
         \includegraphics[width=\textwidth]{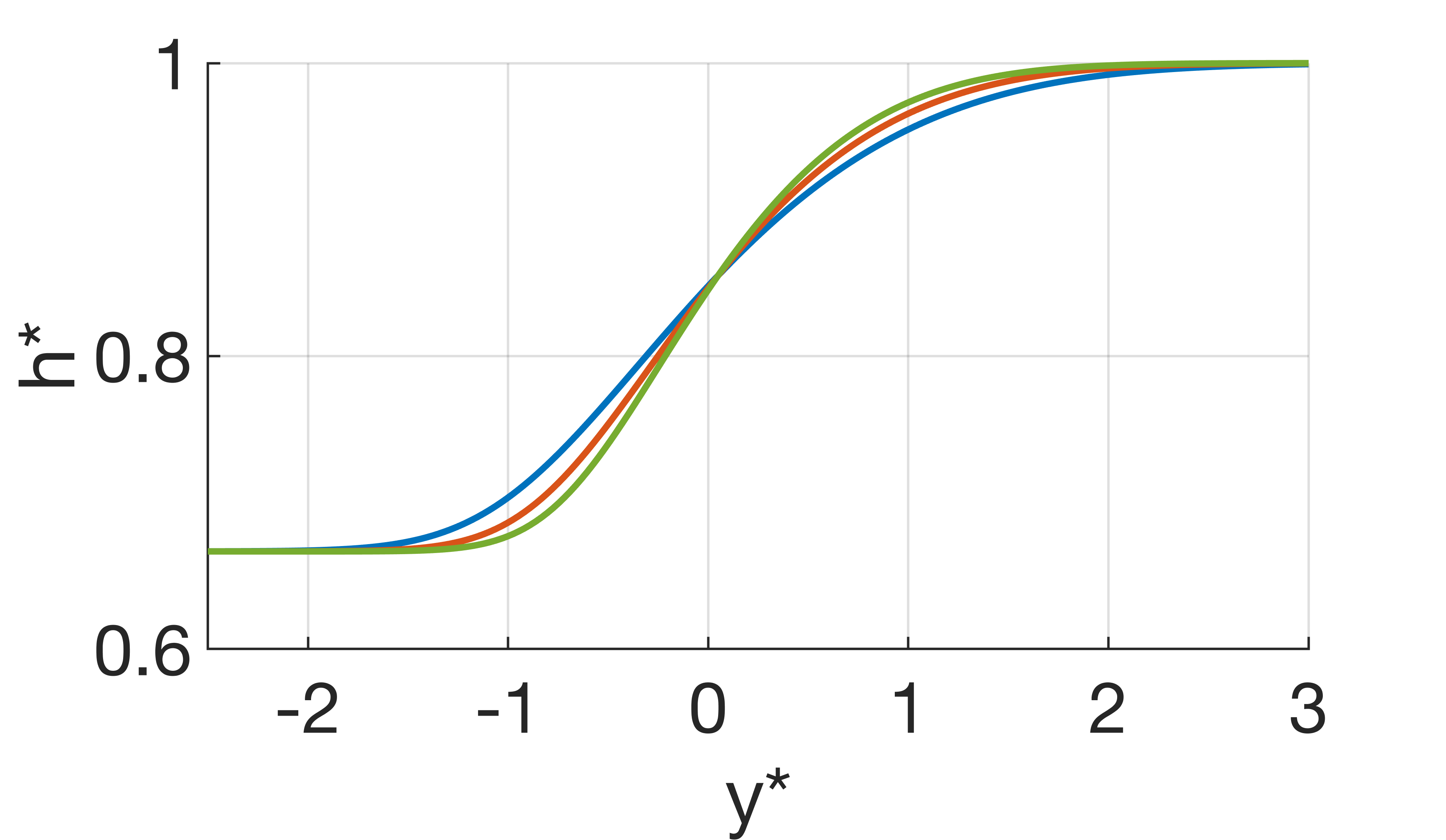}
         \caption{\(h^*(y^*)\)}
         \label{fig:five over x}
     \end{subfigure}
     \begin{subfigure}[b]{0.49\textwidth}
         \centering
         \includegraphics[width=\textwidth]{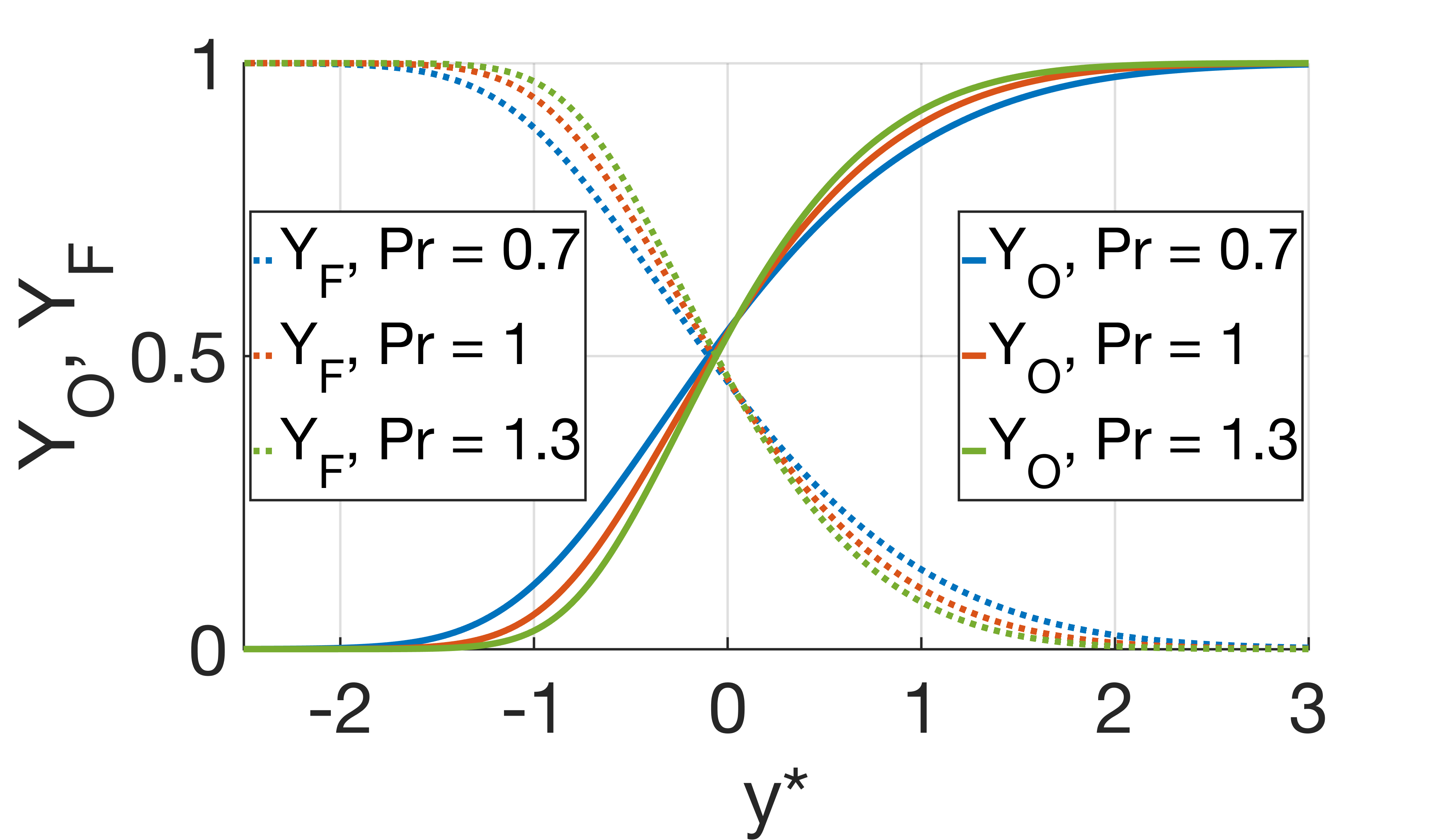}
         \caption{\(Y_O(y^*), Y_F(y^*)\)}
         \label{fig:five over x}
     \end{subfigure}
     \begin{subfigure}[b]{0.49\textwidth}
         \centering
         \includegraphics[width=\textwidth]{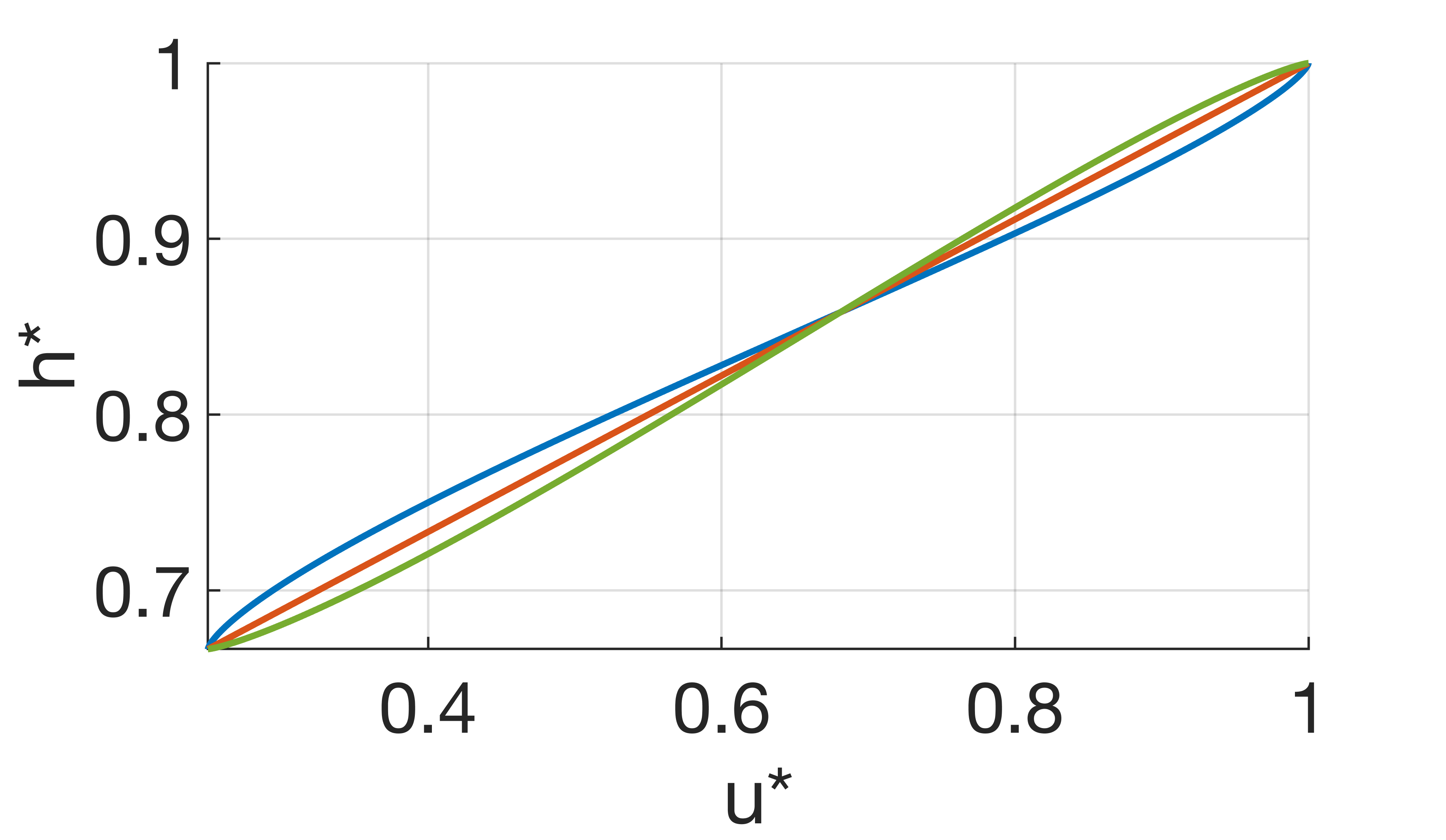}
         \caption{\(h/h_\infty\) as a function of \(u/u_\infty\)}
         \label{fig:five over x}
     \end{subfigure}
        \caption{\(u^*\), \(v^*\), \(h^*\), \(\kappa^*\), \(Y_O\), and \(Y_F\) at \(x^*=1\) for non-reactive Cases 4a, 1, and 4b. Prandtl number varies from $0.7$ to $1.3$. Ambient counterflow strain rate is constant with $x^*$.}
        \label{fig:three graphs}
\end{figure*}

\clearpage
\subsection{Non-reactive Layer with $\kappa_\infty$ and $\kappa_{-\infty}$ as a Function of $x^*$}

Now, we solve the case where the strain rate \(\kappa^*\) is allowed to vary as \(1/x^*\).  The analytical solution for \(A_\infty\) and \(A_{-\infty}\) yielded two solutions (one positive pair and one negative pair); thus, here both cases are examined.
In the case where the constants are positive, there is inflow with compressive normal strain rate in the $y$-direction and outflow with extensional normal strain rate in the $z$-direction. For this case, Fig. 6(a) shows that the inward \(y\)-velocity decreases with \(x^*\). If the imposed compressive strain decreases with $x^*$, then so will the rate of inflow in the mixing layer. Fig. 6(b) shows the $1/x^*$ behavior of $\kappa^*$ in the free-streams.  Clearly, the counterflow strength and its influence decreases here with increasing downstream distance.

 In Fig. 7(b) for the case with the negative sign, the outer flow has inflow with compressive normal strain rate in the $z$-direction and outflow with extensional normal strain rate in the $y$-direction. However, reversal occurs in the interior mixing region.  \(\kappa^*\) is negative in the free-streams and positive in the mixing regions, implying that there is inflow in the \(z\)-direction in the free-streams and outflow in the \(z\)-direction in the mixing region. Nonetheless, the $-1/x^*$ behavior of $\kappa^*$ is clear in the free streams of Fig. 7(b). Fig. 7(a) shows the flow slowing in the \(y\)-direction in the free-streams because the magnitude of the imposed extensive strain is decreasing with $x^*$.

In comparing the results for the two cases with different signs for the constants, it appears that a tendency towards vortex stretching occurs even when the outer flow (with the negative constant) would favor vortex shrinking or compression.  One might question the stability of the case with the negative constant. Stability analysis is left for a future task.

\newpage
\begin{figure*}
     \centering
     \begin{subfigure}[b]{0.49\textwidth}
         \centering
         \includegraphics[width=\textwidth]{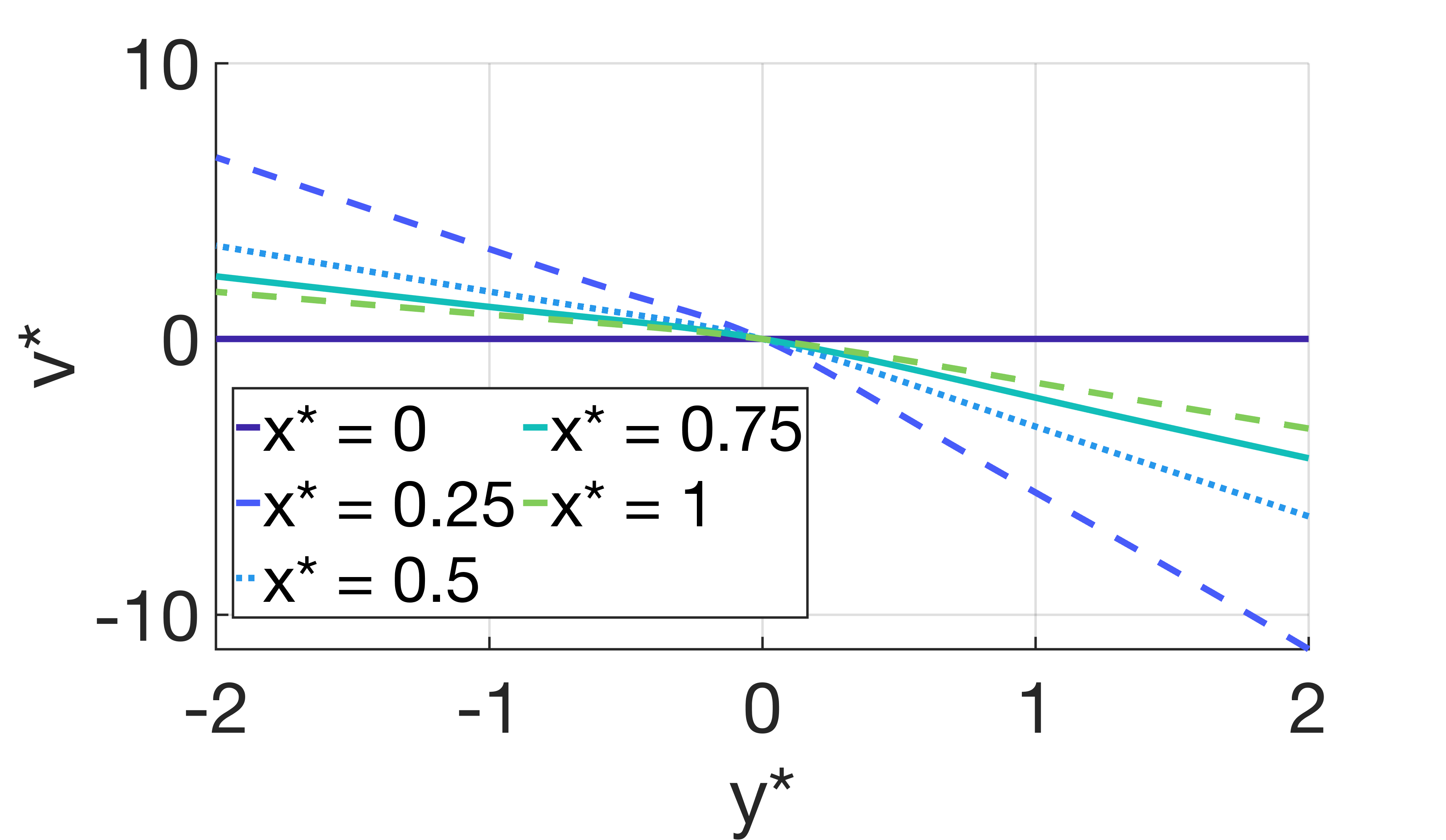}
         \caption{$v^*(y^*)$}
         \label{fig:y equals x}
     \end{subfigure}
     \hfill
     \begin{subfigure}[b]{0.49\textwidth}
         \centering
         \includegraphics[width=\textwidth]{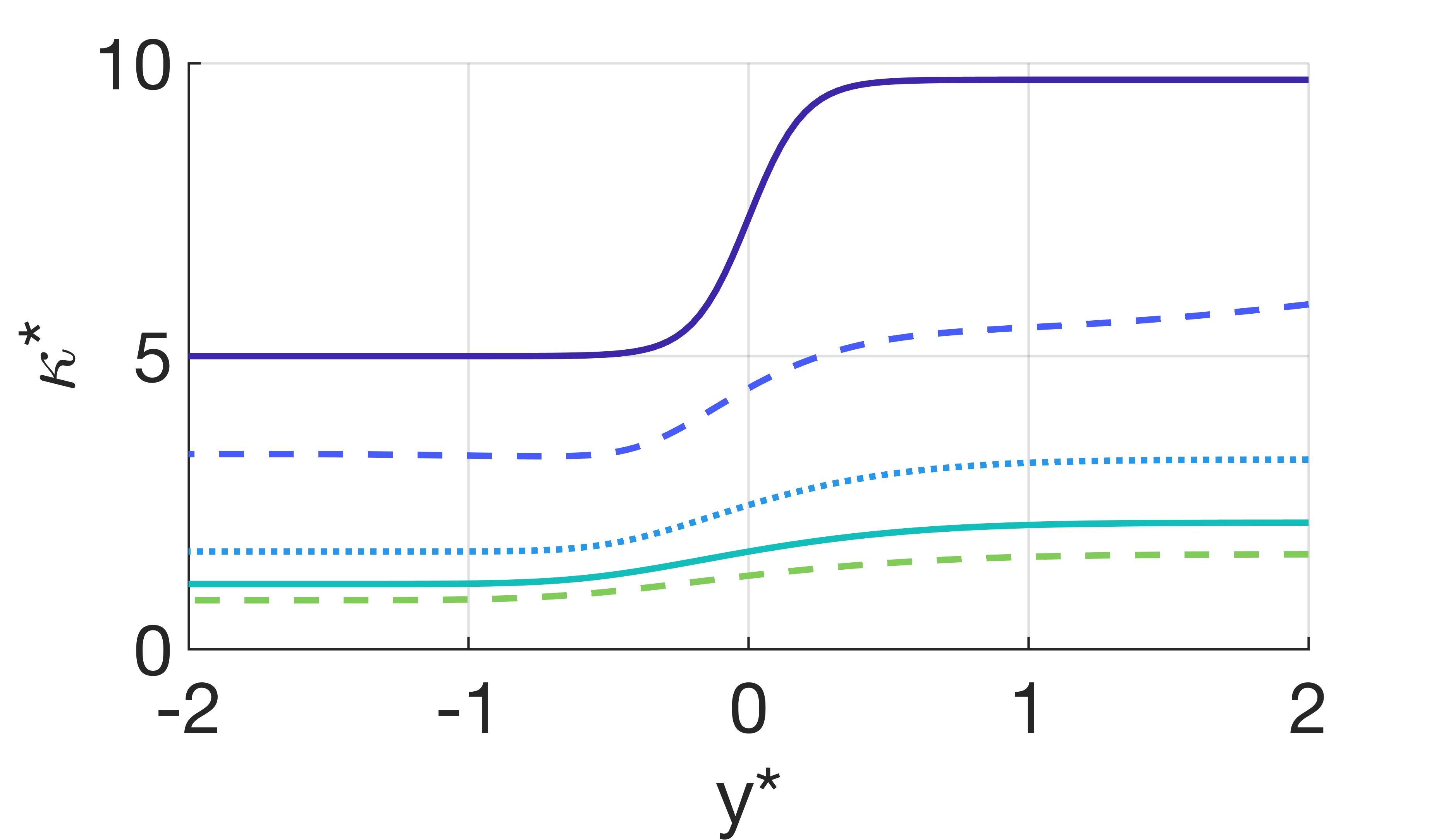}
          \caption{$\kappa^*(y^*)$}
         \label{fig:three sin x}
     \end{subfigure}
     \caption{Results for non-reactive Case 5a at \(x^*=1\). The positive value of \(A_{-\infty}\)} is considered.
        \label{fig:three graphs}
\end{figure*}

\begin{figure*}
     \centering
     \begin{subfigure}[b]{0.49\textwidth}
         \centering
         \includegraphics[width=\textwidth]{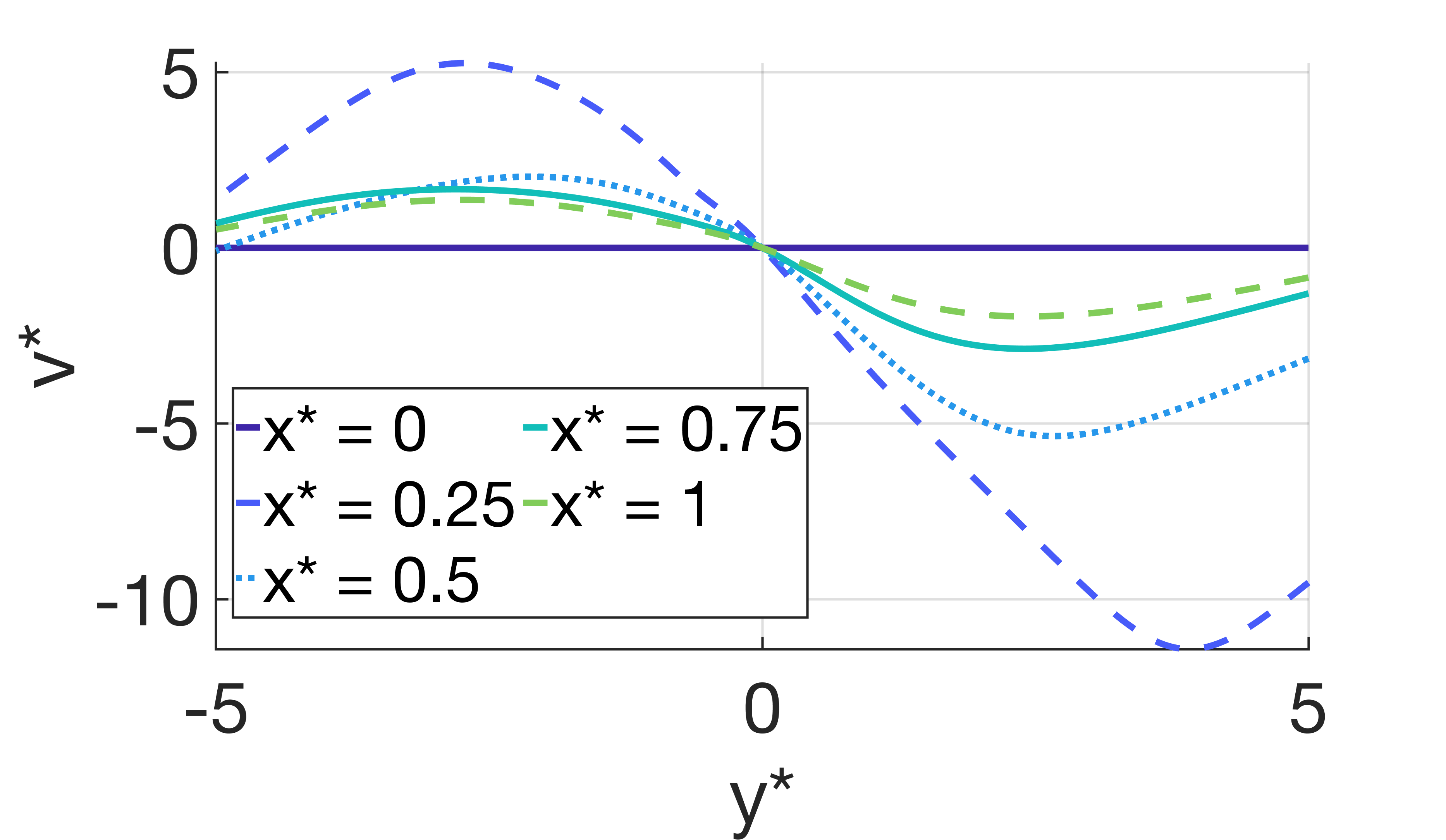}
         \caption{$v^*(y^*)$}
         \label{fig:y equals x}
     \end{subfigure}
     \hfill
     \begin{subfigure}[b]{0.49\textwidth}
         \centering
         \includegraphics[width=\textwidth]{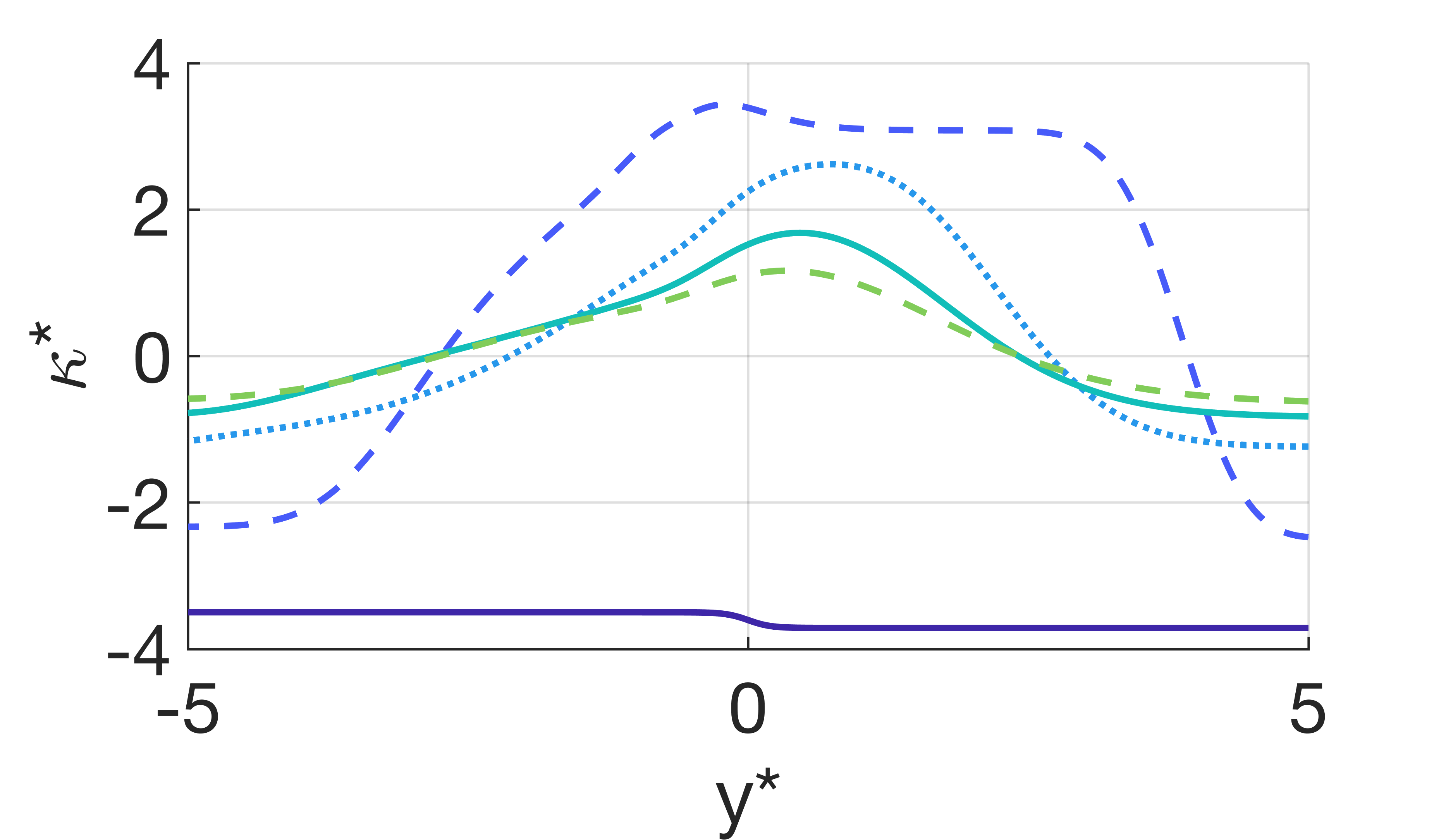}
         \caption{$\kappa^*(y^*)$}
         \label{fig:three sin x}
     \end{subfigure}
     \caption{Results for non-reactive Case 5b at \(x^*=1\). The negative value of \(A_{-\infty}\)} is considered.
        \label{fig:three graphs}
\end{figure*}

Let us compare  the behaviors far downstream for the cases from the prior subsection and this subsection. Figs. 8(a) and (b) show downstream similarity for $u^*$ and $h^*$ with regard to $y^*$. With an imposed normal strain that is constant along the stream, the shear layer's growth is stopped, and the need for a new similarity coordinate is eliminated because $y^*$ becomes the similarity variable downstream. Thus, it is clear that, asymptotically with increasing downstream distance, the shear-layer width becomes constant and a quasi-one-dimensional flow results, although three components of velocity are involved. In the incompressible limit, that asymptote for the non-reactive case is exactly Burgers stretched vortex sheet.

A different observation is shown in Figs. 8(c) and (d) for the case where the imposed normal strain rate varies as the reciprocal of downstream distance. the shear-layer width grows approximately with $\sqrt{x^*}$, allowing a similarity coordinate, $\eta$, as seen in classical shear layer theory.

\begin{figure*}
     \centering
     \begin{subfigure}[b]{0.49\textwidth}
         \centering
         \includegraphics[width=\textwidth]{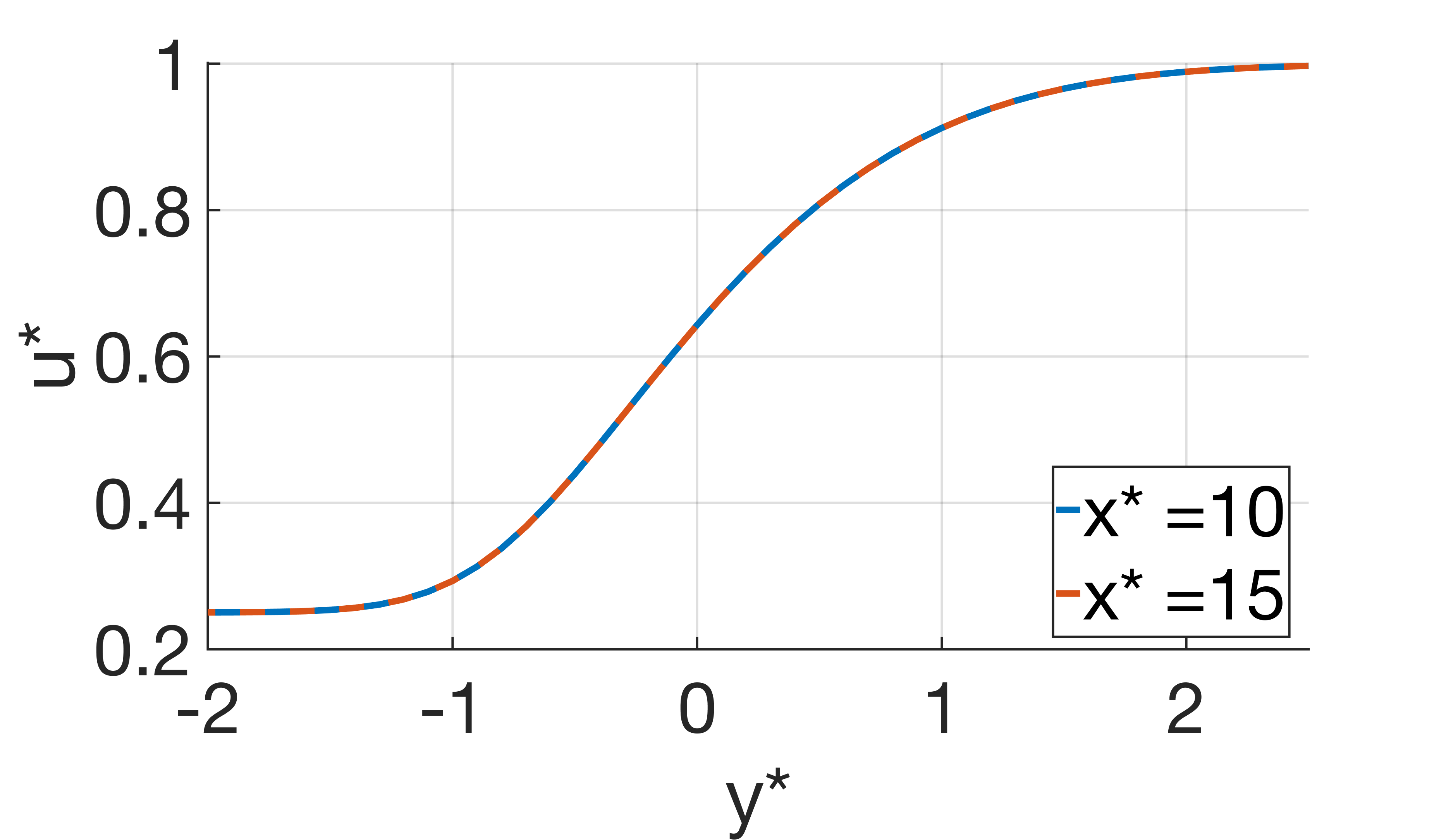}
         \caption{$u^*(y^*)$}
         \label{fig:y equals x}
     \end{subfigure}
     \begin{subfigure}[b]{0.49\textwidth}
         \centering
         \includegraphics[width=\textwidth]{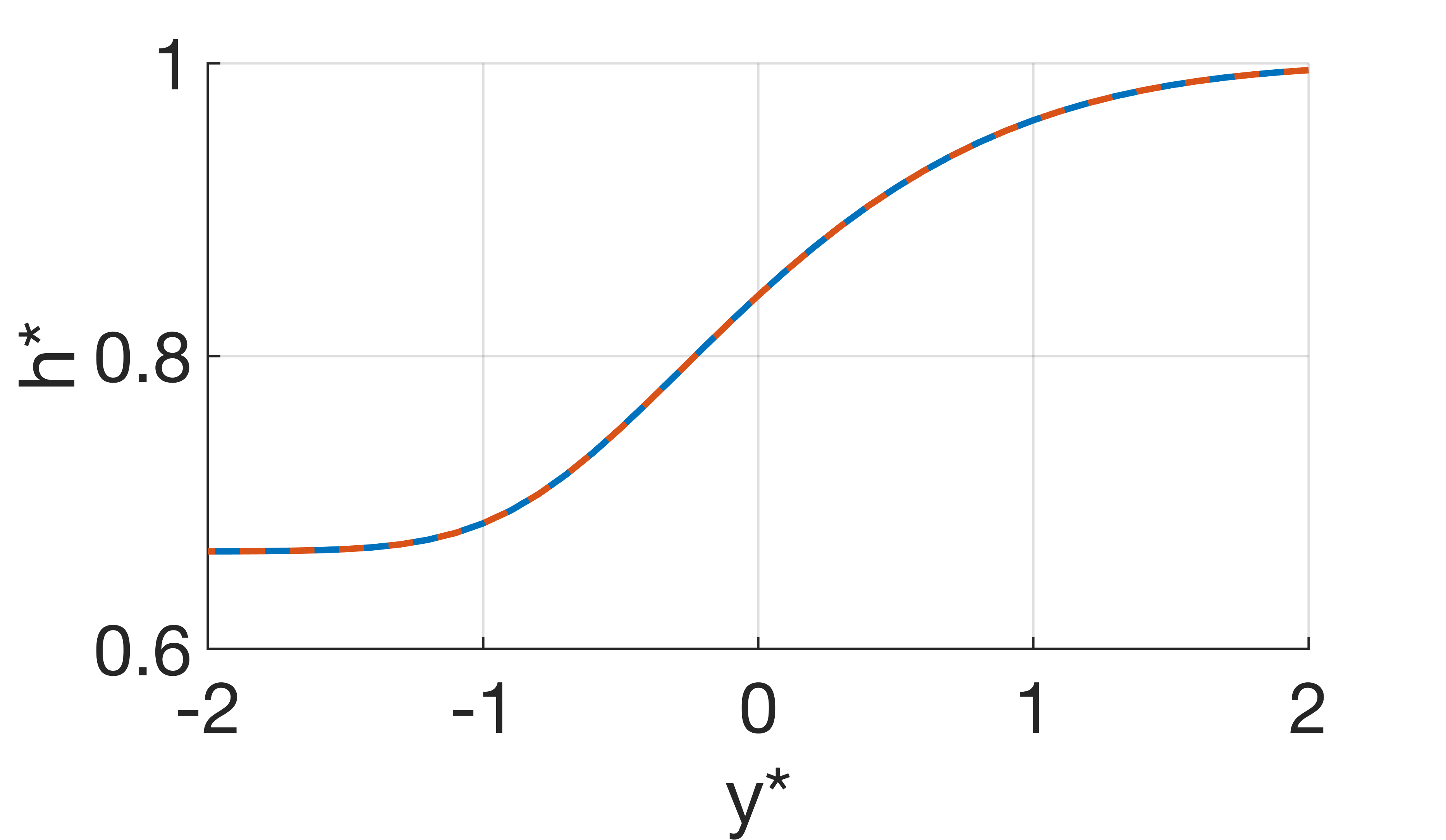}
         \caption{$h^*(y^*)$}
         \label{fig:y equals x}
     \end{subfigure}
     \hfill
     \begin{subfigure}[b]{0.49\textwidth}
         \centering
         \includegraphics[width=\textwidth]{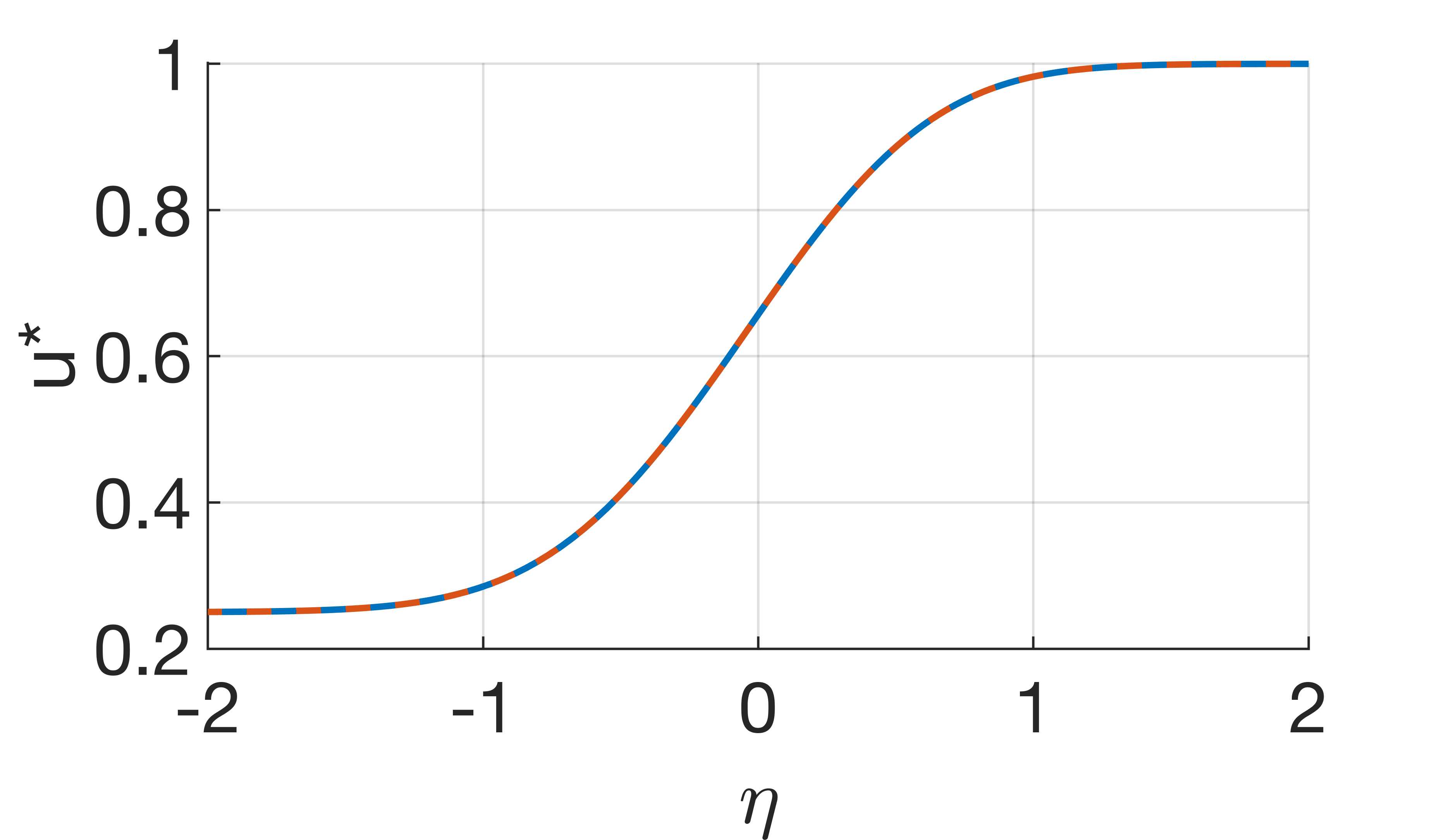}
         \caption{$u^*(\eta)$}
         \label{fig:y equals x}
     \end{subfigure}
     \begin{subfigure}[b]{0.49\textwidth}
         \centering
         \includegraphics[width=\textwidth]{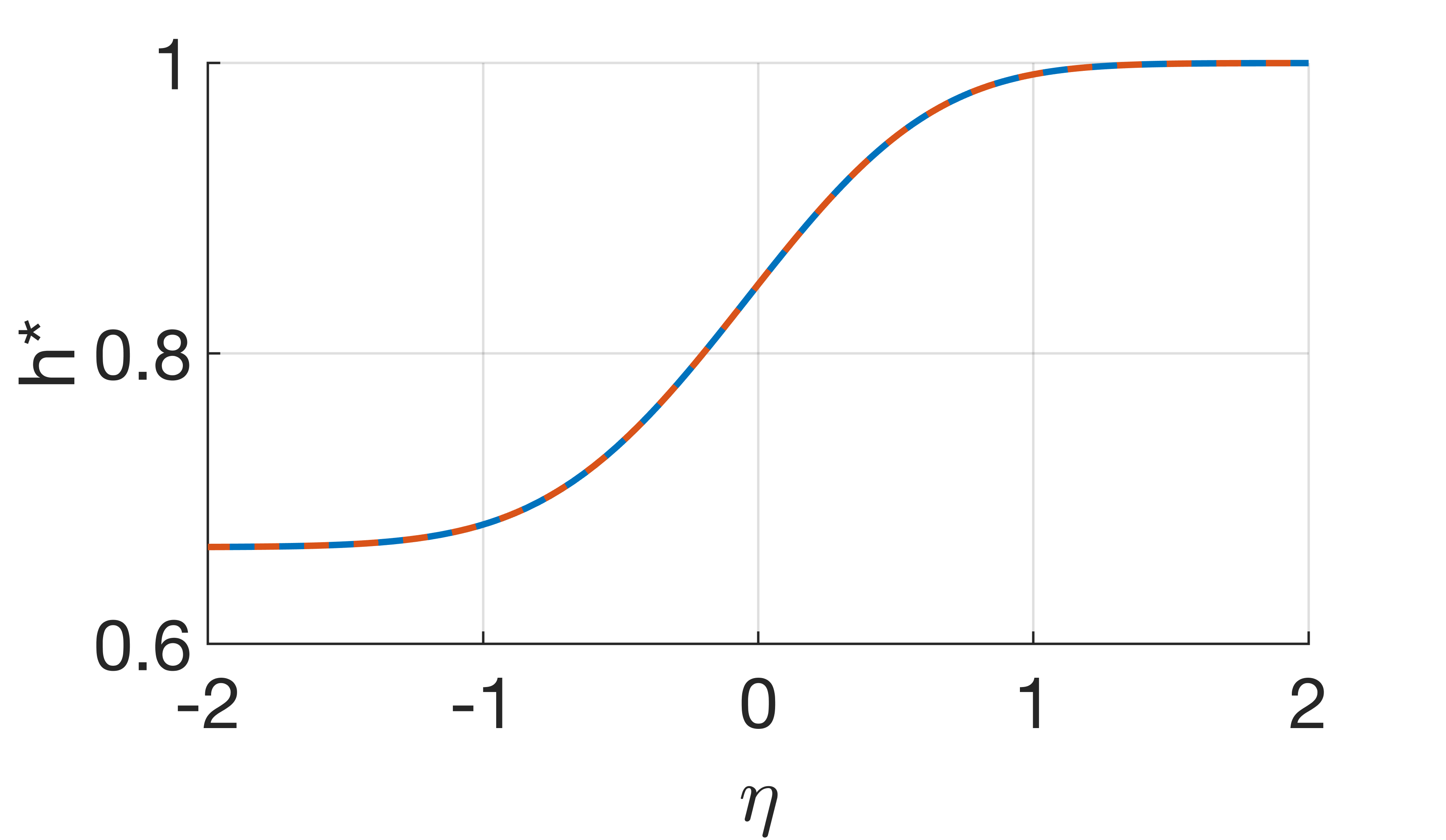}
         \caption{$h^*(\eta)$}
         \label{fig:y equals x}
     \end{subfigure}
     \caption{Downstream similarity for $u^*$ and $h^*$ for the non-reactive case. Subfigures (a) and (b) are of Case 1 where the imposed normal strain is constant with $x^*$. Subfigures (c) and (d) are of Case 5a where the imposed normal strain varies as $1/x^*$.}
        \label{fig:three graphs}
\end{figure*}

\subsection{ Mixing Layer with a Diffusion Flame}

Here, we consider pure propane fuel in the free stream at $y = -\infty$ and pure oxygen in the free stream at $y = \infty$. Upstream, a region of high temperature is provided in the layer. Therefore, we expect mixing of the two reactants in the shear layer, ignition,  and establishment of a diffusion flame.

Fig. 9 depicts the base reactive case (Case 6) at several stations in $x^*$ from $0$ to $5$. The results show that, if the imposed strain is constant with $x^*$, the reacting shear layer asymptotically reaches a constant width with increasing $x^*$. Here, we see an analogy with Burgers stretched vortex sheet with the resulting asymptotic quasi-one-dimensional flow and albeit that three velocity components exist.

Fig. 9(c) shows an increase in peak temperature followed by a converging decrease with increasing downstream distance. The flame moves in the positive $y$-direction, towards the oxygen-rich, higher-speed stream. This temperature peak causes a local decrease in density, which increases the outflow velocity in the $z$-direction as seen in Fig. 9(d). That is, the flow experiences a greater extensional strain in the $z$-direction near the flame. Where the density has decreased, the increased $w^*$-component of velocity maintains the imposed counterflow mass flux. Note that there is no increase of mass flux in the x-direction. The $y$-velocity above and below this region is locally increased near the reaction zone in order to maintain the mass flux at a lower density value, as seen in Fig. 9(b). Fig. 9(f) shows that the reaction rate of fuel decreases substantially downstream while the reaction zone shifts towards the oxygen-rich stream. The difference of magnitude indicates that a decreasing burning rate is required with downstream distance. The mixing-layer thickness varies to accommodate the reduced heat flux from the flame.

\begin{figure*}
     \centering
     \begin{subfigure}[b]{0.49\textwidth}
         \centering
         \includegraphics[width=\textwidth]{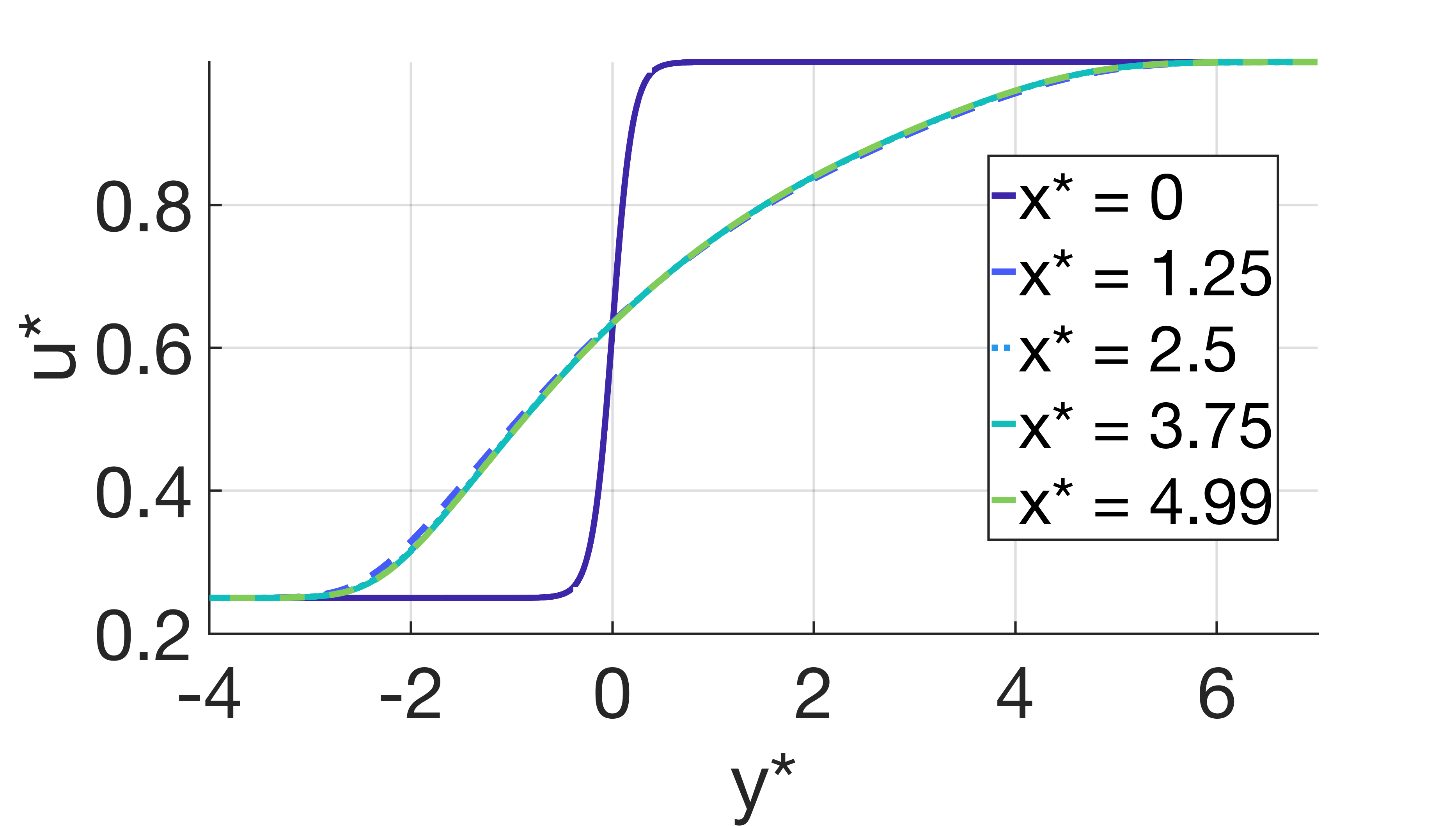}
         \caption{\(u^*(y^*)\)}
         \label{fig:y equals x}
     \end{subfigure}
     \hfill
     \begin{subfigure}[b]{0.49\textwidth}
         \centering
         \includegraphics[width=\textwidth]{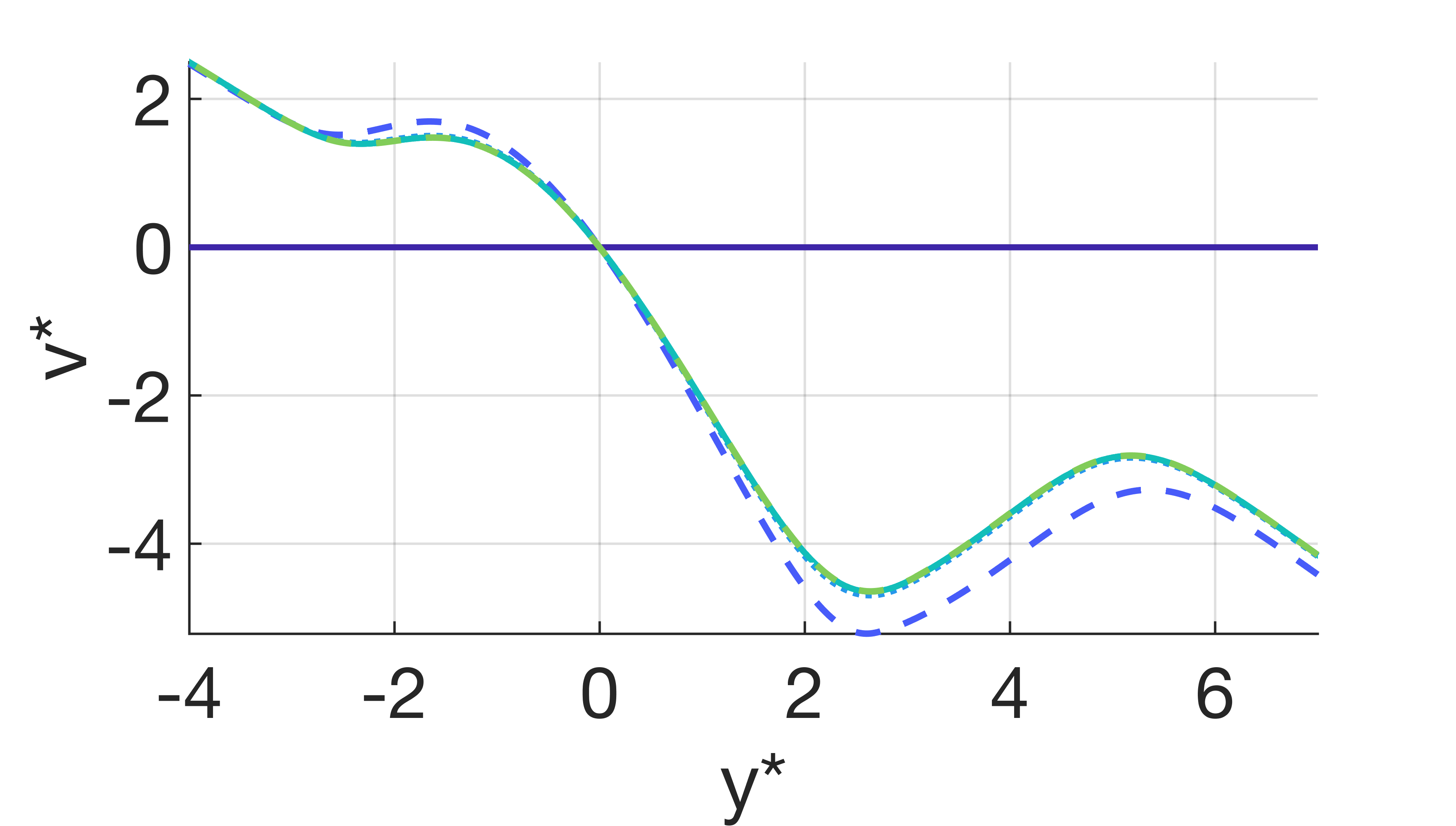}
         \caption{\(v^*(y^*)\)}
         \label{fig:three sin x}
     \end{subfigure}
     \hfill
     \begin{subfigure}[b]{0.49\textwidth}
         \centering
         \includegraphics[width=\textwidth]{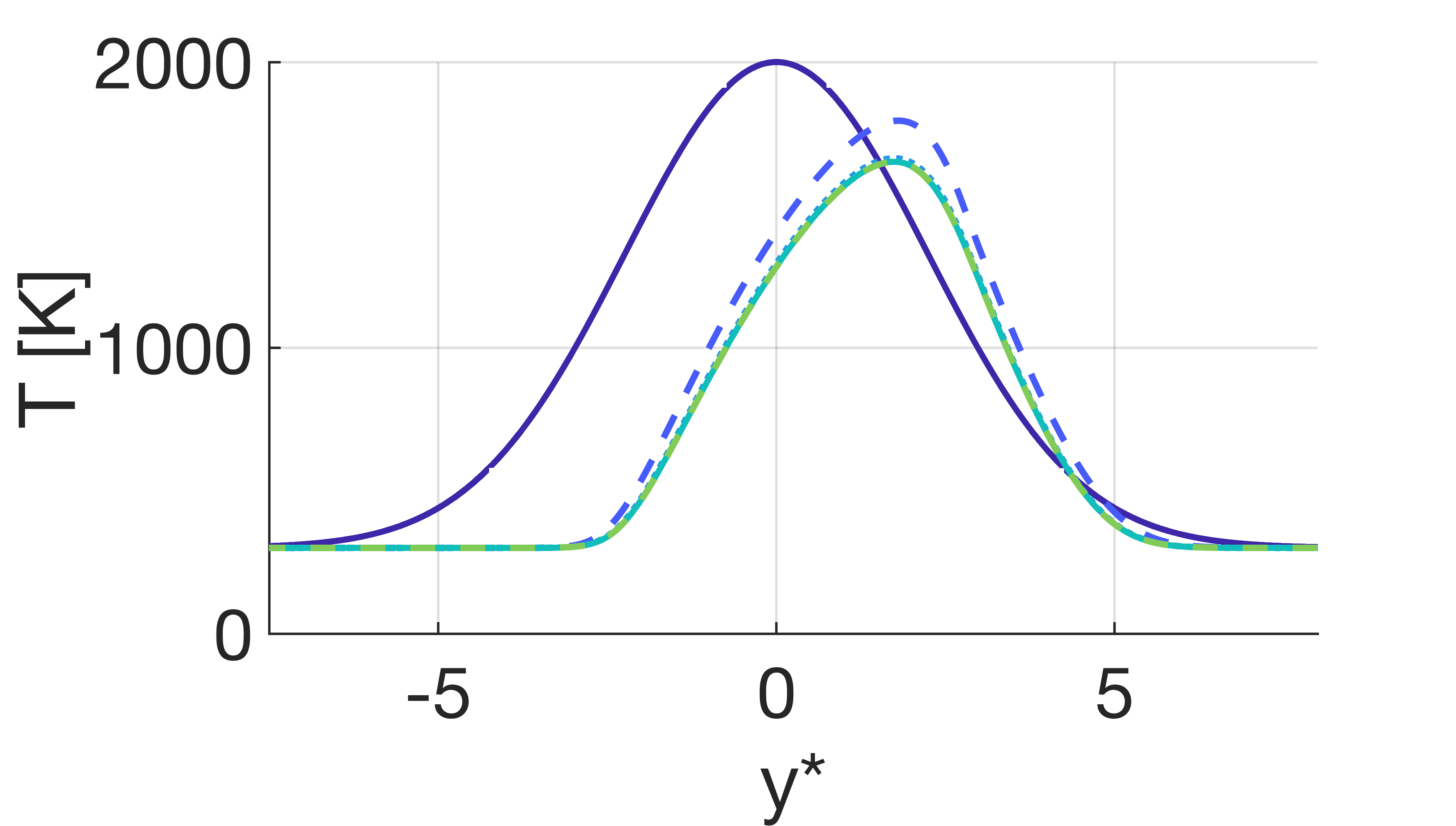}
         \caption{\(T(y^*)\)}
         \label{fig:five over x}
     \end{subfigure}
     \begin{subfigure}[b]{0.49\textwidth}
         \centering
         \includegraphics[width=\textwidth]{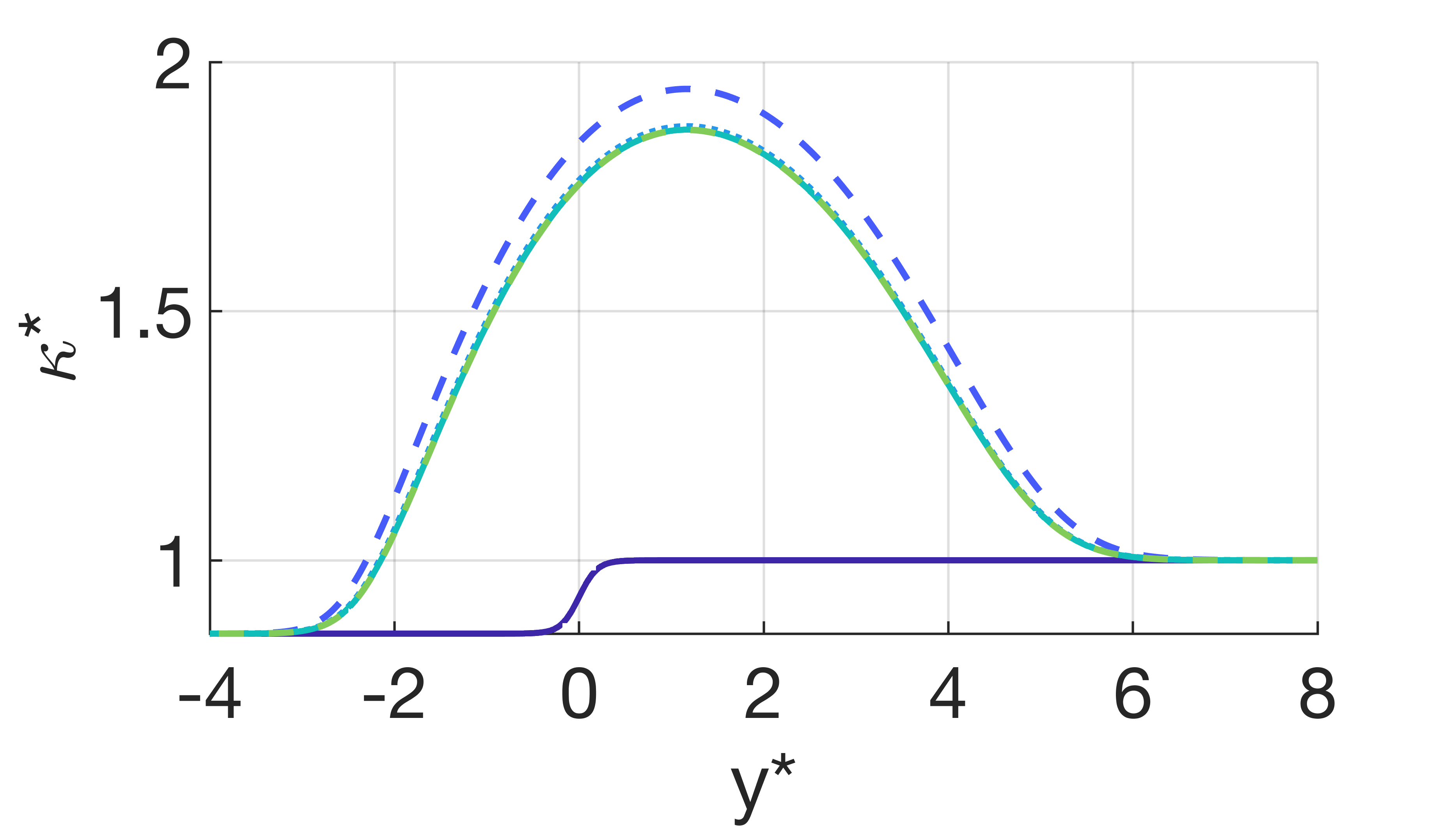}
         \caption{\(\kappa^*(y^*)\)}
         \label{fig:five over x}
     \end{subfigure}
     \begin{subfigure}[b]{0.49\textwidth}
         \centering
         \includegraphics[width=\textwidth]{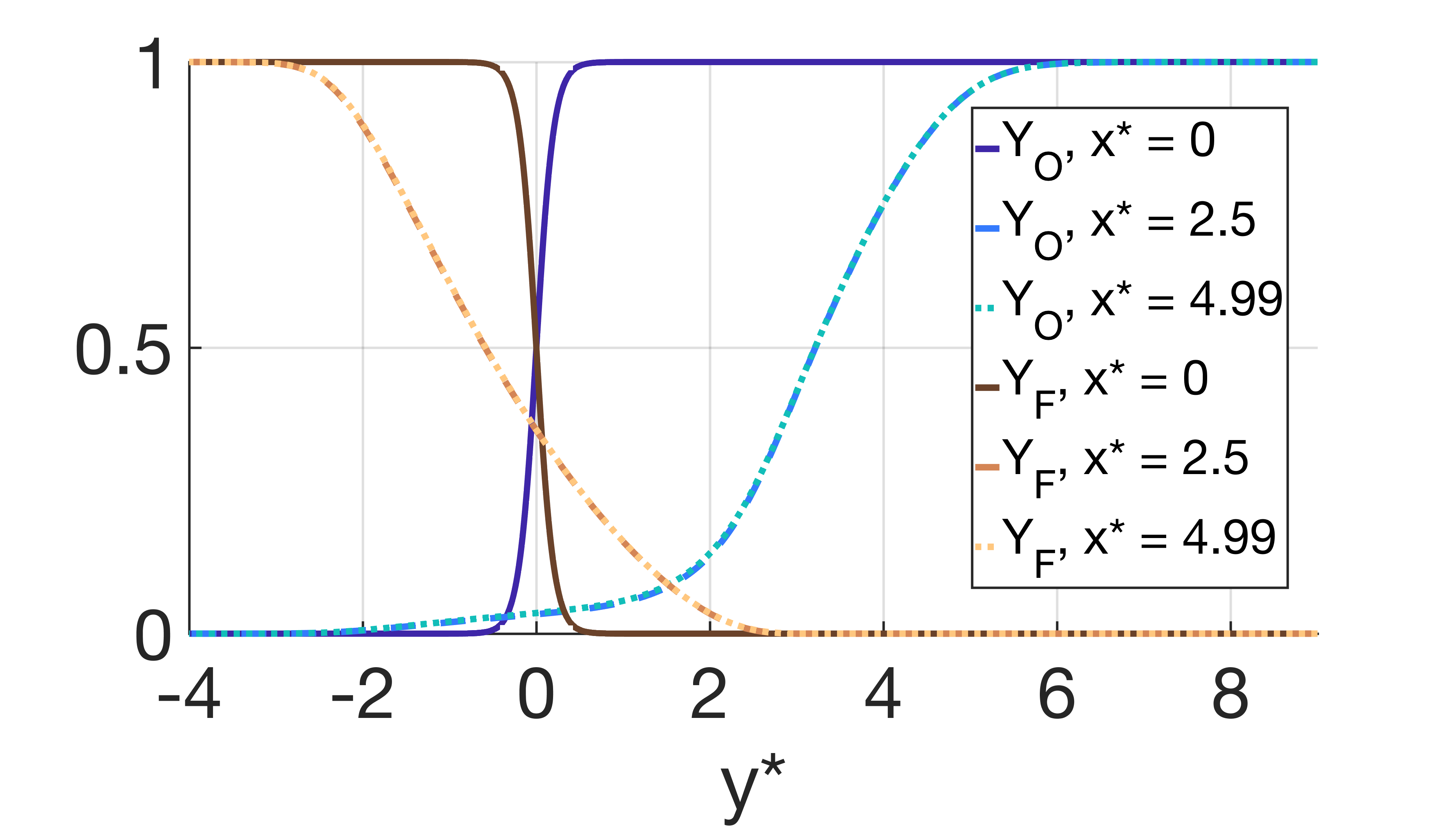}
         \caption{\(Y_O(y^*)\), \(Y_F(y^*)\)}
         \label{fig:five over x}
     \end{subfigure}
     \begin{subfigure}[b]{0.49\textwidth}
         \centering
         \includegraphics[width=\textwidth]{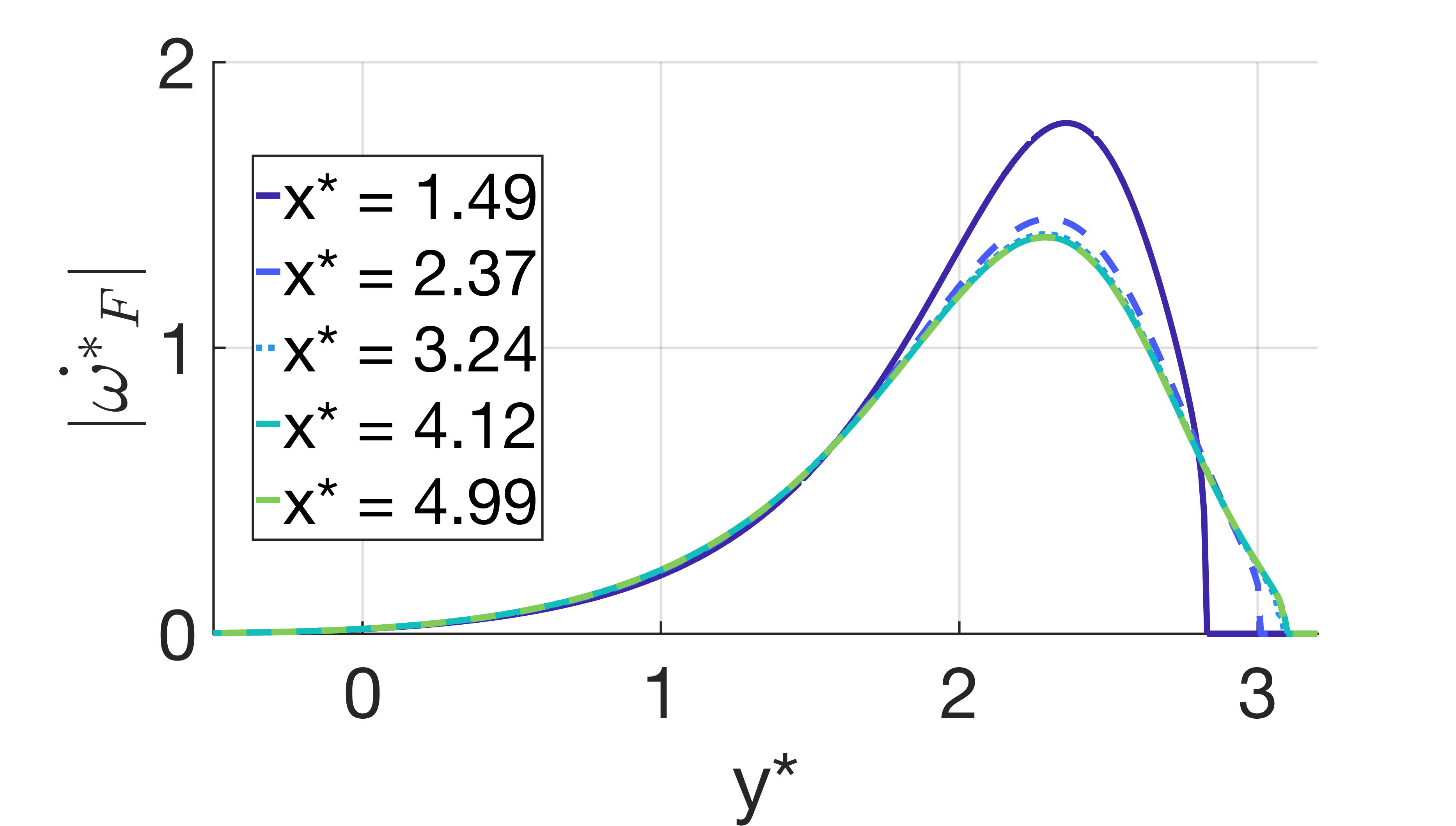}
         \caption{\(\abs{\dot{\omega}_F^*(y^*)}\)}
         \label{fig:five over x}
     \end{subfigure}
     \caption{Results for reactive Case 6 from $x^*=0$ to $x^*=3$ where $f^*=1$.}
\end{figure*}

Fig. 10 compares the effects of two-dimensional flow (\(f^*=0\)) to three-dimensional flow (\(f^* \neq 0\)) for the reactive case. Fig. 10(d) shows that, in the two-dimensional case, the flame grows to be much wider and hotter than what was initialized upstream. When counterflow is imposed, the flame is constricted to smaller widths. This compression increases temperature gradients in the $y$-direction, resulting in faster heat transfer from the reaction zone. A sufficient amount of counterflow ($f^*=2$) will extinguish the flame, altogether. Residence time is essentially the reciprocal of strain rate. Although the flame with counterflow is much smaller and cooler than without, Fig. 10(f) shows that the magnitude of reaction rate between the ($f^*=0$) and ($f^*=1$) cases are very similar. The difference is that the absence of counterflow allows the reaction zone to drift to a greater $y$-value, towards the oxygen stream. Fig. 11(f) shows that the reaction zone will still drift towards the oxygen stream, even if the oxygen stream is slower than the fuel stream.

\begin{figure*}
     \centering
     \begin{subfigure}[b]{0.49\textwidth}
         \centering
         \includegraphics[width=\textwidth]{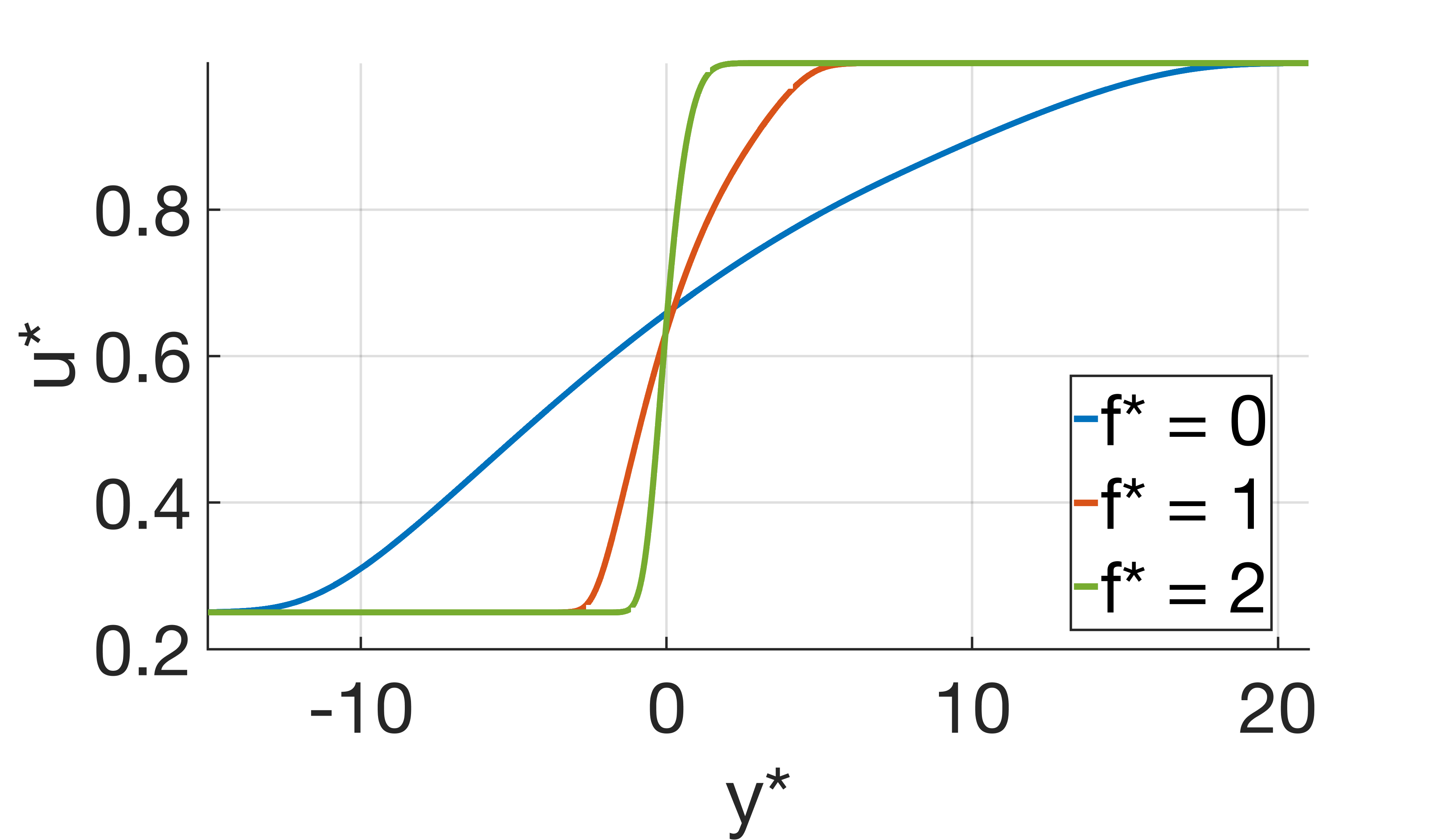}
         \caption{\(u^*(y^*)\)}
         \label{fig:y equals x}
     \end{subfigure}
     \hfill
     \begin{subfigure}[b]{0.49\textwidth}
         \centering
         \includegraphics[width=\textwidth]{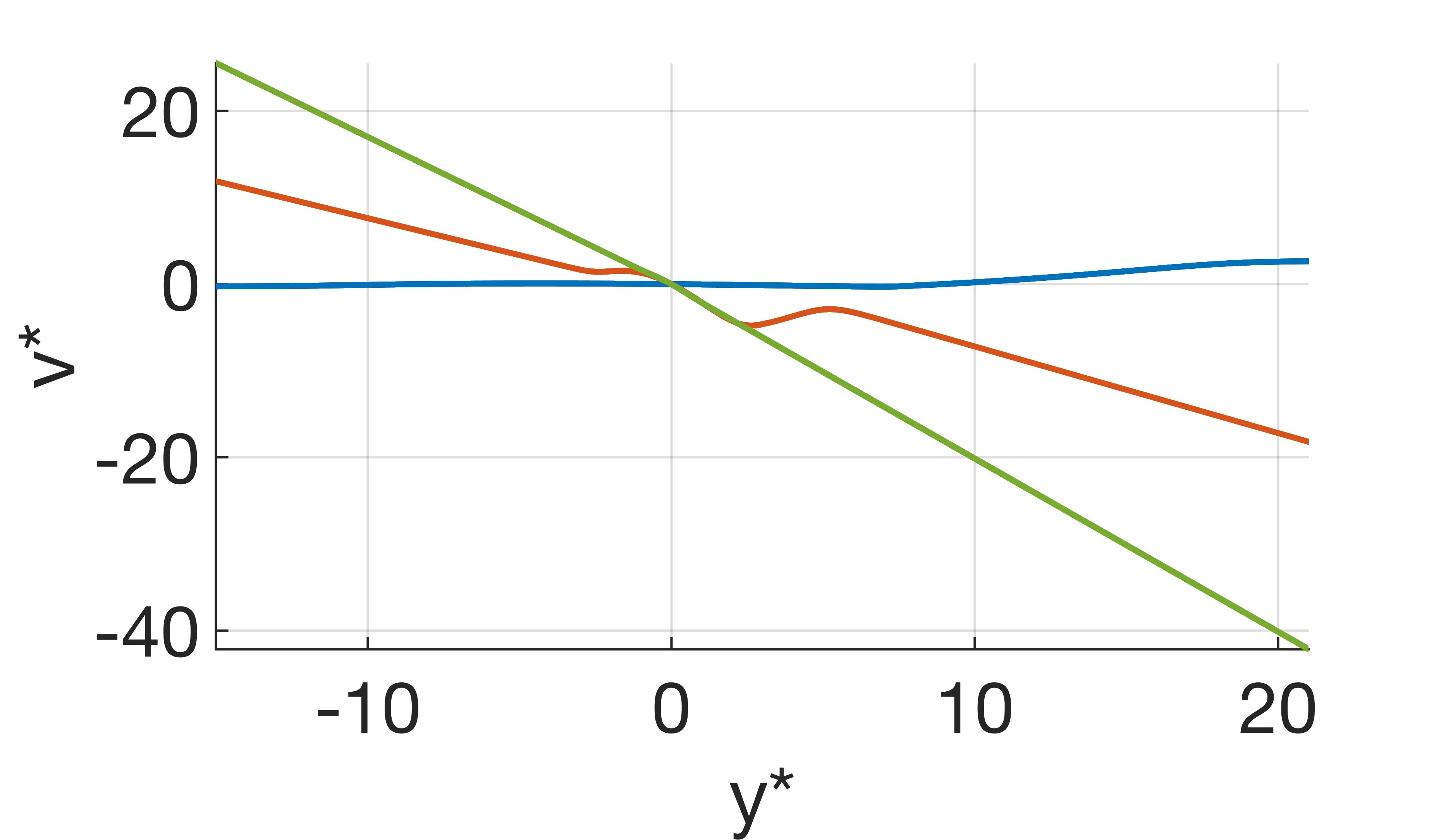}
         \caption{\(v^*(y^*)\)}
         \label{fig:three sin x}
     \end{subfigure}
     \hfill
     \begin{subfigure}[b]{0.49\textwidth}
         \centering
         \includegraphics[width=\textwidth]{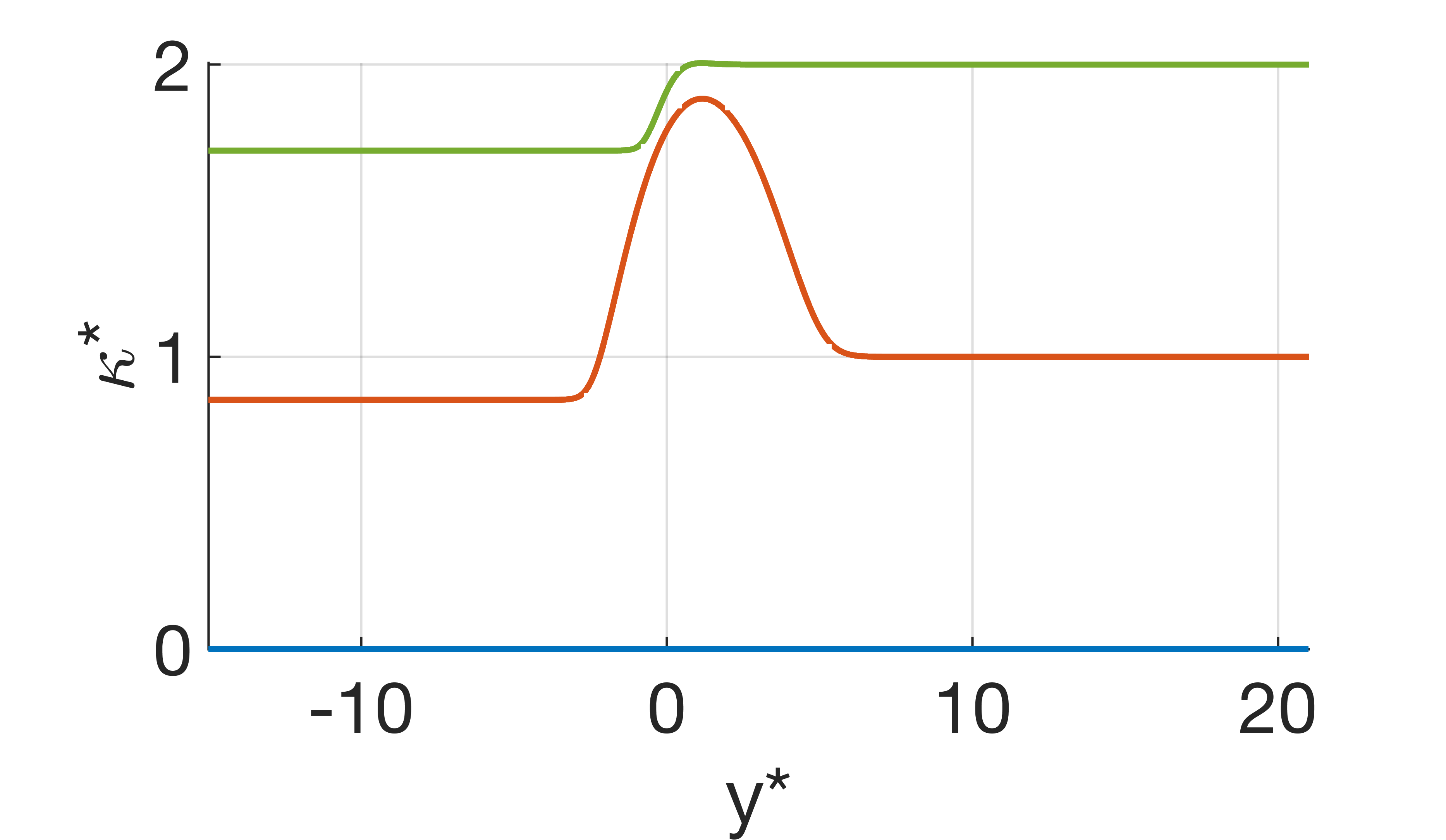}
         \caption{\(\kappa^*(y^*)\)}
         \label{fig:five over x}
     \end{subfigure}
     \begin{subfigure}[b]{0.49\textwidth}
         \centering
         \includegraphics[width=\textwidth]{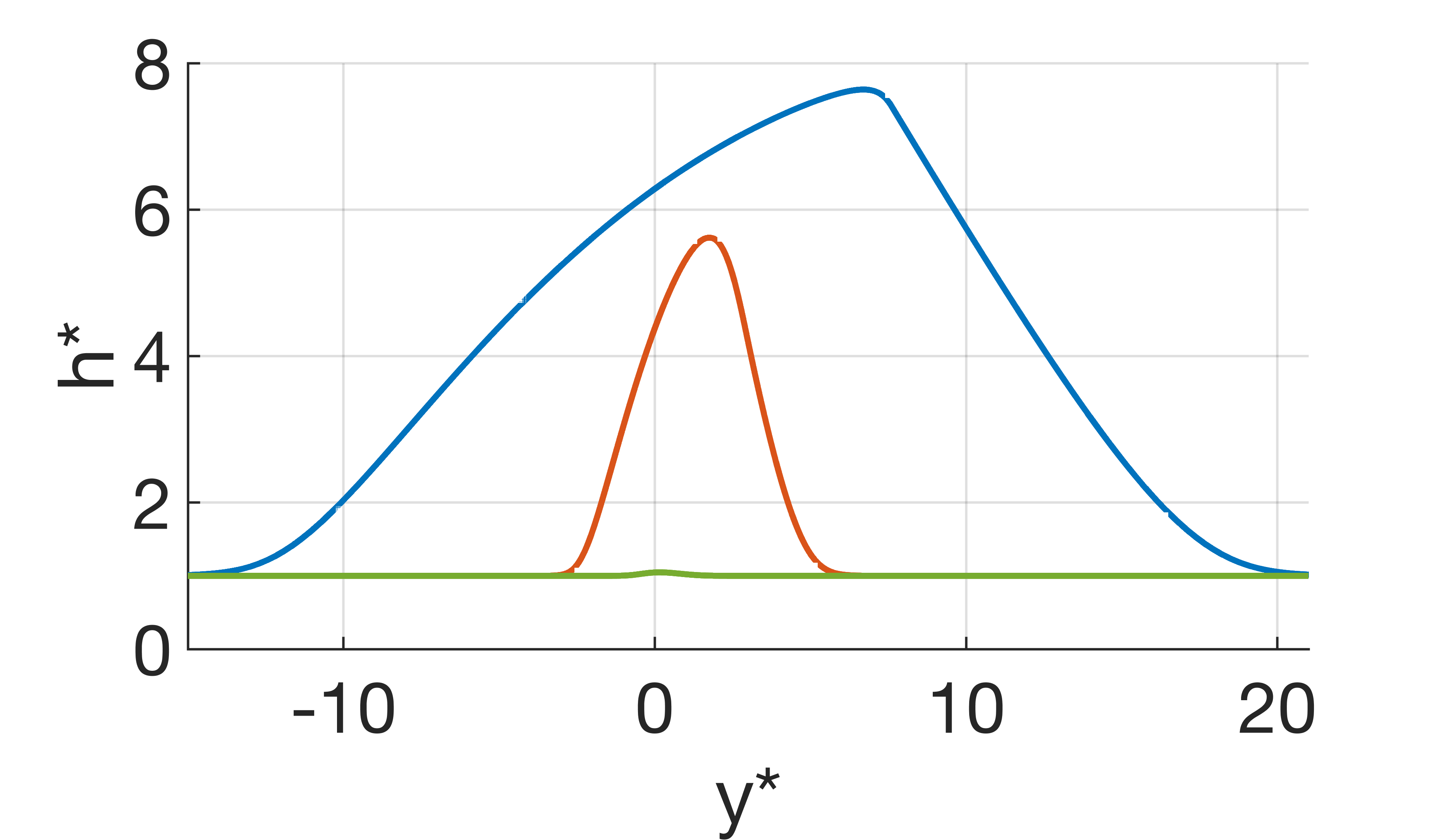}
         \caption{\(h^*(y^*)\)}
         \label{fig:five over x}
     \end{subfigure}
     \begin{subfigure}[b]{0.49\textwidth}
         \centering
         \includegraphics[width=\textwidth]{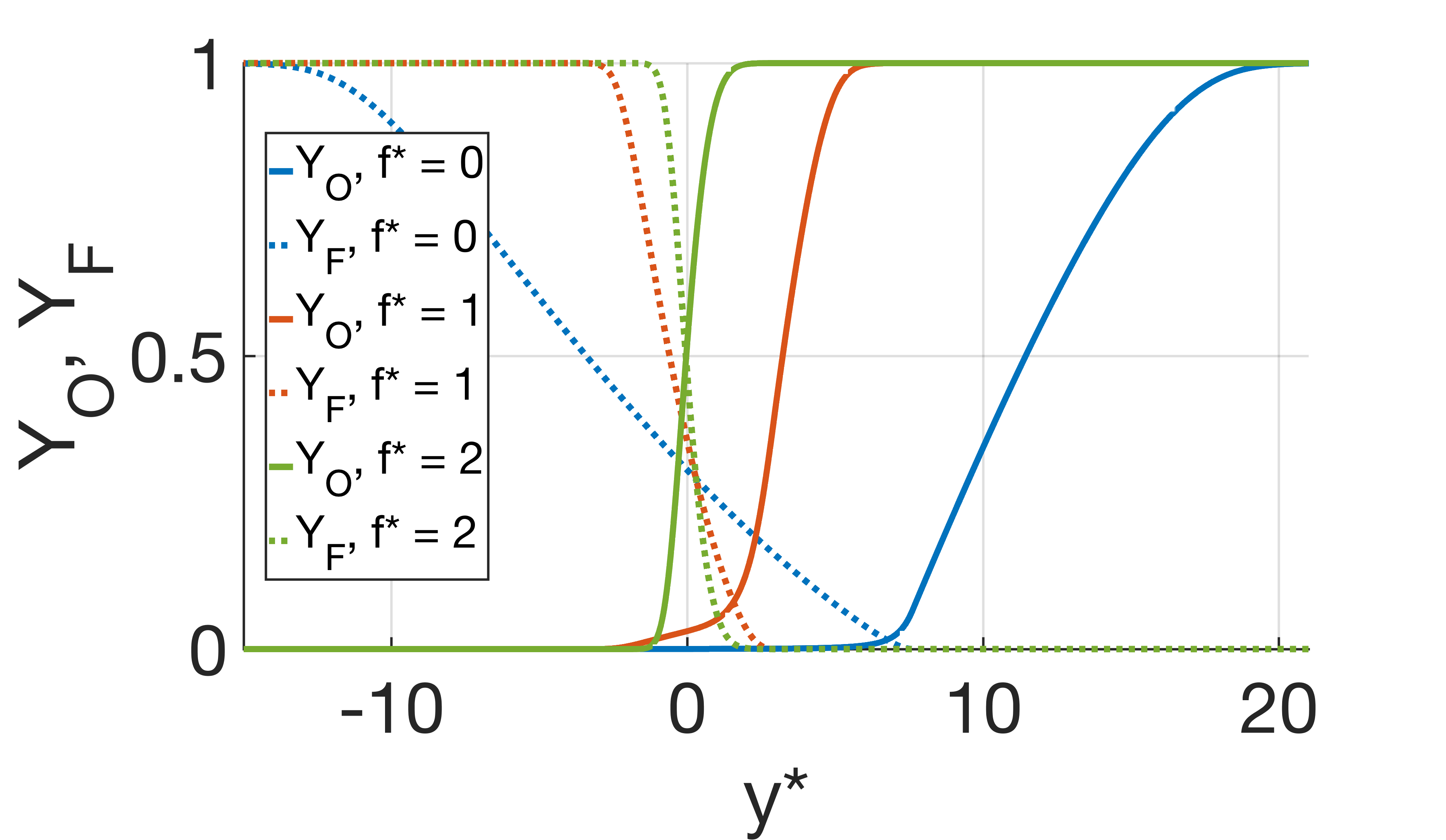}
         \caption{\(Y_O(y^*), Y_F(y^*)\)}
         \label{fig:five over x}
     \end{subfigure}
     \begin{subfigure}[b]{0.49\textwidth}
         \centering
         \includegraphics[width=\textwidth]{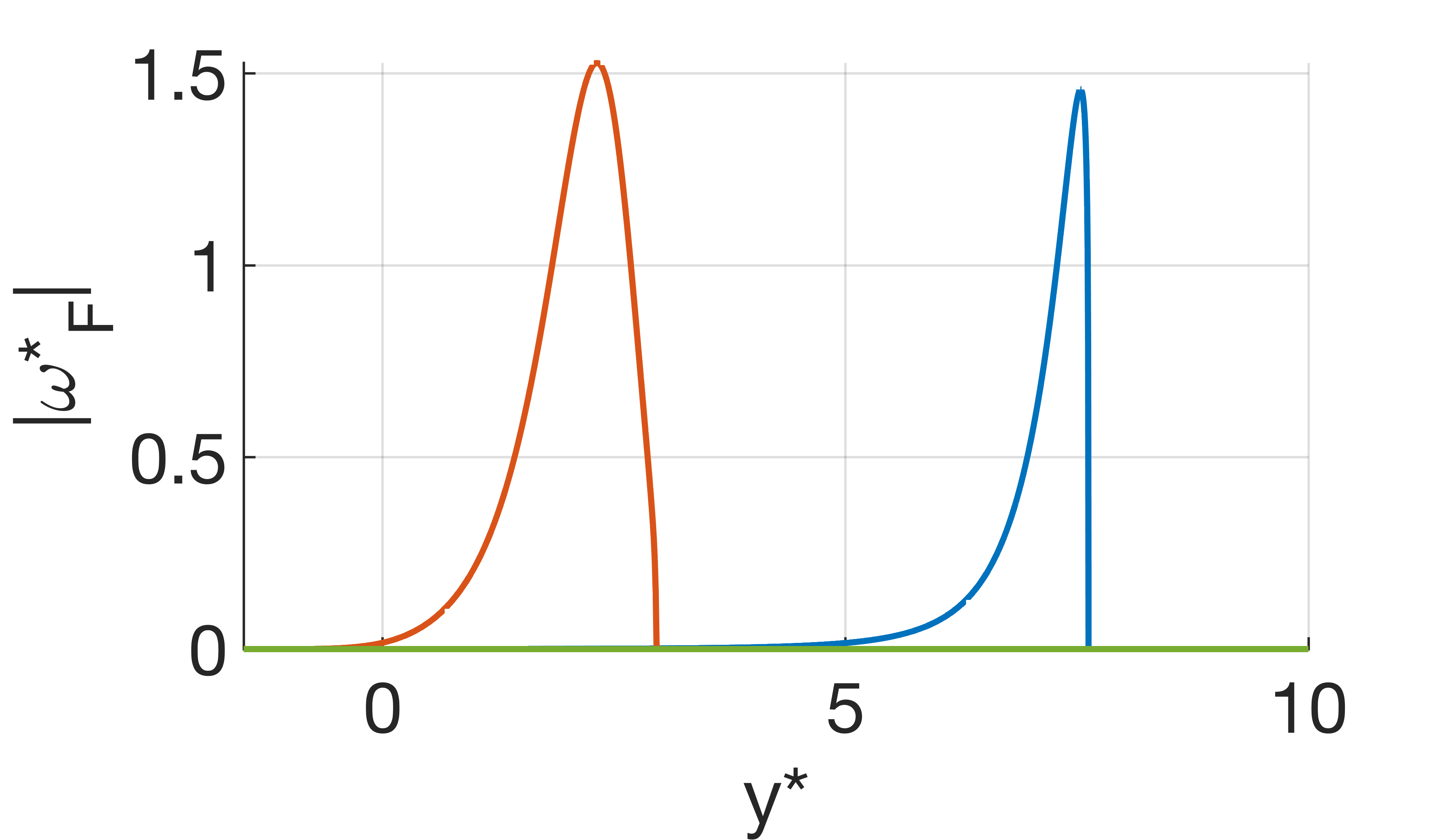}
         \caption{\(\abs{\dot{\omega}_F^*(y^*)}\)}
         \label{fig:five over x}
     \end{subfigure}
        \caption{\(u^*\), \(v^*\), \(h^*\), \(\kappa^*\), \(Y_O\), \(Y_F\), and $\dot{\omega}^*_F$ at \(x^*=2\) for reactive Cases 7a, 6, and 7b. Counterflow strain rate varies from $0$ to $2$.}
        \label{fig:three graphs}
\end{figure*}

\begin{figure*}
     \centering
     \begin{subfigure}[b]{0.49\textwidth}
         \centering
         \includegraphics[width=\textwidth]{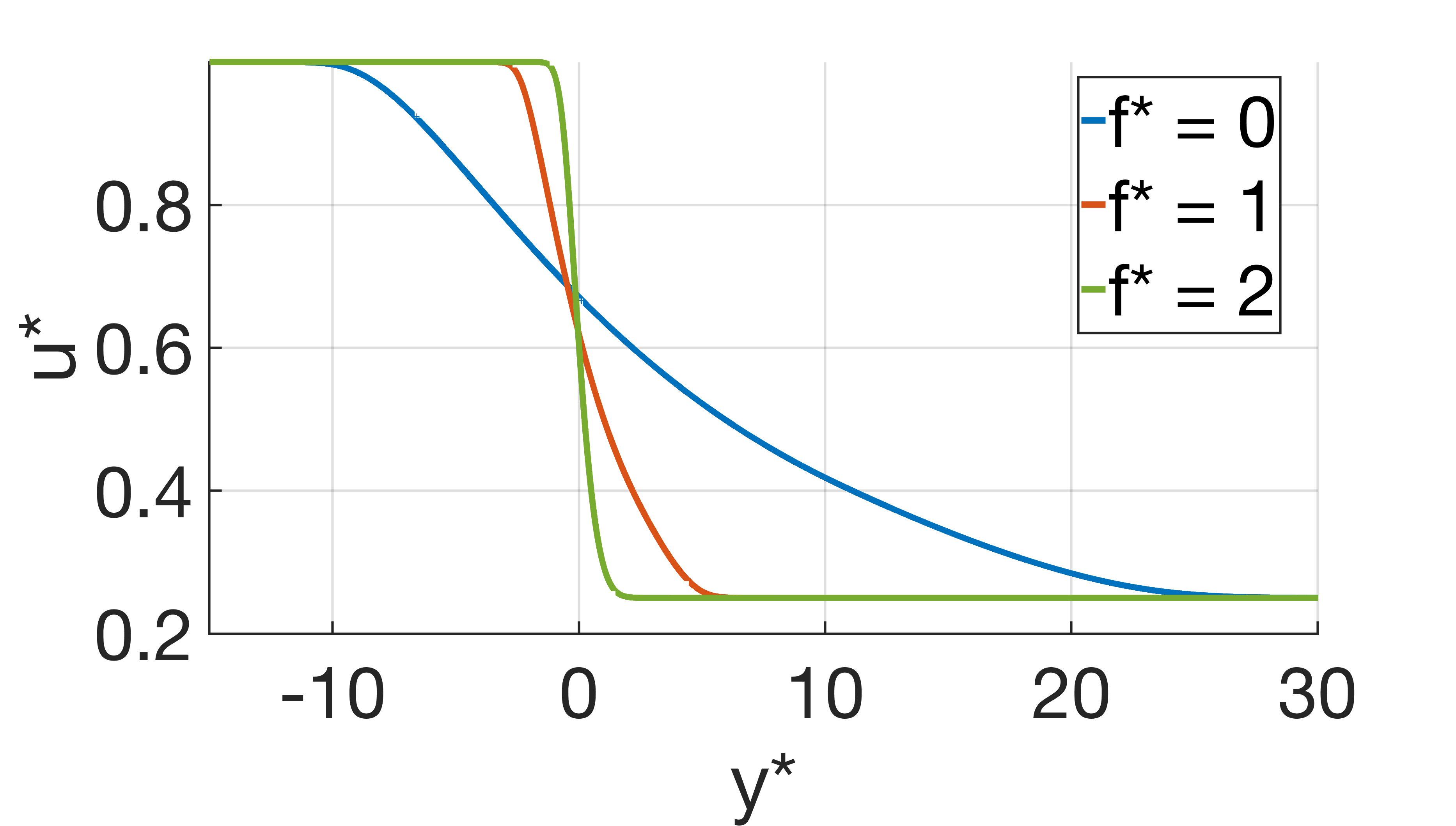}
         \caption{\(u^*(y^*)\)}
         \label{fig:y equals x}
     \end{subfigure}
     \hfill
     \begin{subfigure}[b]{0.49\textwidth}
         \centering
         \includegraphics[width=\textwidth]{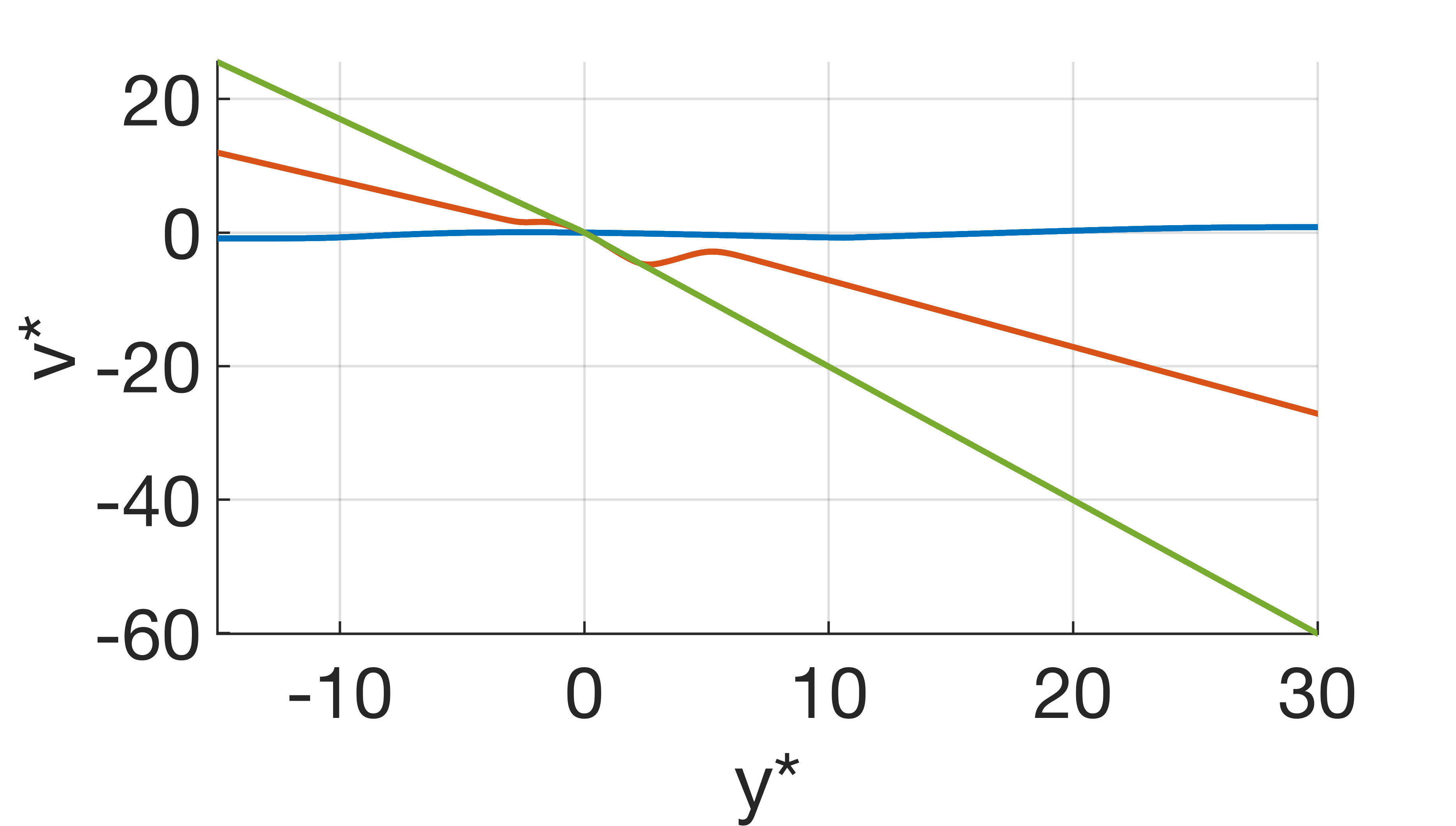}
         \caption{\(v^*(y^*)\)}
         \label{fig:three sin x}
     \end{subfigure}
     \hfill
     \begin{subfigure}[b]{0.49\textwidth}
         \centering
         \includegraphics[width=\textwidth]{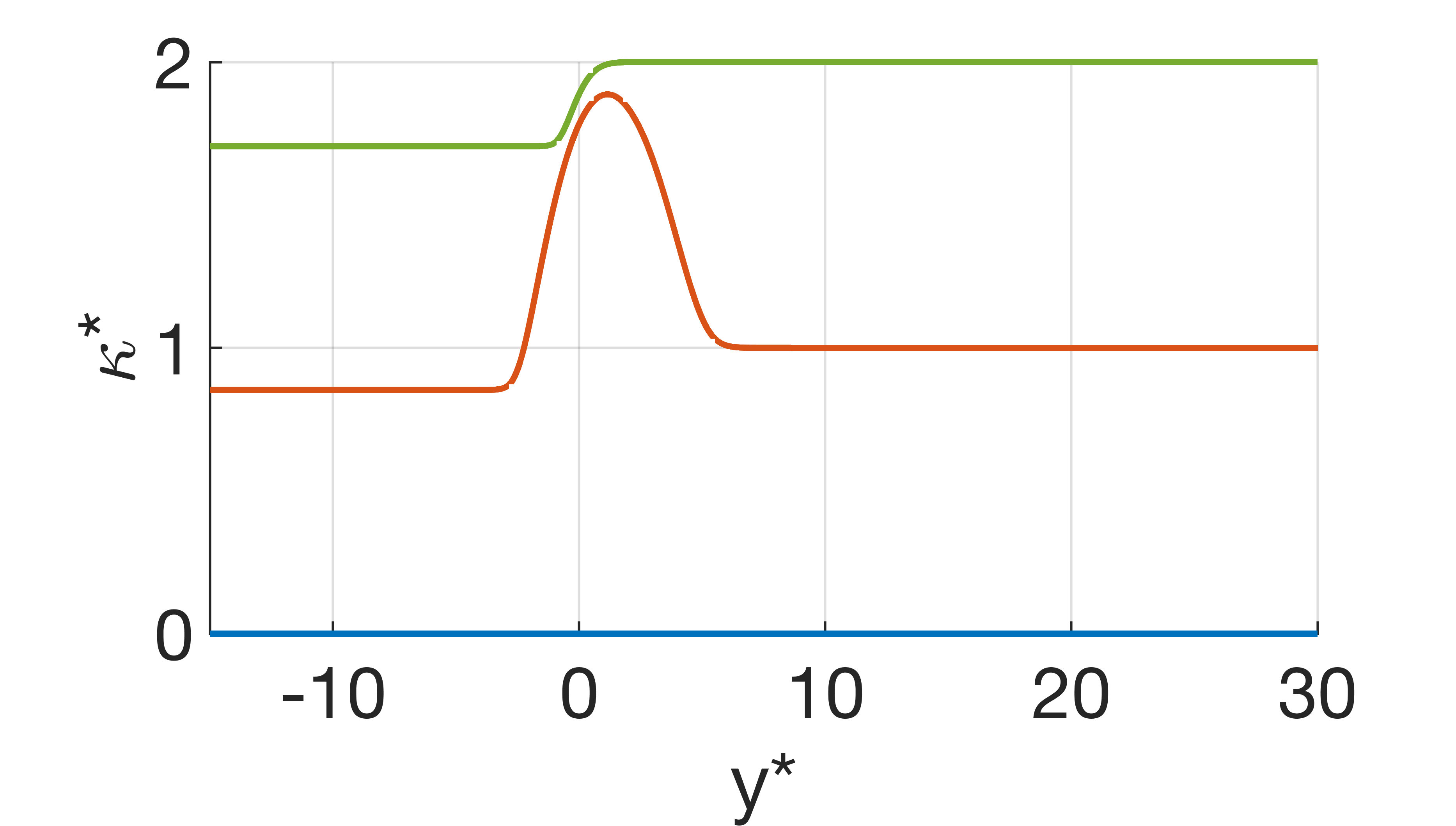}
         \caption{\(\kappa^*(y^*)\)}
         \label{fig:five over x}
     \end{subfigure}
     \begin{subfigure}[b]{0.49\textwidth}
         \centering
         \includegraphics[width=\textwidth]{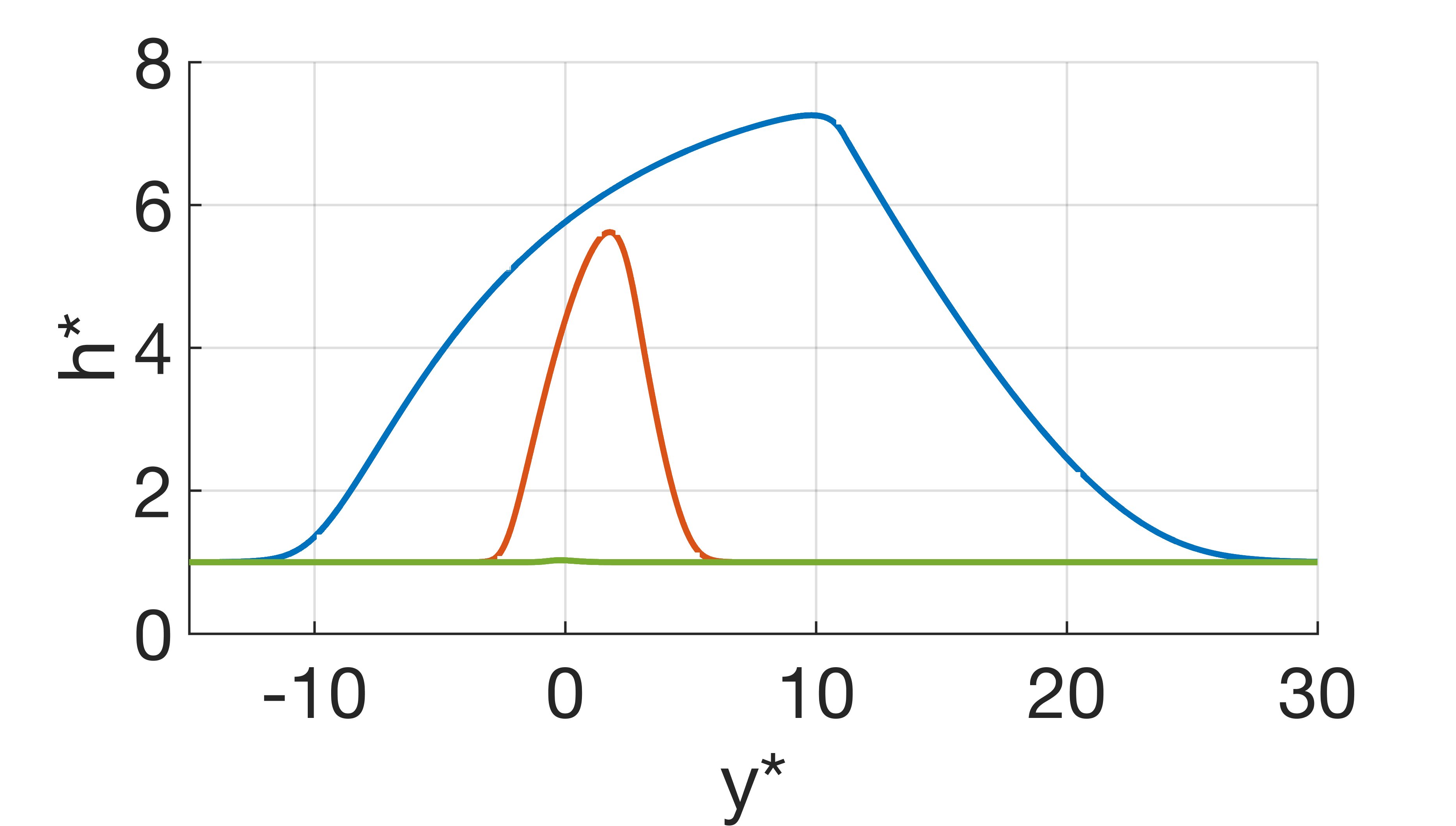}
         \caption{\(h^*(y^*)\)}
         \label{fig:five over x}
     \end{subfigure}
     \begin{subfigure}[b]{0.49\textwidth}
         \centering
         \includegraphics[width=\textwidth]{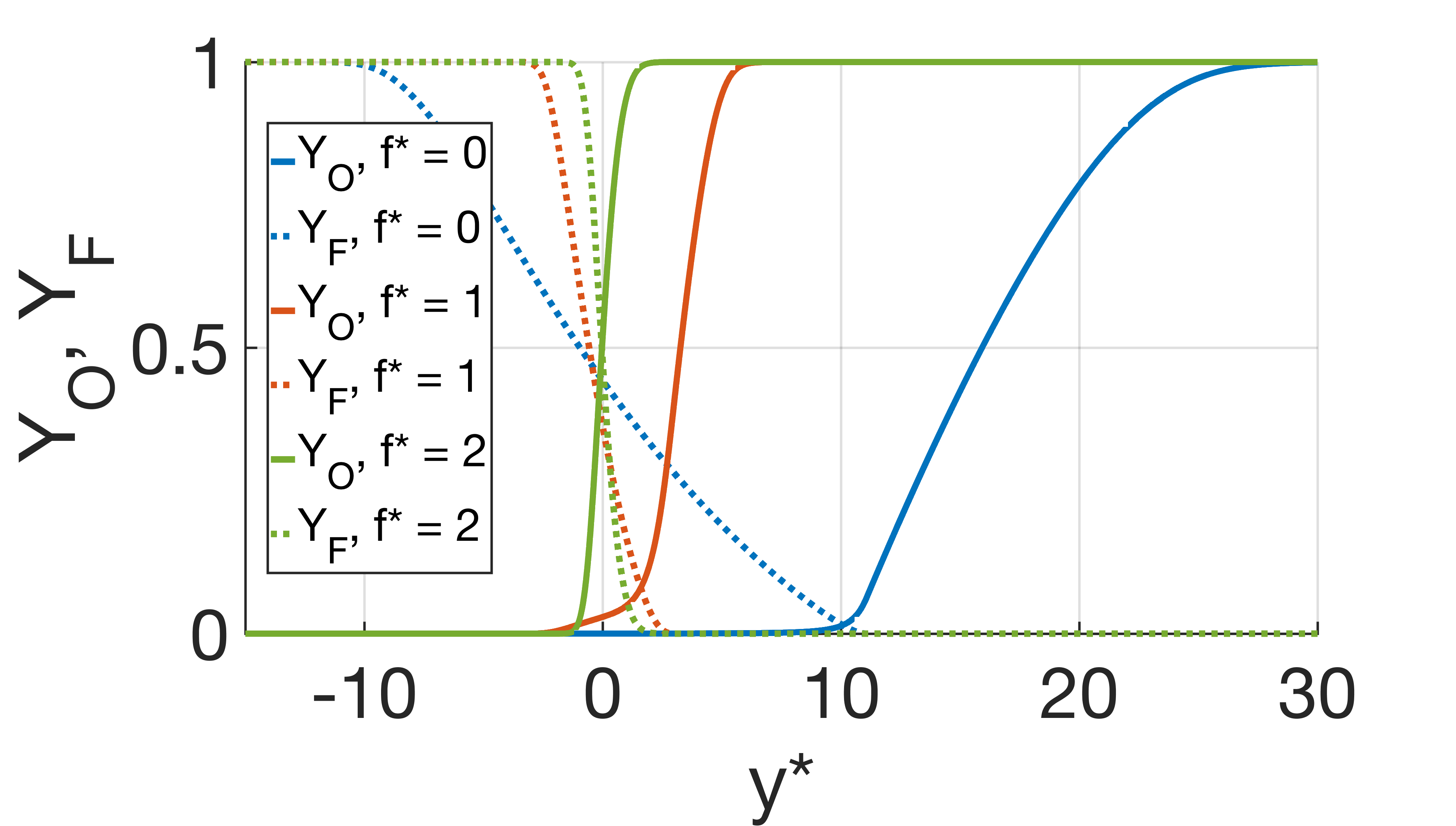}
         \caption{\(Y_O(y^*), Y_F(y^*)\)}
         \label{fig:five over x}
     \end{subfigure}
     \begin{subfigure}[b]{0.49\textwidth}
         \centering
         \includegraphics[width=\textwidth]{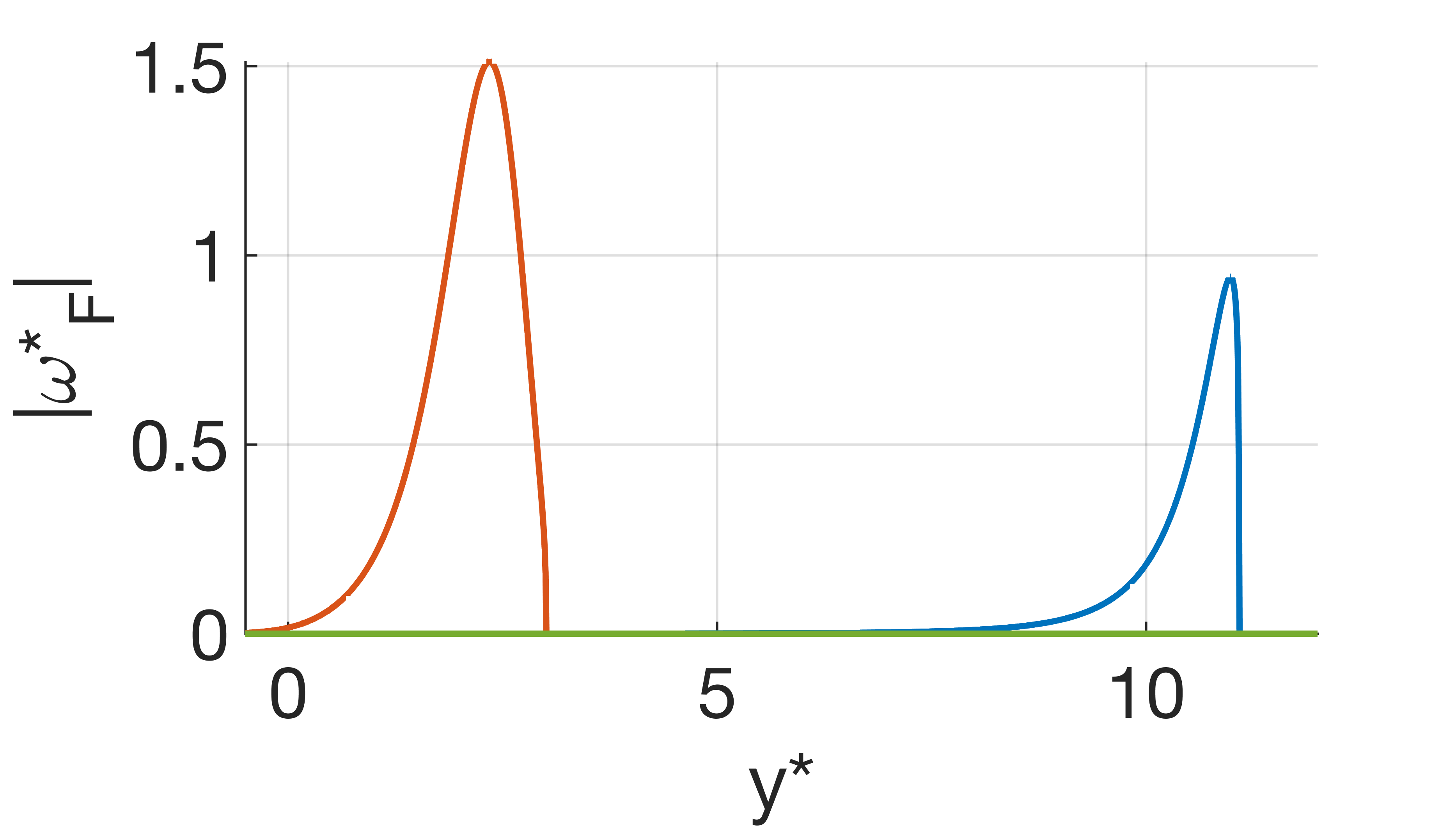}
         \caption{\(\abs{\dot{\omega}_F^*(y^*)}\)}
         \label{fig:five over x}
     \end{subfigure}
        \caption{\(u^*\), \(v^*\), \(h^*\), \(\kappa^*\), \(Y_O\), \(Y_F\), and $\dot{\omega}^*_F$ at \(x^*=2\) for the inverted-velocity reactive Cases 11a, 11b, and 11c. Counterflow strain rate varies from $0$ to $2$.}
        \label{fig:three graphs}
\end{figure*}

For this reacting case as well as the prior non-reactive case, the ratio of the free-stream \(x\)-component velocities has a very minor effect on the downstream behavior of the scalar quantities when all else is held constant. A slight increase in burning rate is observed in Fig. 12(f) when the difference in speed between the two streams is lessened. As the velocity of the slower stream is increased, the velocity throughout the layer increases and less burning occurs near the high-speed stream as shown in Fig. 12(f). Nevertheless, mass diffusion brings the fuel to a region of larger reaction rate.

\begin{figure*}
     \centering
     \begin{subfigure}[b]{0.49\textwidth}
         \centering
         \includegraphics[width=\textwidth]{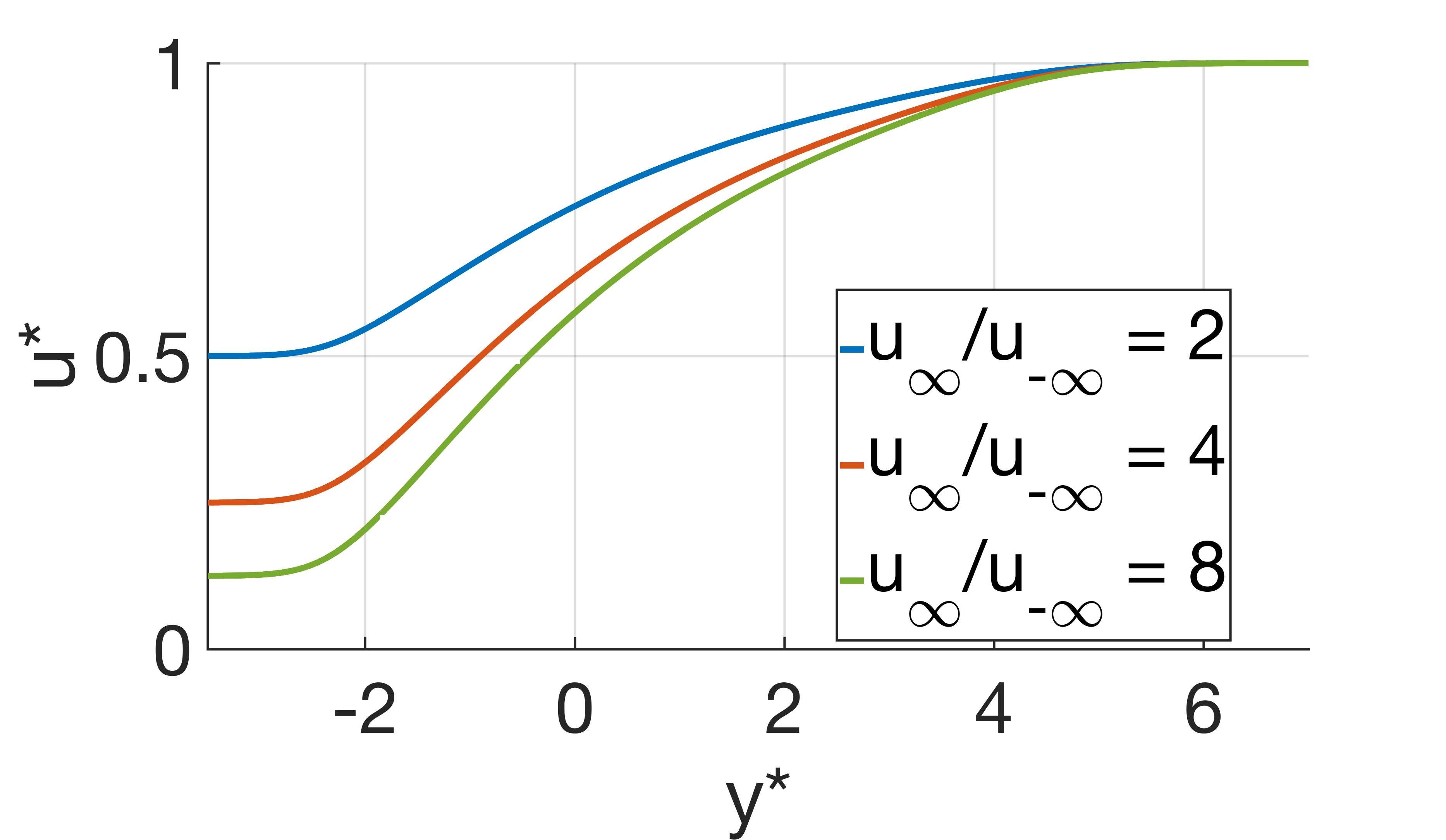}
         \caption{\(u^*(y^*)\)}
         \label{fig:y equals x}
     \end{subfigure}
     \begin{subfigure}[b]{0.49\textwidth}
         \centering
         \includegraphics[width=\textwidth]{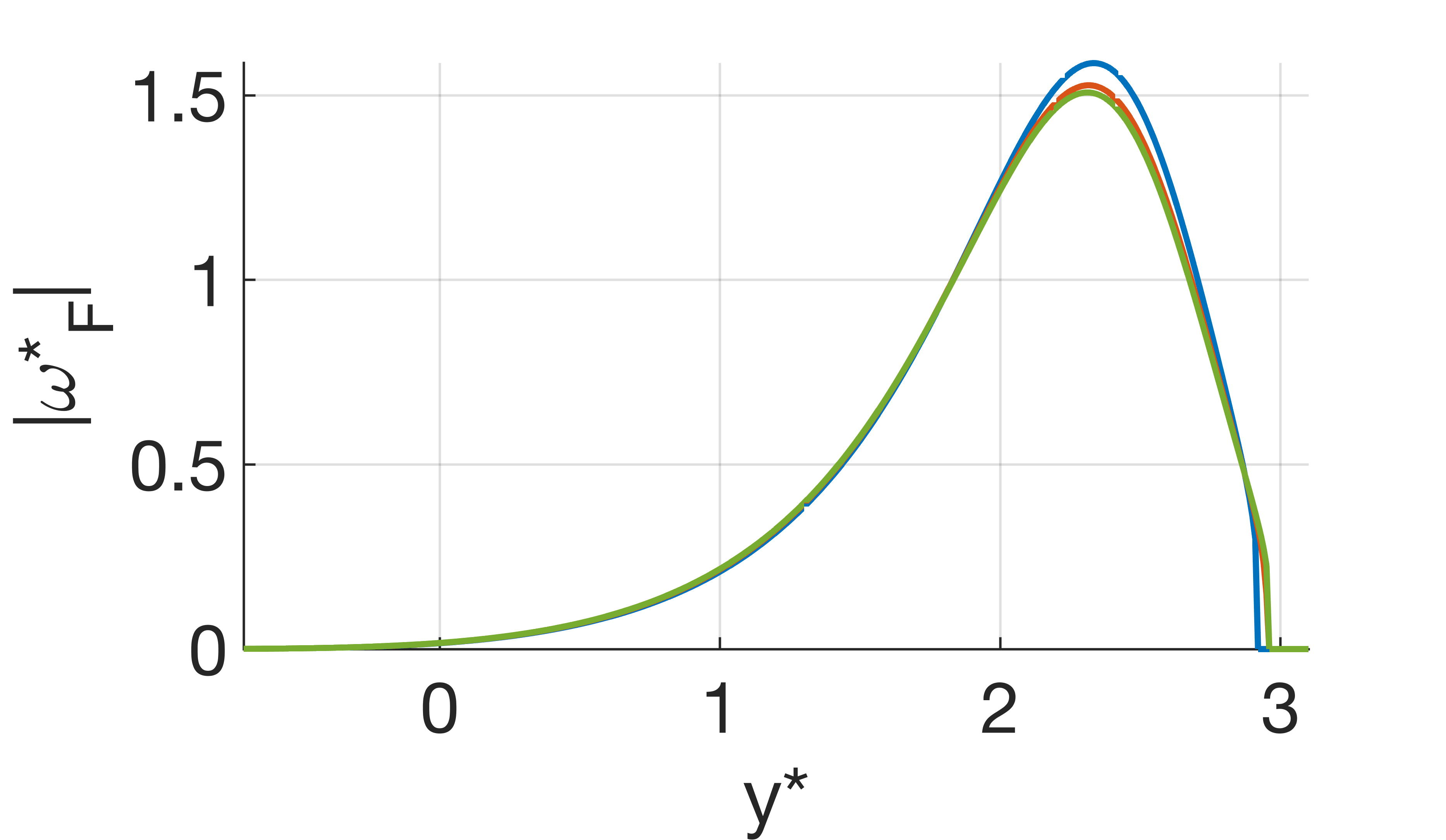}
         \caption{\(\abs{\dot{\omega}_F^*(y^*)}\)}
         \label{fig:five over x}
     \end{subfigure}
        \caption{\(u^*\) and $\dot{\omega}^*_F$ at \(x^*=2\) for reactive Cases 8a, 6, and 8b. The free-stream velocity ratio varies from $2$ to $8$.}
        \label{fig:three graphs}
\end{figure*}

Figs. 13(c) through 13(e) reveal that as \(Pr\) increases, the mixing layers for the scalar quantities become thinner for the same reasons discussed in the non-reactive case. In Fig. 13(f), the reaction zone does not drift as far in the positive $y$-direction with increased Prandtl number.

\begin{figure*}
     \centering
     \begin{subfigure}[b]{0.49\textwidth}
         \centering
         \includegraphics[width=\textwidth]{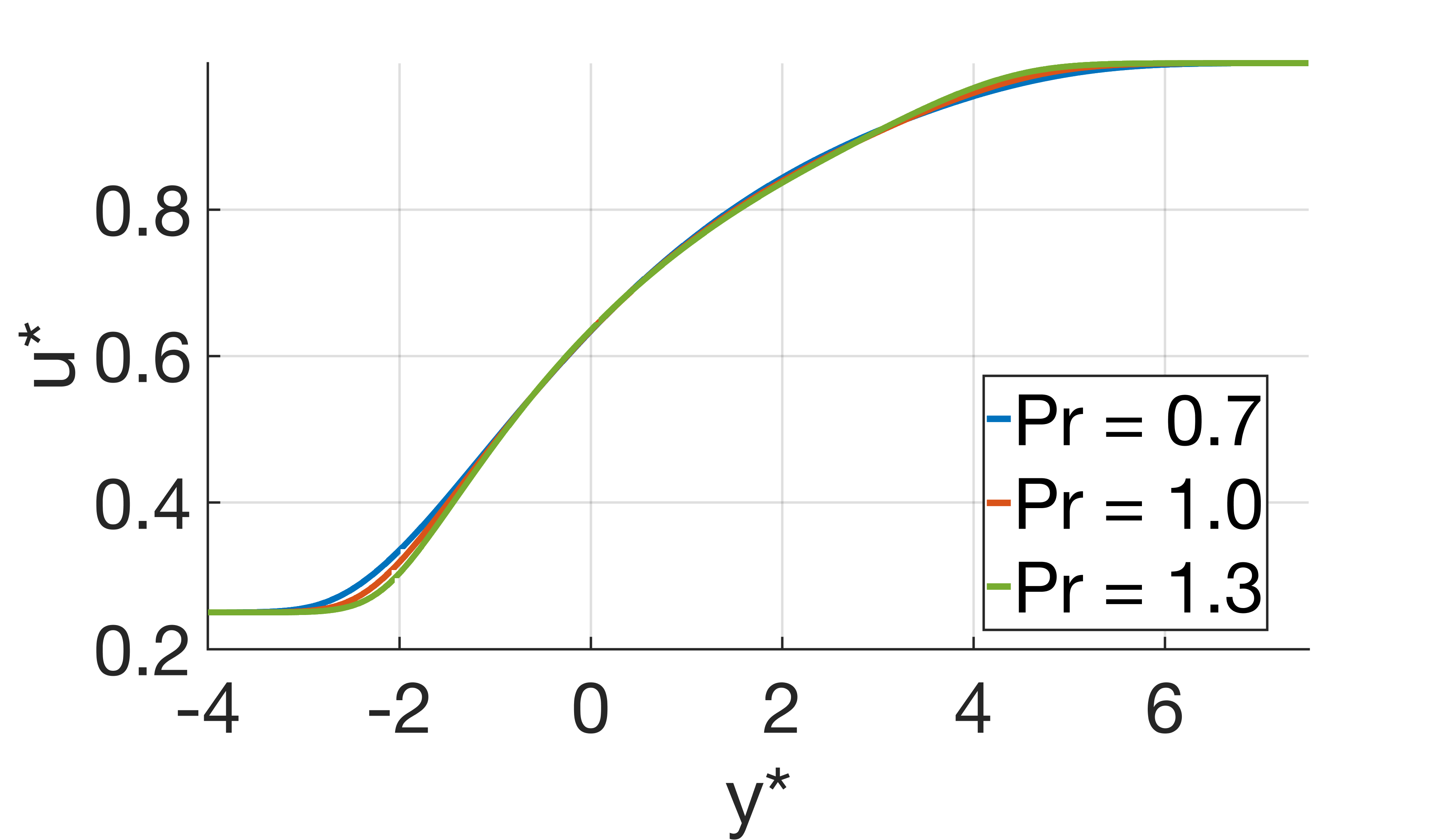}
         \caption{\(u^*(y^*)\)}
         \label{fig:y equals x}
     \end{subfigure}
     \hfill
     \begin{subfigure}[b]{0.49\textwidth}
         \centering
         \includegraphics[width=\textwidth]{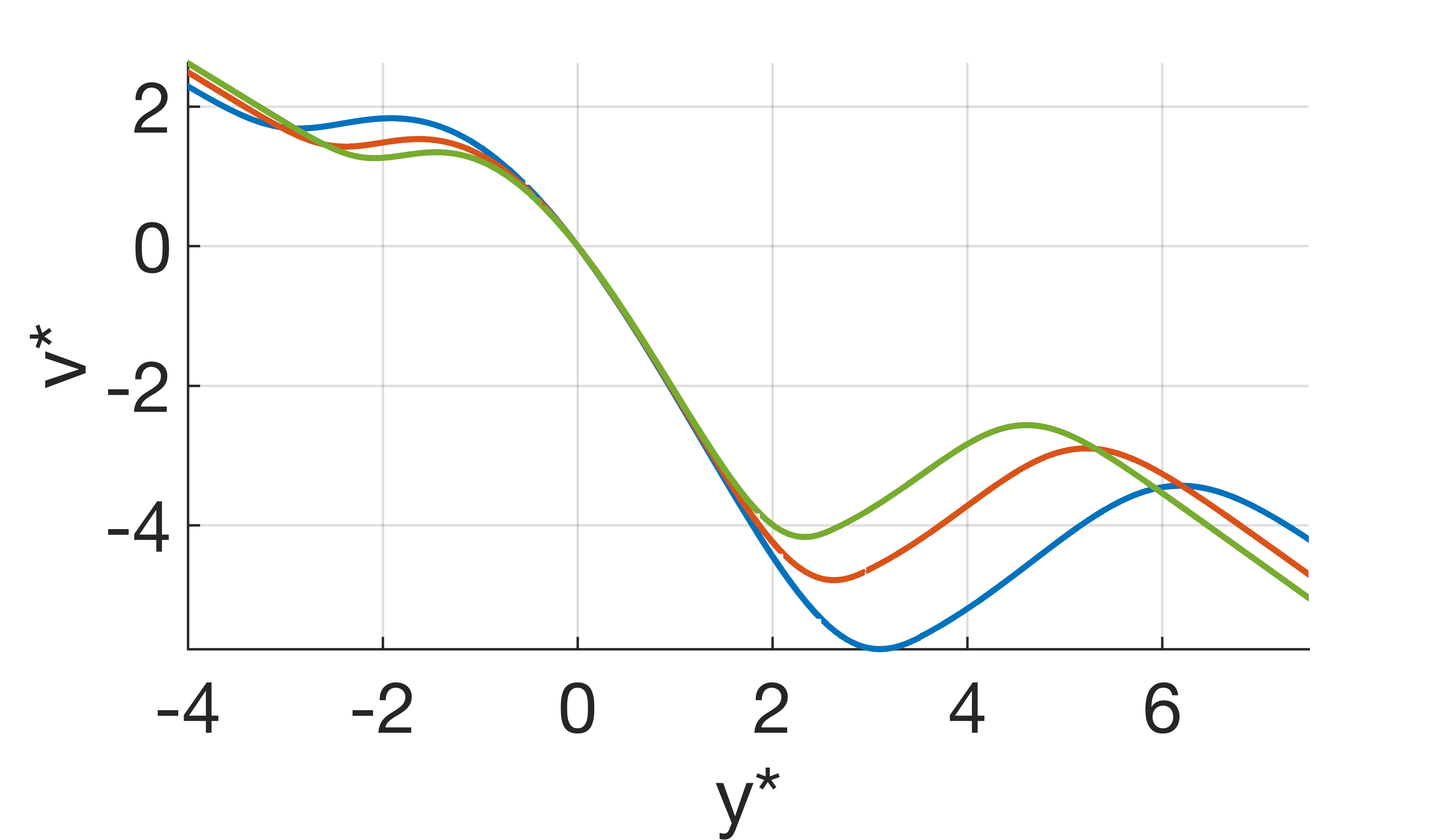}
         \caption{\(v^*(y^*)\)}
         \label{fig:three sin x}
     \end{subfigure}
     \hfill
     \begin{subfigure}[b]{0.49\textwidth}
         \centering
         \includegraphics[width=\textwidth]{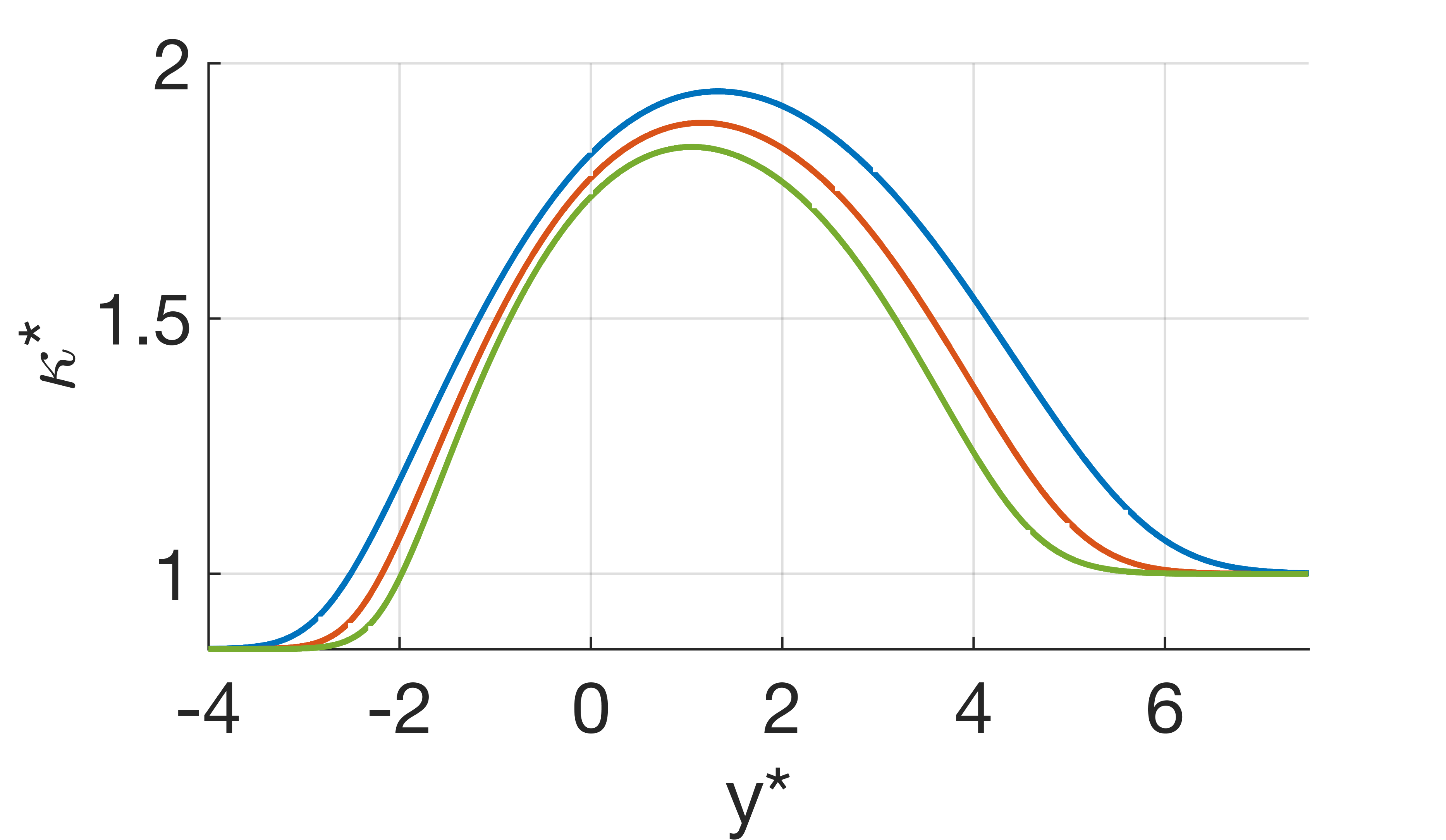}
         \caption{\(\kappa^*(y^*)\)}
         \label{fig:five over x}
     \end{subfigure}
     \begin{subfigure}[b]{0.49\textwidth}
         \centering
         \includegraphics[width=\textwidth]{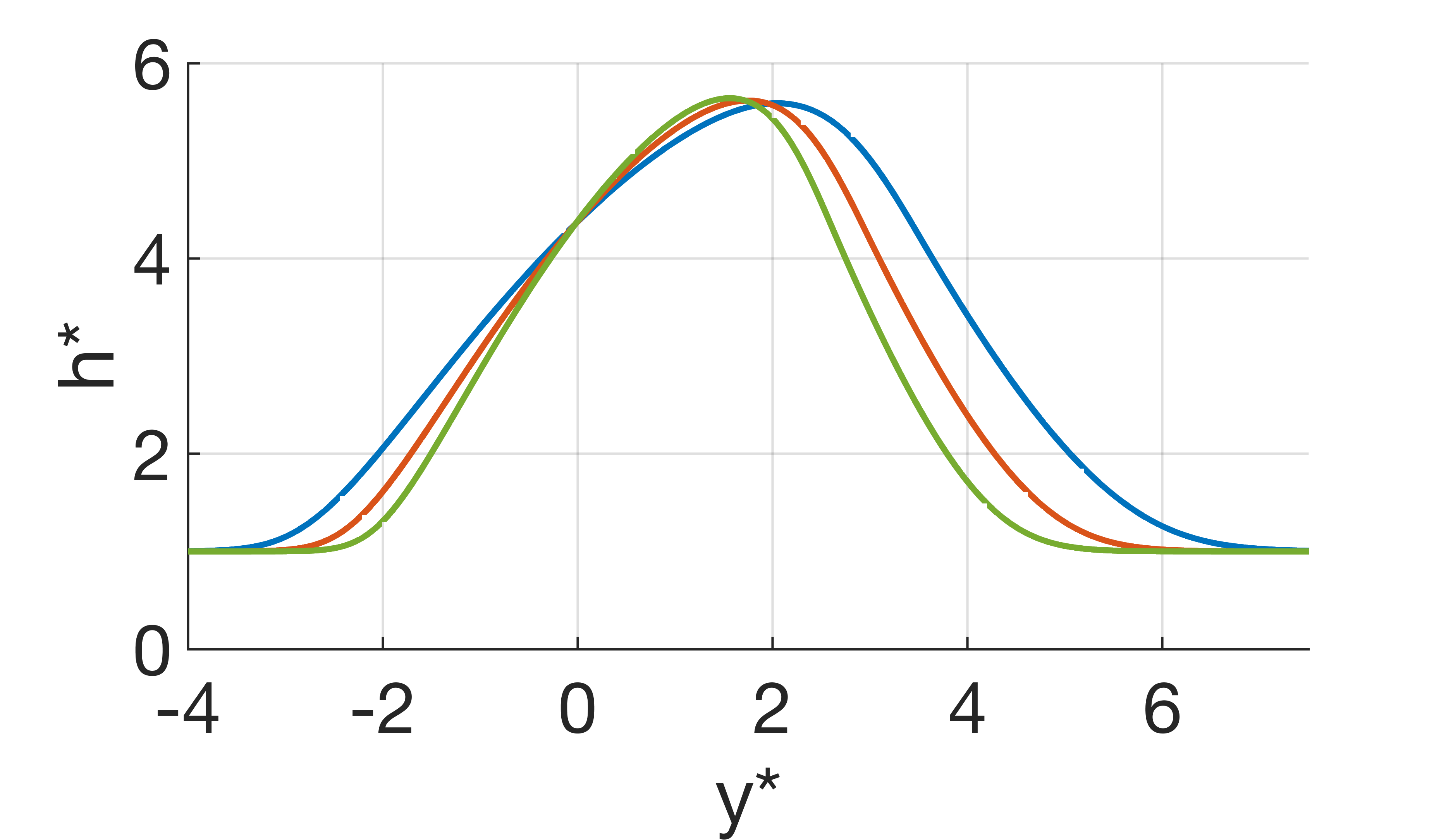}
         \caption{\(h^*(y^*)\)}
         \label{fig:five over x}
     \end{subfigure}
     \begin{subfigure}[b]{0.49\textwidth}
         \centering
         \includegraphics[width=\textwidth]{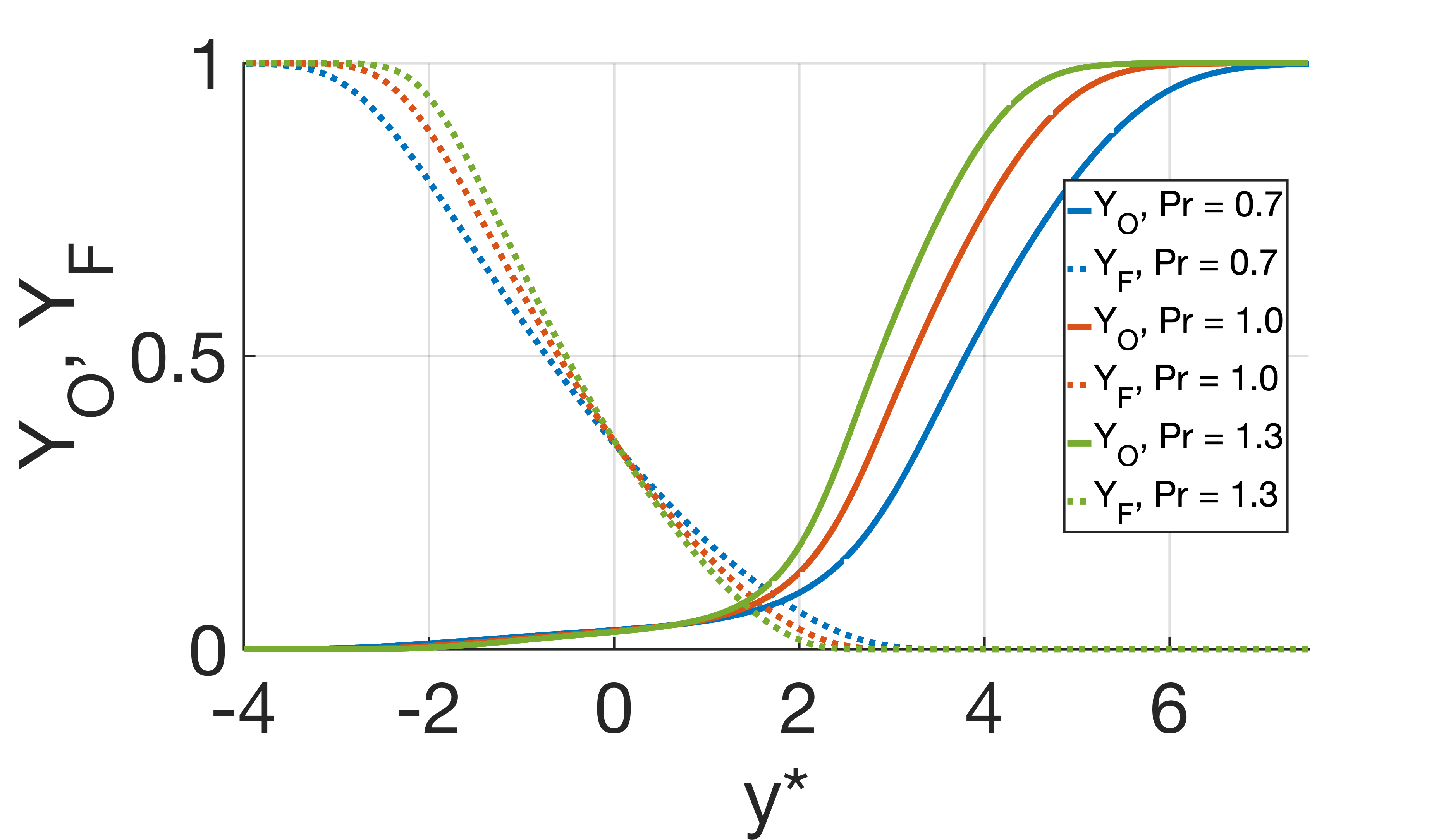}
         \caption{\(Y_O(y^*), Y_F(y^*)\)}
         \label{fig:five over x}
     \end{subfigure}
     \begin{subfigure}[b]{0.49\textwidth}
         \centering
         \includegraphics[width=\textwidth]{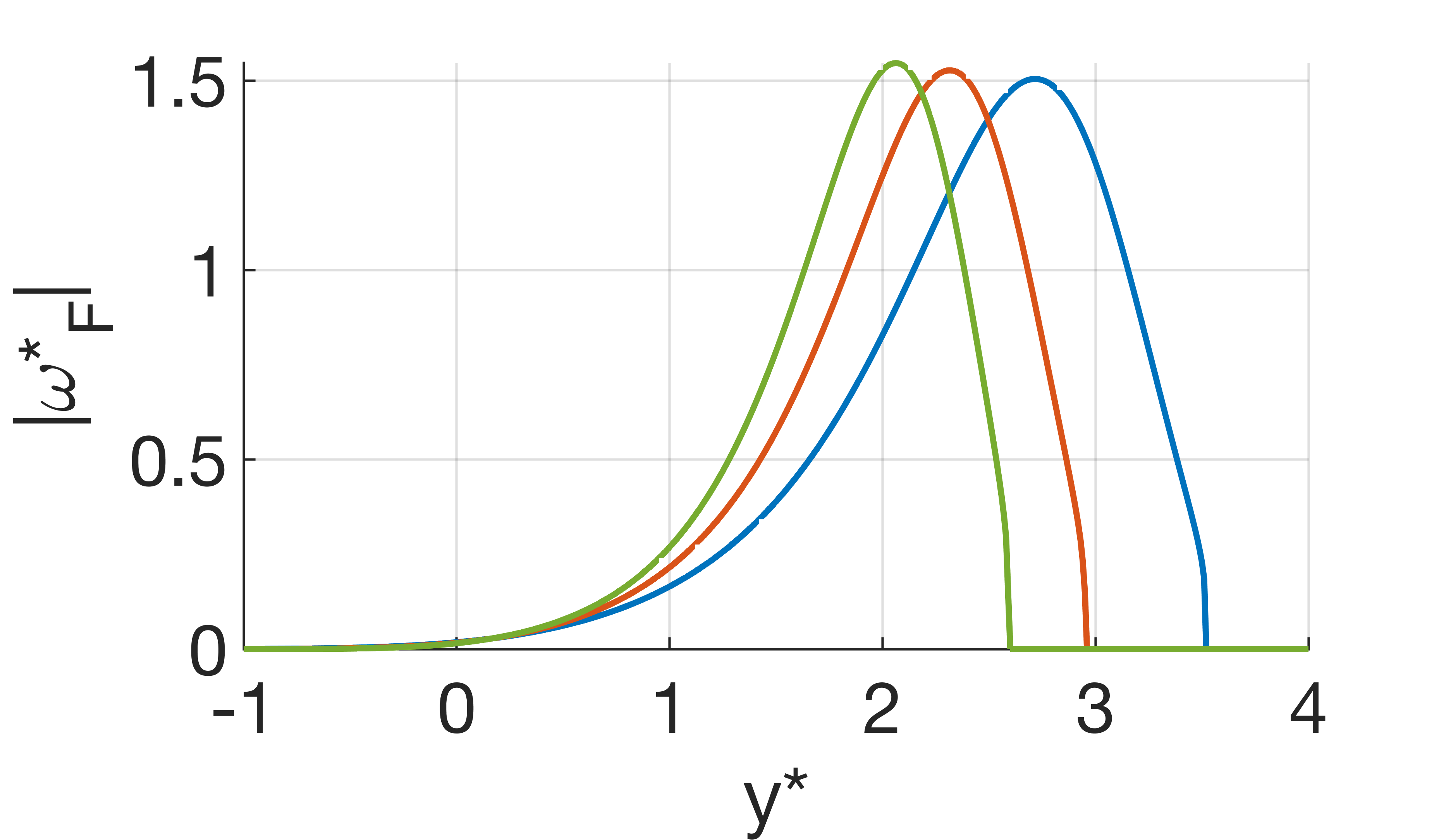}
         \caption{\(\abs{\dot{\omega}_F^*(y^*)}\)}
         \label{fig:five over x}
     \end{subfigure}
        \caption{\(u^*\), \(v^*\), \(h^*\), \(\kappa^*\), \(Y_O\), \(Y_F\), and $\dot{\omega}^*_F$ at \(x^*=2\) for reactive Cases 9a, 6, and 9b. Prandtl number varies from $0.7$ to $1.3$.}
        \label{fig:three graphs}
\end{figure*}

Fig. 14 compares the reactive Cases 6 and 10b to the non-reactive Case 10a. The effects of chemical reaction discussed for Case 6 in Fig. 9 are lessened with a decreased Damk\"{o}hler number in Case 10b. When no reaction occurs, the mixing layer for the scalar quantities remains centered at $y^*=0$ rather than drifting towards the oxygen-rich stream.

\begin{figure*}
     \centering
     \begin{subfigure}[b]{0.49\textwidth}
         \centering
         \includegraphics[width=\textwidth]{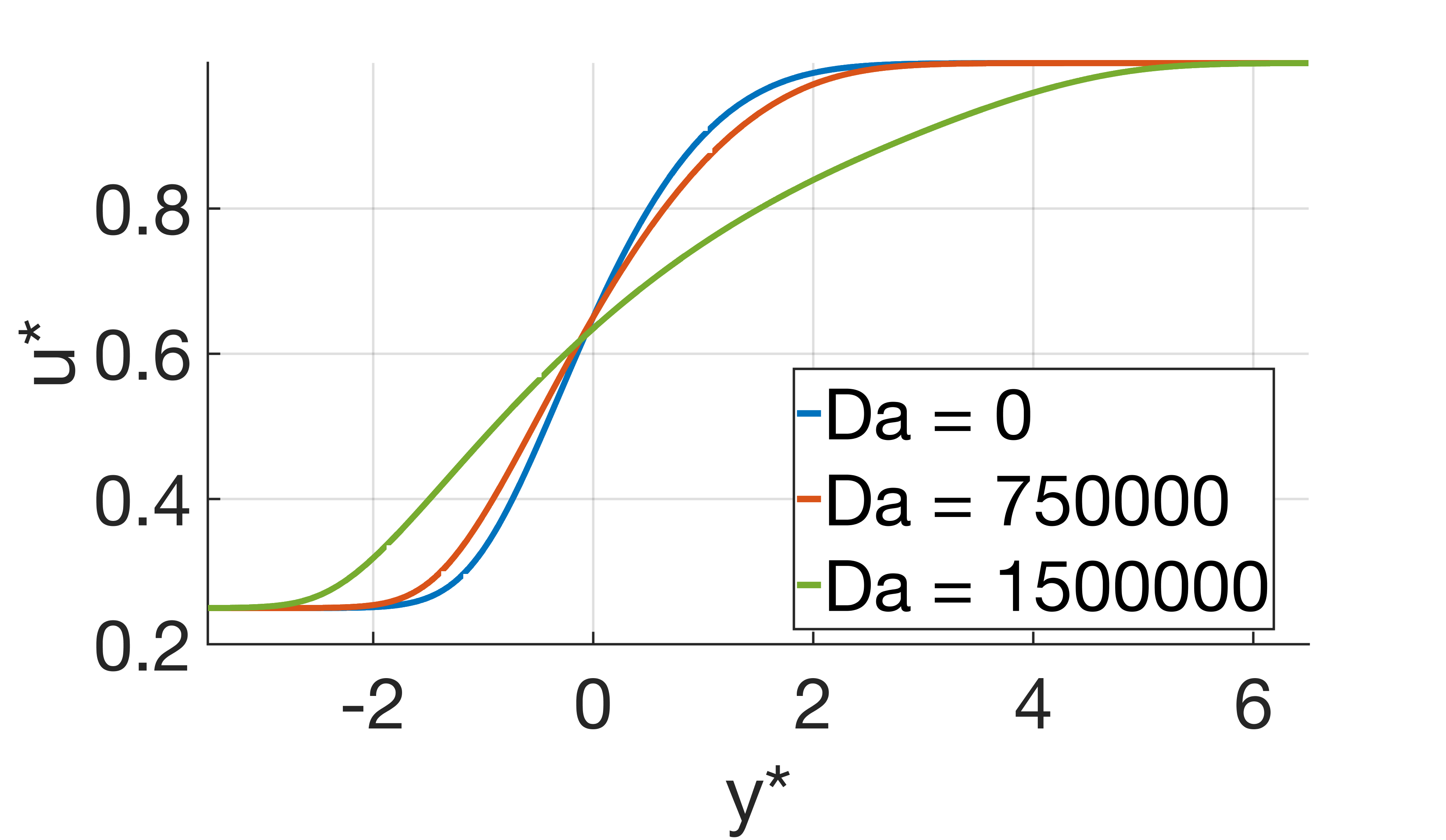}
         \caption{\(u^*(y^*)\)}
         \label{fig:y equals x}
     \end{subfigure}
     \hfill
     \begin{subfigure}[b]{0.49\textwidth}
         \centering
         \includegraphics[width=\textwidth]{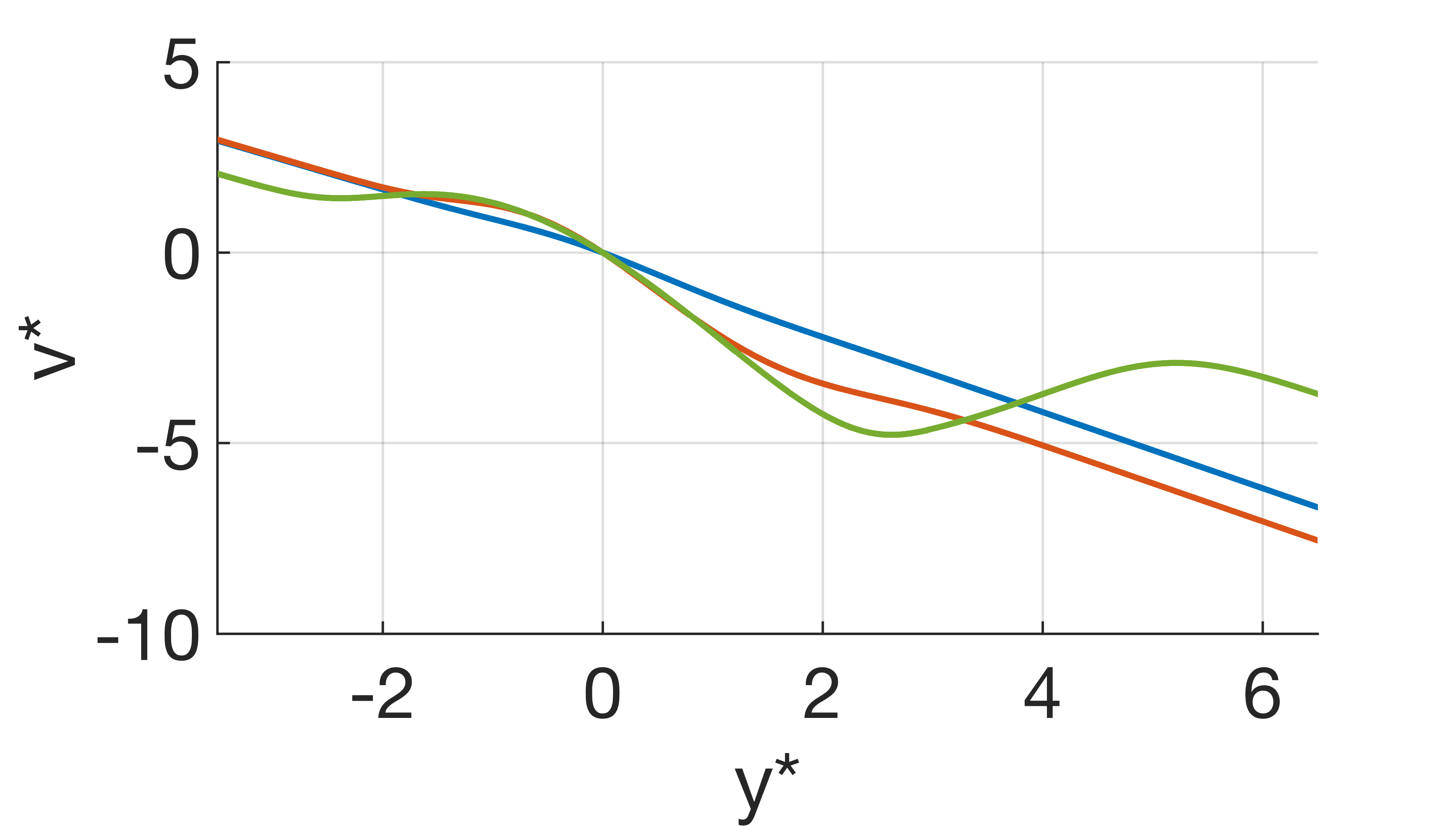}
         \caption{\(v^*(y^*)\)}
         \label{fig:three sin x}
     \end{subfigure}
     \hfill
     \begin{subfigure}[b]{0.49\textwidth}
         \centering
         \includegraphics[width=\textwidth]{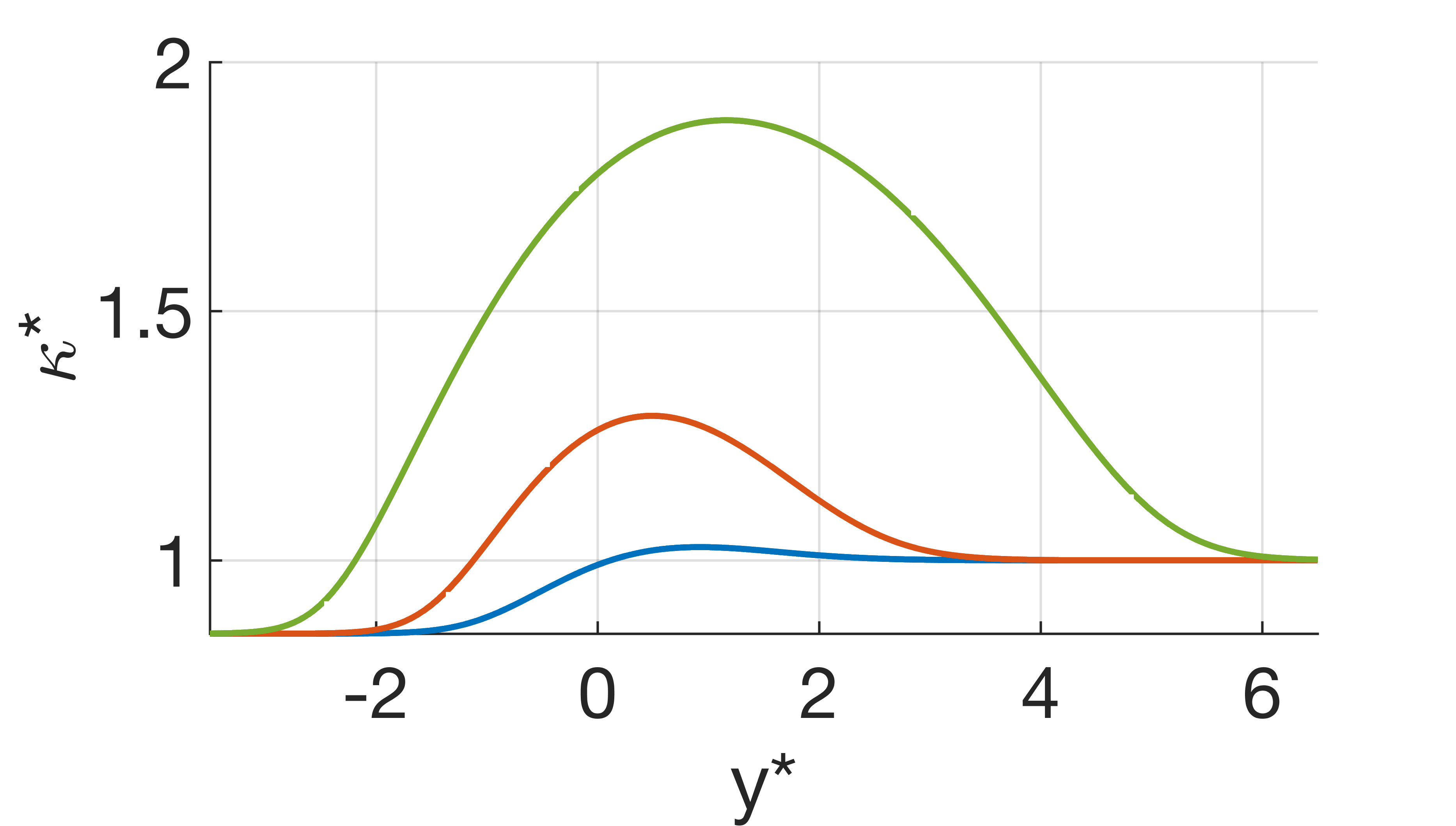}
         \caption{\(\kappa^*(y^*)\)}
         \label{fig:five over x}
     \end{subfigure}
     \begin{subfigure}[b]{0.49\textwidth}
         \centering
         \includegraphics[width=\textwidth]{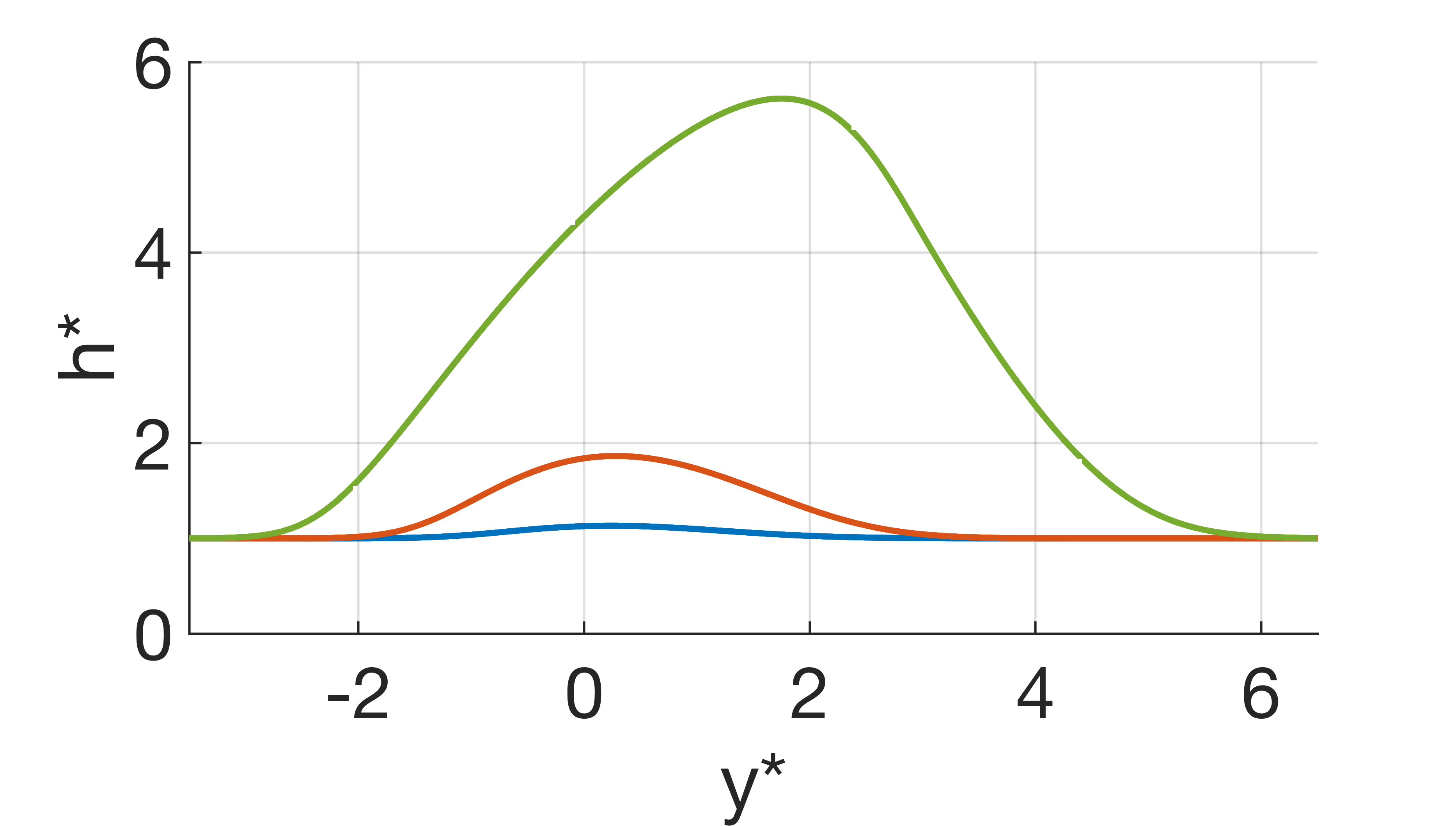}
         \caption{\(h^*(y^*)\)}
         \label{fig:five over x}
     \end{subfigure}
     \begin{subfigure}[b]{0.49\textwidth}
         \centering
         \includegraphics[width=\textwidth]{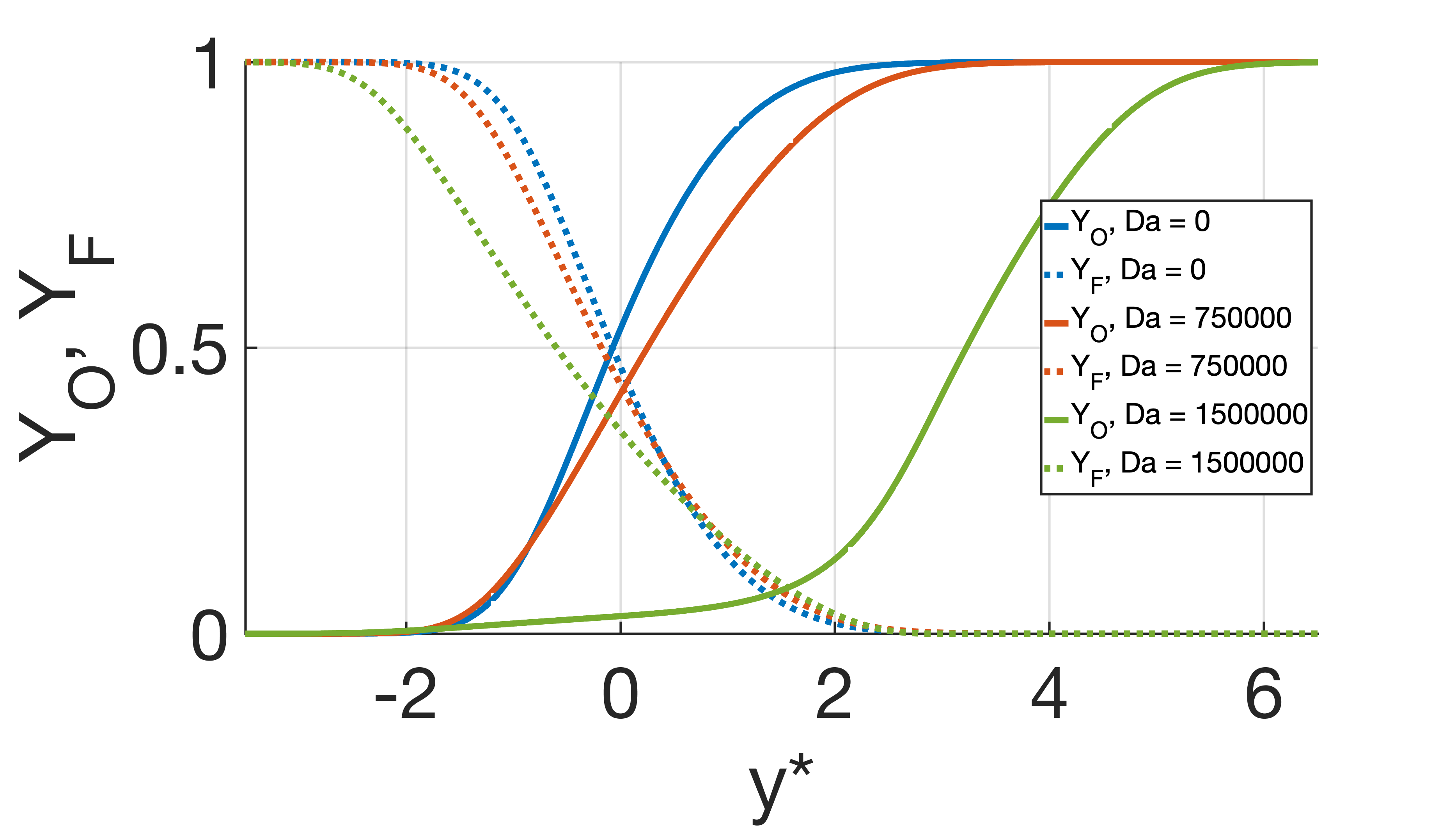}
         \caption{\(Y_O(y^*), Y_F(y^*)\)}
         \label{fig:five over x}
     \end{subfigure}
     \begin{subfigure}[b]{0.49\textwidth}
         \centering
         \includegraphics[width=\textwidth]{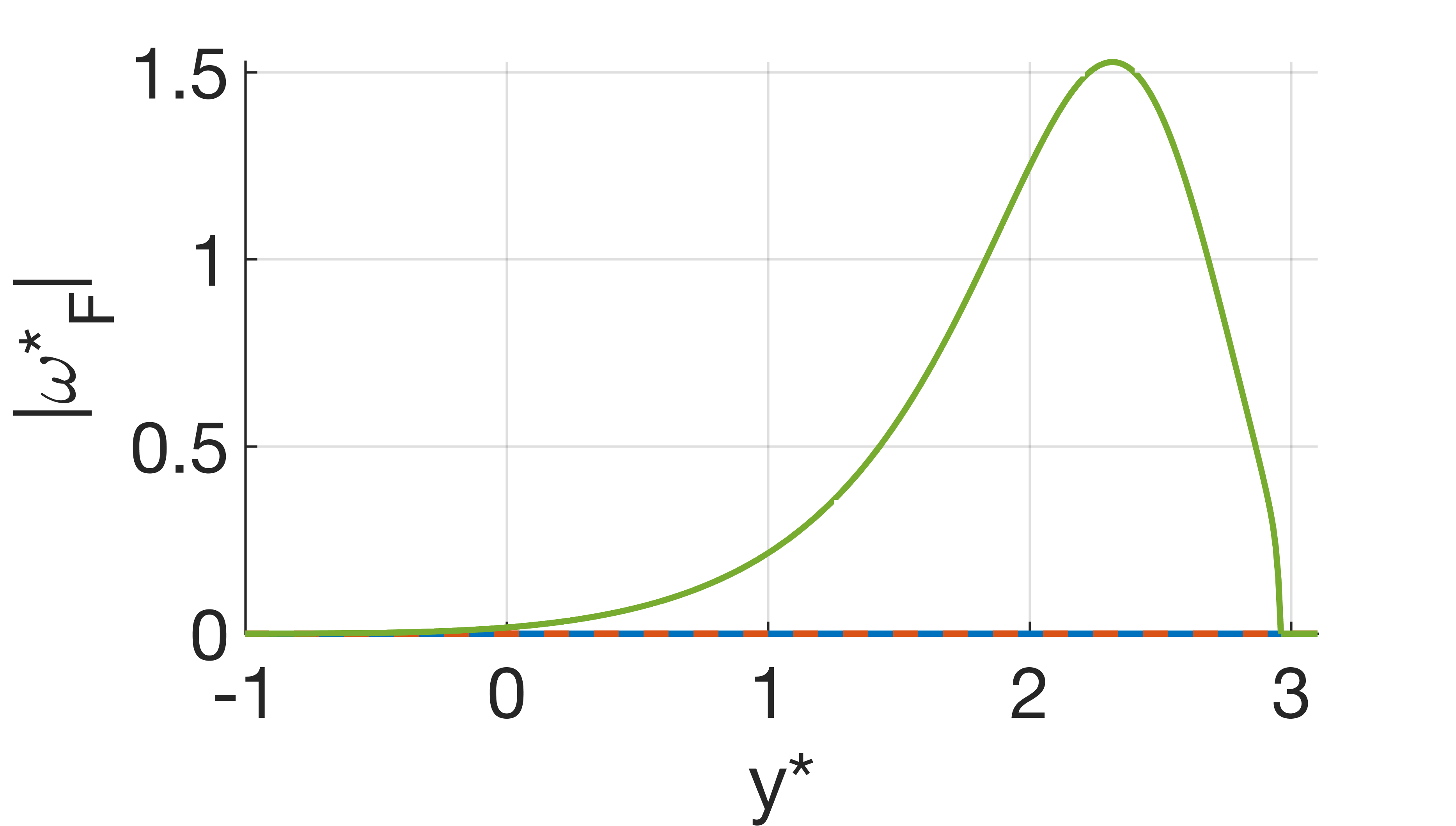}
         \caption{\(\abs{\dot{\omega}_F^*(y^*)}\)}
         \label{fig:five over x}
     \end{subfigure}
        \caption{\(u^*\), \(v^*\), \(h^*\), \(\kappa^*\), \(Y_O\), \(Y_F\), and $\dot{\omega}^*_F$ at \(x^*=2\) for reactive Cases 10a, 6, and 10b. Damk\"{o}hler number varies from $0$ to $1,500,000$.}
        \label{fig:three graphs}
\end{figure*}

\clearpage
\subsection{Multi-flame structures}
A premixed fuel-rich flame and premixed fuel-lean flame can co-exist with the primary diffusion flame. This multi-flame Case 12 is achieved by changing the ambient mass fraction of oxidizer in the fast stream from $1$ to $11/12$ with propane added in the fast stream. The ambient mass fraction of fuel in the slow stream is decreased from $1$ to $\frac{2}{3}$ with oxygen added. Fig. 15(f) clearly shows multiple flames forming a steady structure far downstream.

In addition to the diffusion flame, a premixed fuel-lean flame is clearly visible in the temperature profiles of Fig. 15(c) as a downward concavity on the positive $y$-side of the diffusion flame. The premixed fuel-rich flame is not obvious in Fig. 15(c), but Fig. 15(f) shows that a fuel-rich premixed reaction is occuring, albeit at a slower rate than the diffusion and fuel-lean premixed flames. Consistent with previous findings, \cite{sirignano2021combustion, sirignano2021diffusion, sirignano2021mixing} the premixed flames depend on heat flux from the stronger diffusion flame. At the particular Damk\"{o}hler number, a premixed flame would not survive independently.

\begin{figure*}
     \centering
     \begin{subfigure}[b]{0.49\textwidth}
         \centering
         \includegraphics[width=\textwidth]{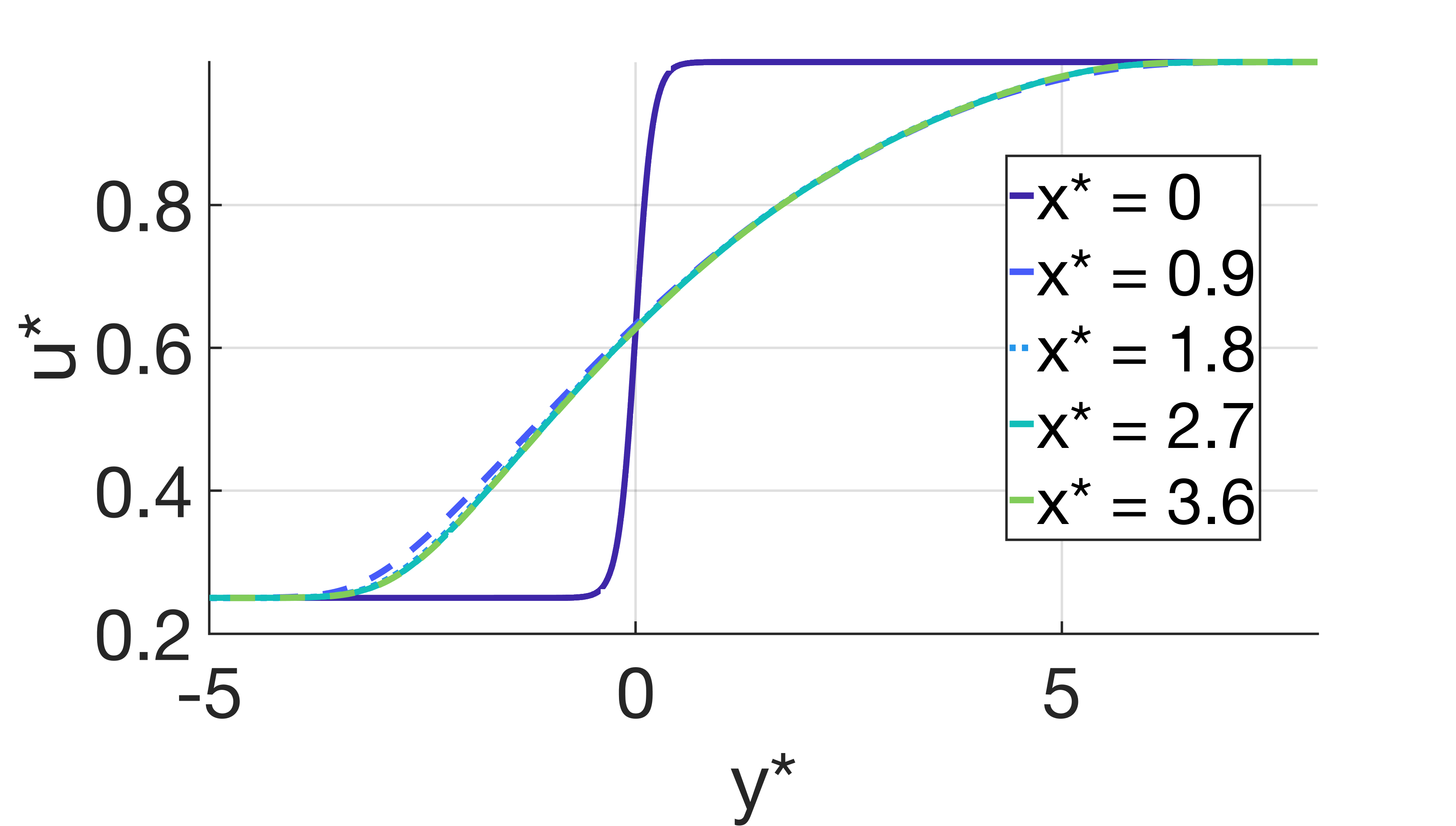}
         \caption{\(u^*(y^*)\)}
         \label{fig:y equals x}
     \end{subfigure}
     \hfill
     \begin{subfigure}[b]{0.49\textwidth}
         \centering
         \includegraphics[width=\textwidth]{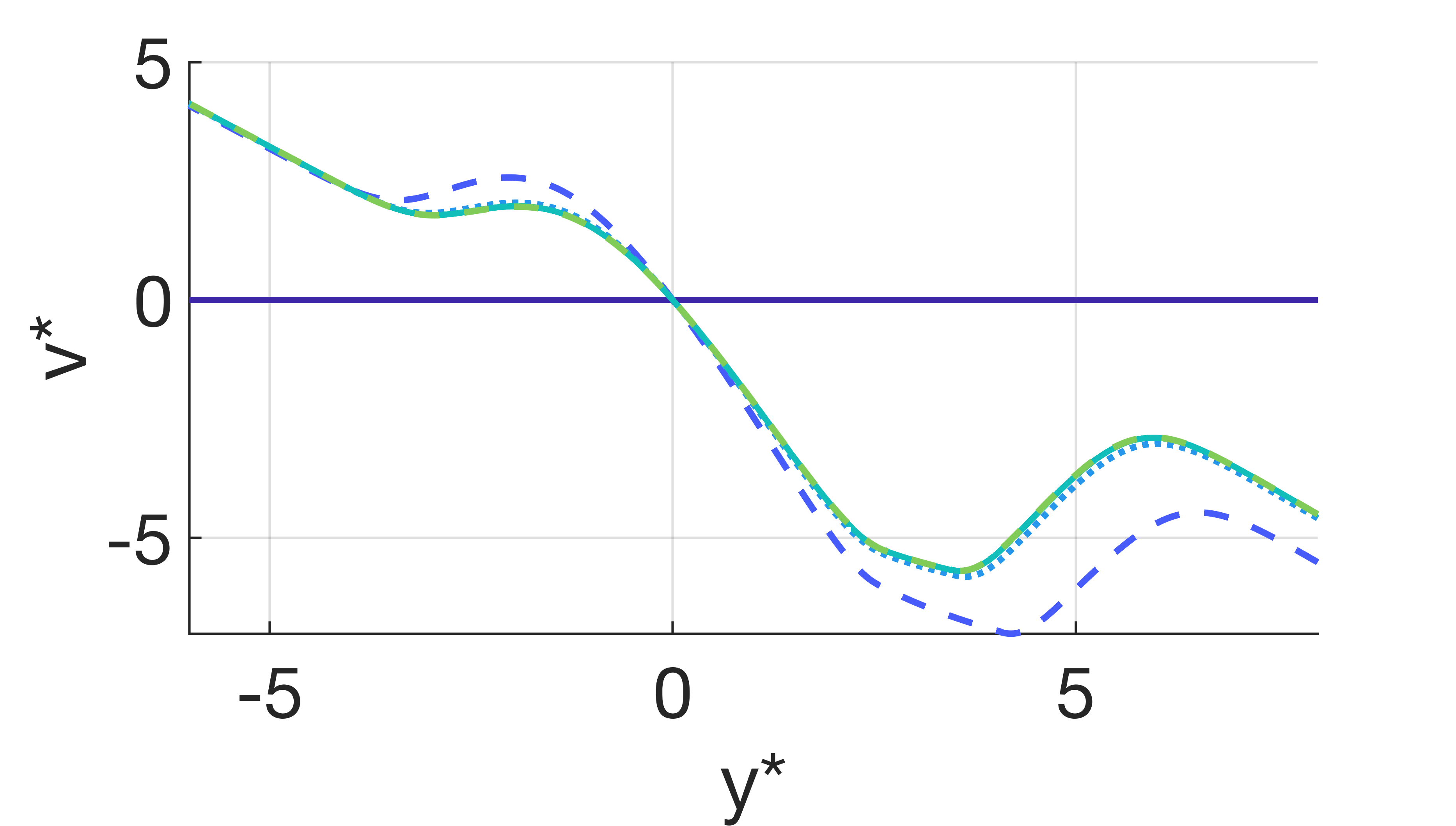}
         \caption{\(v^*(y^*)\)}
         \label{fig:three sin x}
     \end{subfigure}
     \hfill
     \begin{subfigure}[b]{0.49\textwidth}
         \centering
         \includegraphics[width=\textwidth]{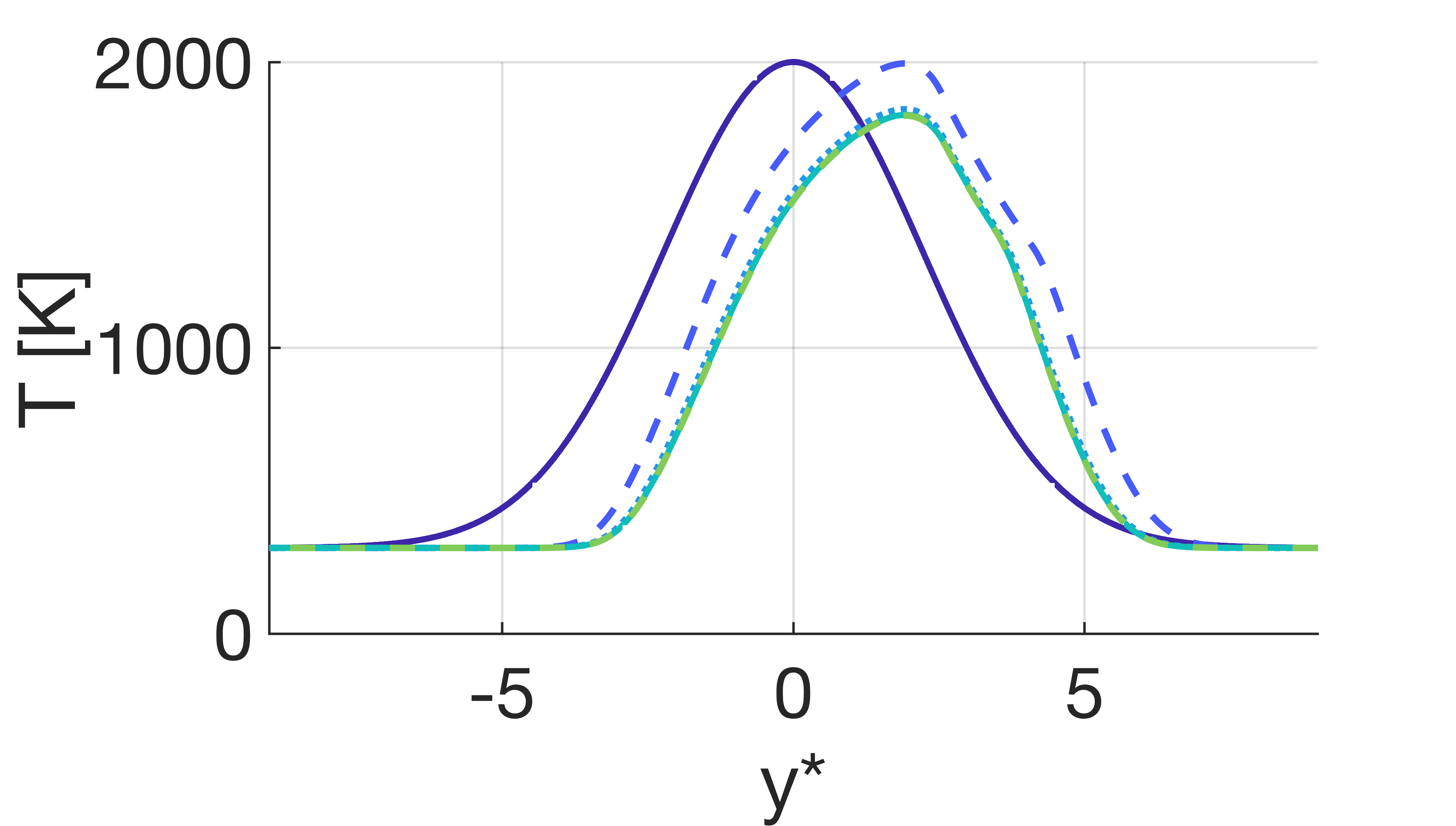}
         \caption{\(T(y^*)\)}
         \label{fig:five over x}
     \end{subfigure}
     \begin{subfigure}[b]{0.49\textwidth}
         \centering
         \includegraphics[width=\textwidth]{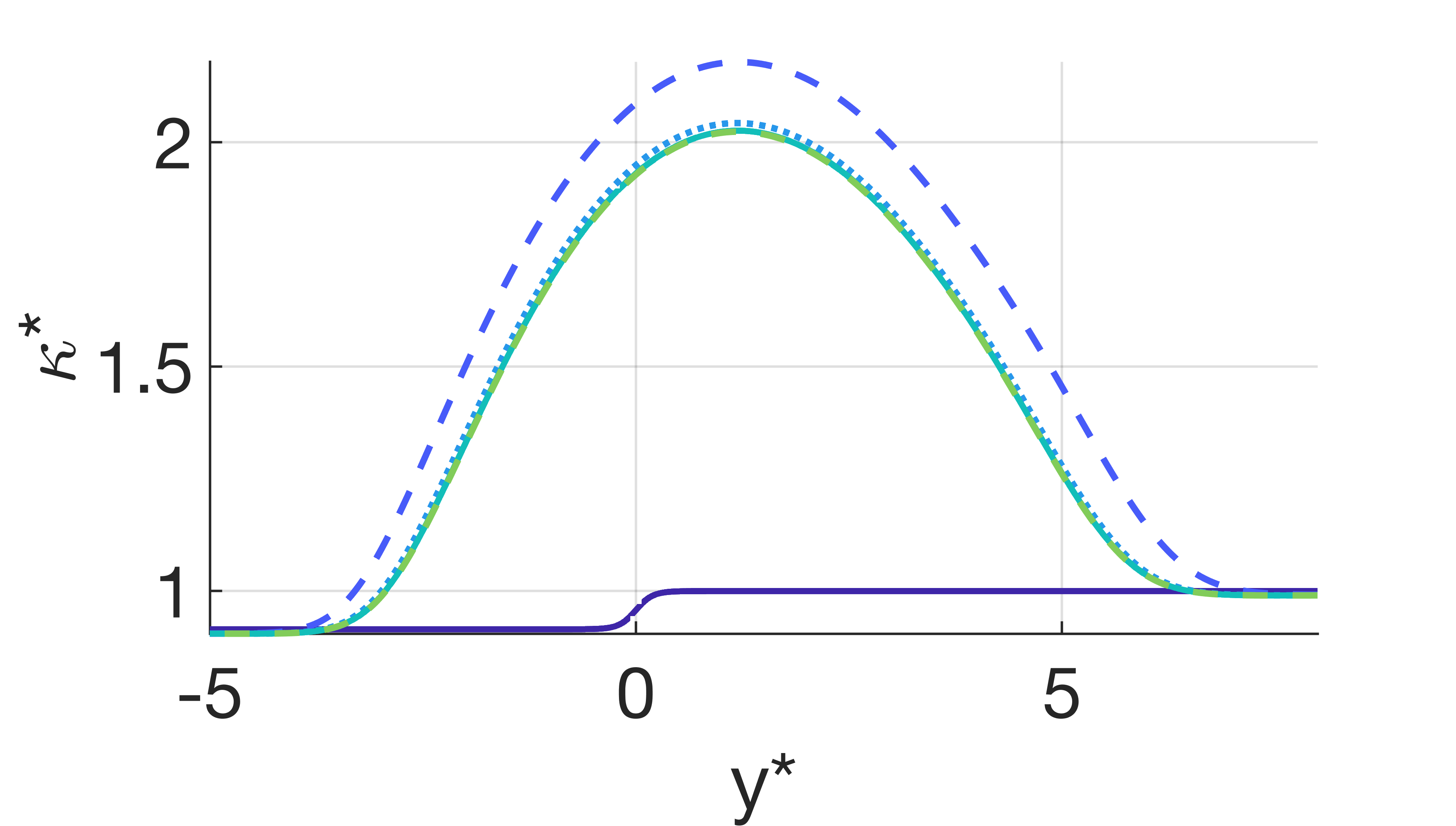}
         \caption{\(\kappa^*(y^*)\)}
         \label{fig:five over x}
     \end{subfigure}
     \begin{subfigure}[b]{0.49\textwidth}
         \centering
         \includegraphics[width=\textwidth]{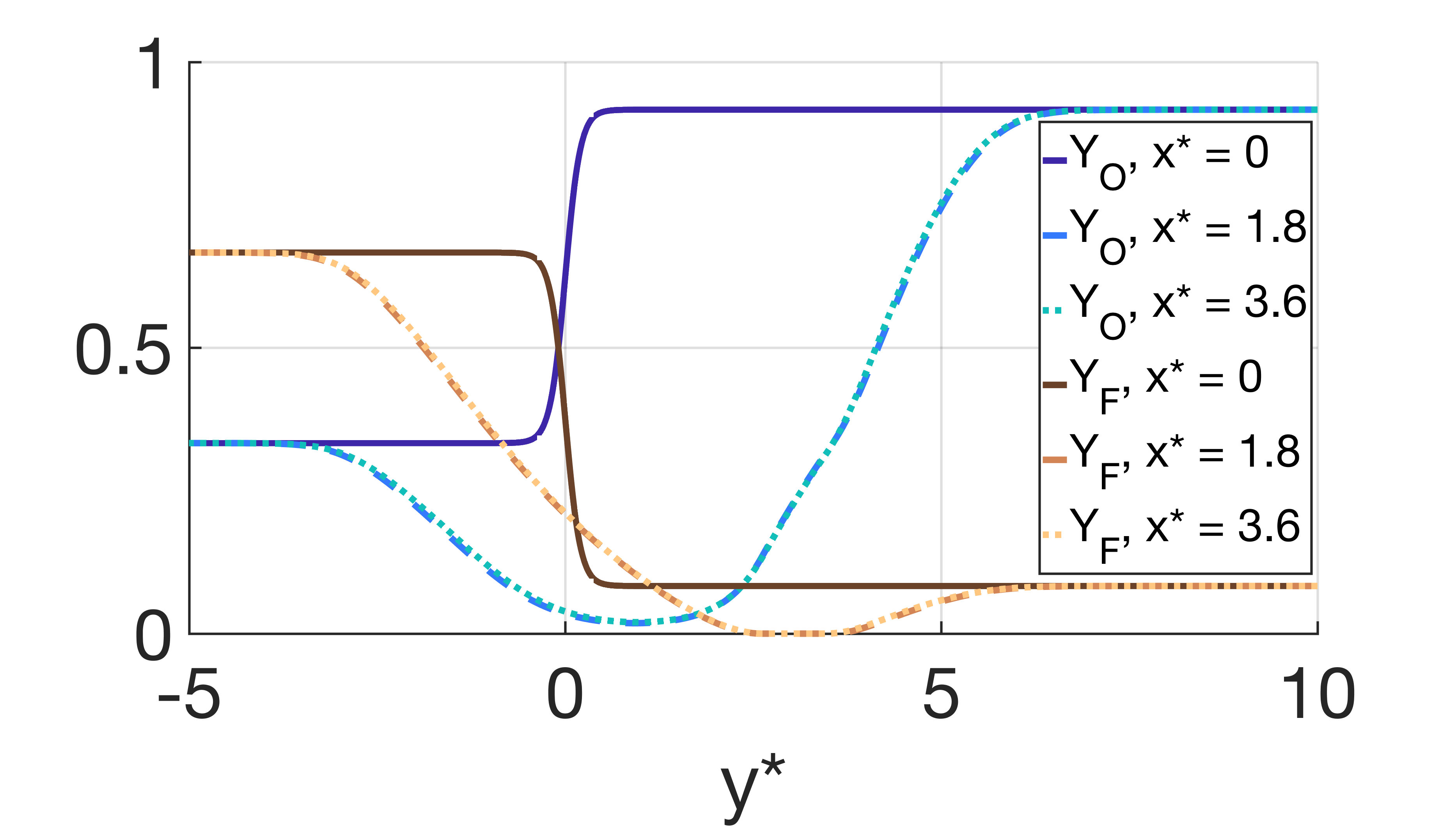}
         \caption{\(Y_O(y^*)\), \(Y_F(y^*)\)}
         \label{fig:five over x}
     \end{subfigure}
     \begin{subfigure}[b]{0.49\textwidth}
         \centering
         \includegraphics[width=\textwidth]{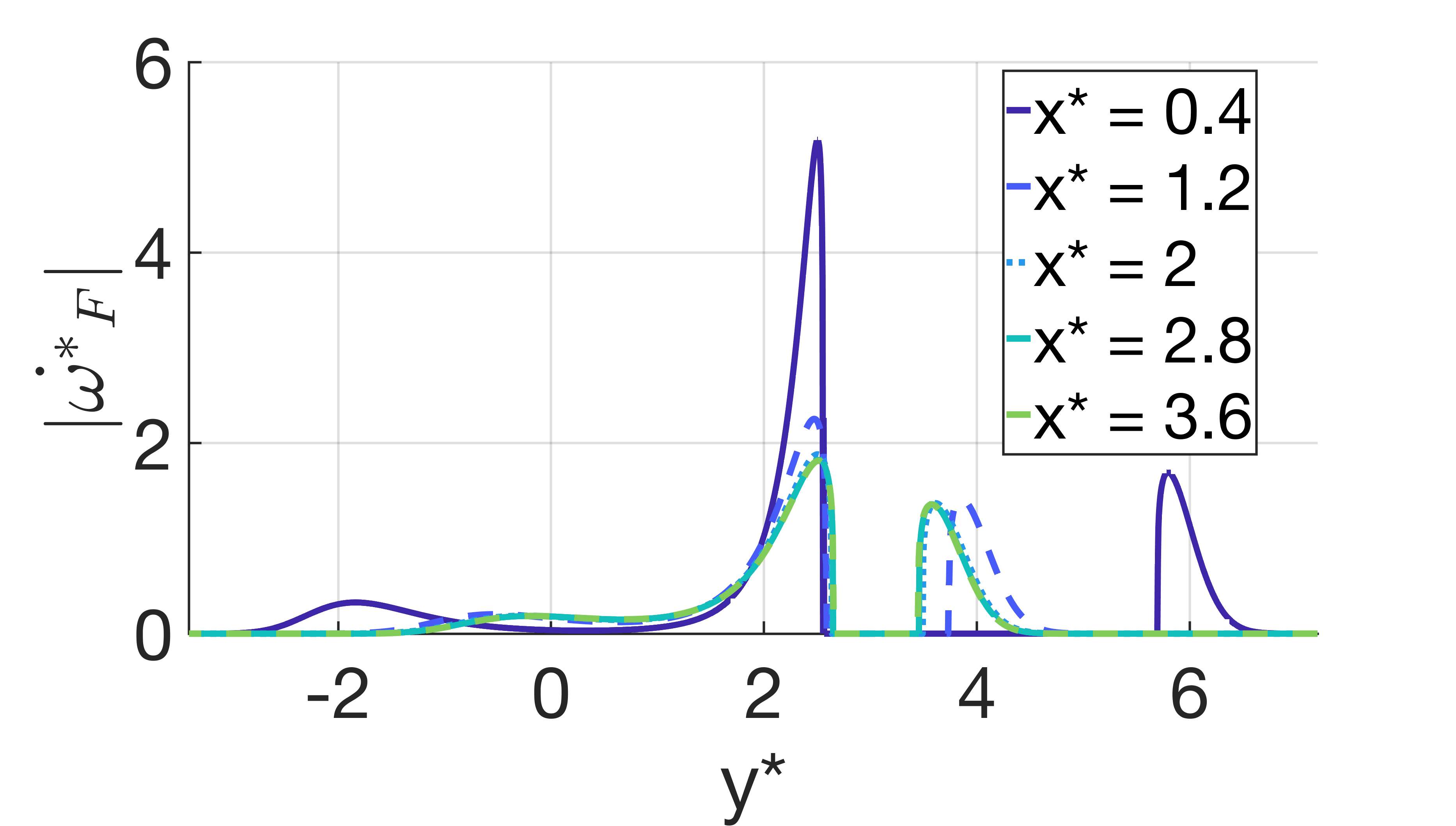}
         \caption{\(\abs{\omega_F^*(y^*)}\)}
         \label{fig:five over x}
     \end{subfigure}
     \caption{Results for reactive multi-flame Case 12 from $x^*=0$ to $x^*=3.6$ where $f^*=1$.}
\end{figure*}

\section{Conclusion}

A fundamental problem for configurations commonly found with turbulent mixing and combustion has been addressed. Specifically, mixing and reaction in a shear layer with vortex stretching has been examined analytically. The downstream asymptotes  yield reduced-order behavior which can be helpful in developing flamelet models for turbulent combustion.

A constantly imposed counterflow has significant effects on the width of a shear layer downstream. Rather than growing as the square root of downstream distance, as found in two-dimensional shear considerations, the shear-layer width will reach a constant width when counterflow is imposed with constant strain rate at all streamwise positions. Far downstream, similarity is achieved as all flow variables, including three velocity components, exhibit one-dimensional behavior varying only with $y^*$ across the shear layer. When this counterflow varies as $1/x^*$, similarity is still observed with $\eta$, which agrees with classical shear-layer theory.

When chemical reactions are introduced, downstream similarity with $y^*$ is still observed for constant strain rate imposed along the layer. For sufficient $y$-direction compression, a flame will extinguish downstream. Here, outward, spanwise flow decreases the residence time of combustion reactants and the thinner mixing layer results in faster heat transfer. Likewise, there is a range of imposed strain rate where a flame does not extinguish and maintains its form similarly downstream. Furthermore, multi-flame structures can become similar downstream, depending only on the $y$-position.

Future studies might use more rigorous multi-step chemical-reaction models. Gasses might be modeled as real for a high-pressure domain. Rather than selecting the linear behavior of the flow in the $z$-direction, a three-dimensional numerical solution might be developed, testing for higher-order effects for the spanwise flow.

The asymptotic downstream quasi-one-dimensional behavior with three meaningful velocity components offers interesting possibilities for future computational and experimental research. This fully developed downstream behavior retains the major features of the flow: shear and vorticity, mixing, applied strain and counterflow, and three velocity components. Researchers often seek to examine new concepts in configurations that include key physics but reduce the dimensionality of the problem. Here, we have qualitative similarity to the fully developed
Poiseuille channel flow where only one-dimension is needed to describe variation in the flow field. However, in this case,  it retains three
meaningful  velocity components and is better suited for studies where
heat and mass transport are important.

\providecommand{\noopsort}[1]{}\providecommand{\singleletter}[1]{#1}%

\end{document}